# RADIATIVE CORRECTIONS TO THE SUBPROCESS OF ANNIHILATION OF QUARK-ANTIQUARK PAIR OF PROMPT PHOTON PRODUCTION IN PROTON-PROTON COLLISIONS AT NICA ENERGIES $\sqrt{s}$= 10 GEV

**M.R.Alizada[1], A.I.Ahmadov[1,2]**

[1]Department of Theoretical Physics, Baku State University
Z.Khalilov 33, Az-1148 Baku, Azerbaijan, mohsunalizade@gmail.com
[2]Department of Theoretical Physics, Institute of Physical Problems,
Baku State University Z.Khalilov 33, Az-1148 Baku, Azerbaijan

## ABSTRACT

In this work, a comprehensive theoretical and numerical study of the quantum-chromodynamic and quantum-electrodynamic radiative corrections: 1. $q\bar{q} \to q\bar{q}\gamma$, 2. $q\bar{q} \to gg\gamma$, and 3. $q\bar{q} \to g\gamma\gamma$ to the annihilation of a quark-antiquark pair $q\bar{q} \to g\gamma$ of prompt photon production in proton-proton collisions at NICA energies $\sqrt{s}$ = 10 GeV was performed. The differential cross-sections and double spin asymmetries $A_{LL}$ of the radiative corrections on the sum of energies of colliding protons $\sqrt{s}$, transverse momentum $p_T$, cosine of scattering angle $Cos(\theta)$, rapidity y of photon and $x_T$were investigated analytically using the FeynCalc.

Radiative corrections are significant at low collision energies of protons $\sqrt{s}$ and small transverse momentum of photons $p_T$. It was shown that radiative corrections to the annihilation of a quark-antiquark pair process satisfy the following relations:

$$2.\ \sigma_{\sqrt{s}}(q\bar{q} \to gg\gamma) > 1.\ \sigma_{\sqrt{s}}(q\bar{q} \to q\bar{q}\gamma) > 3.\ \sigma_{\sqrt{s}}(q\bar{q} \to g\gamma\gamma)$$

$$1.\ \frac{d\sigma(q\bar{q}\to q\bar{q}\gamma)}{dp_T^2} > 3.\ \frac{d\sigma(q\bar{q}\to g\gamma\gamma)}{dp_T^2} > 2.\ \frac{d\sigma(q\bar{q}\to gg\gamma)}{dp_T^2}.$$

The polarization of colliding protons has almost the same effect on processes 1. $q\bar{q} \to q\bar{q}\gamma$, 2. $q\bar{q} \to gg\gamma$, and 3. $q\bar{q} \to g\gamma\gamma$ and in this case above indicated relation is also satisfied.

Numerical modulations of prompt photon production in radiative corrections 1. $q\bar{q} \to q\bar{q}\gamma$, 2. $q\bar{q} \to gg\gamma$ and 3. $q\bar{q} \to g\gamma\gamma$ were performed using the PYTHIA 8.316 event generator. A direct comparison between strict leading order analytical calculations and PYTHIA 8.316 modulations revealed excellent agreement in the central kinematic regions.

## 1. INTRODUCTION

Photons are one of the main products in high energy proton-proton collisions. Photons play a special role in the study of high-energy hadron and nuclear interactions, as they play the role of penetrating probes. Being color neutral, they have a rather long free run not only in very dense hadron matter, but also in the environment of deconfinmed quarks and gluons, quark-gluon plasma (QGP). This ensures that if they are created at any time, they will exit the interaction area unchanged and be detected. The study of photon radiation is also of interest because, in most hadron interactions, photons fly out as a background to other signals being studied.

Photons produced in the proton-proton collisions with energies in the range from 1.5 GeV to several GeV carry information about the formation of about the distribution of partons in nucleons as they are produced by hard scattering of partons, such as Compton scattering of quark-gluon $qg \to q\gamma$ and annihilation of quark-antiquark pair $q\bar{q} \to g\gamma$, as well as bremsstrahlung of quarks $qq \to qq\gamma$, and the quark-gluon phase. These processes was described by perturbative quantum chromodynamics (pQCD) in the leading order (LO), next to leading order (NLO), and next-to-next-to-leading order (NNLO) [1-3].

Especially significant among them are processes involving antiproton, which give direct access to the distribution functions of valence quarks in the proton without involving additional unknown variables, as, for example, in the case of pion-proton collisions [4]. In proton-antiproton collisions the gluon contribution to attitude is very significant in the high-energy field. Protons and antiprotons under high energies look almost the same, and therefore there is no particular difference that protons collide with protons as in the Large Hadron Collider (LHC) or protons with antiprotons as in the American Tevatron collider [5]. Drell-Yan processes with proton-antiproton $p\bar{p} \to \gamma X$ collisions occupy a special place. Among the many possible Drell-Yan processes, valence antiquark processes in the initial state have an obvious advantage - large cross-sections, compared to Drell-Yan processes with marine annihilating antiquark [6,7].

Calculations in LO are highly dependent on the arbitrary choice of factorization scale $\mu_F$ and renormalization $\mu_R$. Including NLO and NNLO corrections stabilizes the theoretical cross section, making it independent of these artificial parameters. The accurate description of spin observables and differential cross sections at high energies mandates the inclusion of higher-order radiative corrections and logarithmic resummation techniques. Radiative corrections to the production of prompt photons in proton-proton collisions are theoretical predictions of QCD. These corrections can be crucial for accurately describing the production of prompt photons. Radiative corrections can alter the momentum and angle of the observed photon and affect the cross section values. Radiative corrections are typically calculated using methods that include the elimination of infrared divergences and are necessary for comparing experimental results with theoretical predictions [8,9].

The paper [10] provides a comprehensive theoretical study of isolated prompt photon production in hadronic collisions. The authors implement the isolation criteria used by collider experiments into a NLO QCD calculation. This work is pivotal for understanding how experimental photon selection affects theoretical predictions. Precise separation of the "direct" component (where the photon originates from the hard process) and the "fragmentation" component (where the photon is produced from a high-$p_T$ parton). Formal proof of QCD factorization validity for isolated particles, transition from the "small-$R$ approximation" to a full treatment of the isolation cone radius $R$. Proven that isolation effectively reduces the fragmentation component, enhancing the sensitivity to the hard subprocess. Demonstrated that the small-$R$ approximation overestimates the fragmentation contribution by approximately 8–10% for large $R$ values ($\sim 1$). Observed that for very small cones ($R \sim 0.1$), the NLO calculation fails to be physically reliable, leading to results where the isolated cross-section exceeds the inclusive one, thereby signaling the need for higher-order resummation or non-perturbative contributions in that regime.

The article [11] describes a study of the production of inclusive prompt photons in nuclear collisions at the Relativistic Heavy Ion Collider (RHIC) and LHC. The authors use NLO calculations based on nuclear particle distribution functions (nPDFs) to predict modifications in photon production. Their calculations include real radiative corrections along with other higher-order QCD effects. Nuclear modification factors are presented for $d-Au$ collisions at the RHIC and $p-Pb$ collisions at the LHC at different rapidities. The results show that these factors can be used to study the nPDFs over a wide range of Bjorken-x. The authors emphasize that combining data from the RHIC and LHC will allow a more accurate determination of the various existing sets of nPDFs. They also discuss the possible influence of a dense medium in heavy-ion collisions on the production of prompt photons.

The paper [12] presents what appears to be the first calculation of the direct photon production cross section to NNLO accuracy in QCD. The results are compared with ATLAS 8 TeV data. The authors demonstrate that the inclusion of NNLO terms and electroweak corrections provides a highly accurate description of LHC data, significantly reducing previous theoretical discrepancies. Application of the $N$-jettiness slicing procedure to regulate infrared (IR) singularities for a final state involving colored partons at the Born level. Demonstration that combining NNLO QCD with electroweak (EW) Sudakov logarithms is essential for high-energy physics phenomenology.

The paper [13] presents a measurement of the differential cross-section for the production of inclusive isolated (prompt) photons in proton-proton collisions at a center-of-mass energy of 13 TeV. Using data collected by the ATLAS detector at the LHC, the measured cross-sections are compared against NLO and NNLO QCD theoretical predictions. The differential cross-sections were measured over an impressive $E_T^\gamma$ range spanning from 125 GeV up to 2500 GeV. Thanks to the increased statistics, the kinematic region where the measurement is limited by systematic uncertainties (rather than statistical) was extended up to $E_T^\gamma \sim 1$ TeV (an improvement from the previous 600 GeV limit). The NLO pQCD predictions provide an adequate description of the data within the combined experimental and theoretical uncertainties. The NNLO QCD predictions (from NNLOJET) represent a significant step forward, providing an excellent description of the data. Furthermore, NNLO significantly reduces theoretical scale uncertainties by a factor ranging from 2 to 20, confirming the precision of the theoretical models and paving the way for better PDF constraints.

In the article [14], the authors demonstrate several advantages of radiative correction calculations. First, all allowed orders of $\alpha^2\alpha_s$ subprocesses $\gamma q \to \gamma q g$ and $\gamma g \to \gamma q \overline{q}$ can be included; the corresponding calculation for purely hadronic scattering with high $p_T$ is very complex due to the much larger number of subprocesses involved. Second, interactions of different types of corrections can be observed. This provides a potentially useful test of ideas for choosing the optimal factorization scale for deep inelastic processes. Third, the calculation is a pedagogically interesting process. An example of the factorization of hard scattering is the QCD theorem, realized beyond the leading logarithm approximation.

The article [15] presents measurements of prompt photons with $p_T$ >2 GeV/c produced in high-energy proton-proton collisions. It highlights the challenges of collinear pQCD calculations at low $p_T$ and the uncertainties arising from the photon fragmentation function (FF).

The study [16] examines radiative corrections to proton-antiproton and electron-positron annihilations, including virtual and real photon production. The study is relevant for measurements of time-similar form factors planned for the PANDA experiment at FAIR (Facility for Antiproton and Ion Research, GSI, Germany). The paper presents the relevant formulas and develops an autonomous Monte Carlo (MC) integrator based on the calculated radiative cross section.

An analysis of the literature showed that at LHC energies, higher-order radiative corrections are massive. Because $\sqrt{s}$ is so high, the kinematic boundary is nowhere near the typical measured $p_T$ spectrum, meaning threshold log corrections are only relevant at ultra-high $p_T$. Instead, collinear logs from photon fragmentation dominate the radiative corrections if isolation cuts are not tightly applied. The differential cross-section spans many orders of magnitude across a massive $p_T$ range (from tens of GeV up to several TeV). The $K$-factor (the ratio of $\frac{\sigma_{NLO}}{\sigma_{LO}}$) is much larger at the LHC. Because the gluon density inside the proton grows exponentially at small $x$, the Compton scattering channel $qg \to q\gamma$ completely dominates prompt photon production. The annihilation of quark-antiquark pair $q\bar{q} \to g\gamma$ channel drops to a minor fraction (typically $<10\%$) of the total cross-section. Hard bremsstrahlung $q\bar{q} \to g\gamma g$ and new channels opening up at NLO (such as $gg \to q\bar{q}\gamma$ via loops) significantly alter the shape of the differential cross-section. Radiative corrections in the LHC energies indicate a colossal contribution of gluons to the cross-section of prompt photon production in proton-proton collisions. Since antiquarks are rare in proton-proton collisions (they are only "sea quarks"), the actual cross section at the LHC is controlled precisely by the gluon density. Near the production threshold (when all the energy goes into creating the photon and jet), massive emission of soft gluons occurs. Taking these gluons into account (resumming the threshold logarithms, generating the so-called $K$-factor) significantly increases the cross section. Accurate calculation of the corrections made it possible to use photon production to "map" the gluon distribution inside the proton at small and medium momentum fractions $x$ [17-19].

Research into the prompt photons production in proton-proton and proton-antiproton collisions at $\sqrt{s}$=10 GeV energies of colliders: NICA (Nuclotron-based Ion Collider Facility, Russia) and FAIR has its advantages [20,21]. The energy of NICA collider does not allow the production of many elementary particles, which makes it difficult to accurately determine the differential cross-section of prompt photons production. In energies NICA can investigate directly phase transition quark-gluon-plasma and hadronization.

In earlier, we investigated the subprocesses of Compton scattering of quark-gluon $qg \to q\gamma$, annihilation of a quark-antiquark pair $q\bar{q} \to g\gamma$, and bremsstrahlung $qq \to qq\gamma$ of prompt photon production in proton-proton collisions at NICA energies $\sqrt{s}$= 10 GeV without and with taking into account the longitudinal polarization of colliding protons in the LO and NLO approximations [22-24]. It was found that the contributions of the subprocesses of Compton scattering of quark-gluon $qg \to q\gamma$ and annihilation of a quark-antiquark pair $q\bar{q} \to g\gamma$ to the total cross section of the prompt photon production process at an energy in the center-of-mass system of $\sqrt{s}$=10 GeV are comparable in magnitude, more 50%, and $\approx$48% correspondingly, while the contribution of bremsstrahlung $qq \to qq\gamma$ is negligible 0.03%. We have shown that the prompt photons produced in pure electrodynamics Compton scattering $q\gamma \to q\gamma$ constitute 10% of the prompt photons produced in mixed Compton scattering mixed chromo-electrodynamic $qg \to q\gamma$. Contribution of NLO calculations to Compton scattering quark-gluon and annihilation of quark antiquark pair makes up about 8%, 15% of the main process.

Unlike at the LHC energies, the annihilation channel $q\bar{q} \to g\gamma$ is highly competitive with Compton scattering quark-gluon at NICA energies. It accounts for roughly ≈ 48% of the total prompt photon cross section, because valence quarks and antiquarks (from the sea/valence boundary at large $x$) carry a significant fraction of the proton's momentum.

The presented article examines radiative corrections to the annihilation of quark-antiquark pair for an accurate description and calculation of the differential cross section of process of prompt photons production in proton-proton collisions at NICA energies.

## 2. DIFFERENTIAL CROSS SECTION OF THE SUBPROCESSES RADIATIVE CORRECTIONS: 1. $q\bar{q} \to q\bar{q}\gamma$, 2. $q\bar{q} \to gg\gamma$ and 3. $q\bar{q} \to g\gamma\gamma$ TO THE ANNIHILATION OF A QUARK-ANTIQUARK PAIR

### 2.1 Radiative corrections of the 1. $q\bar{q} \to gq\gamma$ 2. $q\bar{q} \to gg\gamma$ and 3. $q\bar{q} \to g\gamma\gamma$ to annihilation of quark-antiquark pair $q\bar{q} \to g\gamma$ at collisions nonpolarized protons

The following subprocesses: 1. $q\bar{q} \to gq\gamma$, 2. $q\bar{q} \to gg\gamma$ and 3. $q\bar{q} \to g\gamma\gamma$ relate to quantum-chromodynamic and quantum-electrodynamic radiative corrections to the annihilation of a quark-antiquark pair $q\bar{q} \to g\gamma$.

The Feynman diagrams and matrix elements of subprocesses 1-3 are presented in the Appendices 1-3.

In the parton level the differential cross sections of subprpcesses are determined by the expression:

$$\frac{d\hat{\sigma}}{d\hat{t}} = \frac{\hat{s}_1}{16\pi\hat{s}^4}|\bar{M}|^2.$$

To calculate the differential cross section at the hadron level, we used the parton distribution function (PDFs) $G_{q_1/h_1}(x_1)$ and $G_{q_2/h_2}(x_2)$ in accordance with [25]. $x_1$ and $x_2$ fractions of the momentum carried away by partons from the total momentum of protons.

The differential cross section at the hadron level is determined as:

$$d\sigma = G_{\frac{q_1}{h_1}}(x_1)G_{\frac{q_2}{h_2}}(x_2)dx_1dx_2\hat{\sigma}, \quad (1)$$

where, $\hat{\sigma} = \int \frac{d\hat{\sigma}}{d\hat{t}} d\hat{t}$ - cross-section of subprocess at the parton level and

$$\frac{d\sigma}{dp_T^2} = \int (\frac{-t}{2p_T^2}) G_{\frac{q_1}{h_1}}(x_1) G_{\frac{q_2}{h_2}}(x_2) dx_1 dx_2 \frac{d\hat{\sigma}}{d\hat{t}} \quad (2)$$

The relation between the Mandelstam invariants for the subprocess - parton level $\hat{s}$, $\hat{t}$ and $\hat{u}$ through the Mandelstam invariants $s$, $t$ и $u$ of the main process $pp \rightarrow \gamma X$ are:

$$\hat{s} = x_1 x_2 s, \; \hat{t} = x_1 t, \; \hat{u} = x_2 u, \; t = -p_T \sqrt{s} e^{-y}, \; u = -p_T \sqrt{s} e^{y}, \quad (3)$$

where, $\sqrt{s}$ - sum of energy of colliding protons, $p_{T,}$ $y$ - transverse momentum and rapidity of prompt photon, correspondingly.

Transition from $\frac{d\hat{\sigma}}{dt}$ to $\frac{d\sigma}{dy}$ and $\frac{d\sigma}{dp_T^2}$ we need to use the Jacobian of the transition from $dt$ to $y$ and $p_T$, which have the following form:

$$dt = -p_T \sqrt{s} e^{-y} dy = -t dy, \quad (4)$$

$$dt = -\sqrt{s} e^{-y} dp_T = \frac{|t|}{2p_T^2} dp_T^2, \quad (5)$$

where, $dp_T^2 = 2p_T dp_T$.

Substituting last obtaining expressions (2) and (3) into expression (1), we obtain:

$$d\sigma = \int (-t) G_{\frac{q_1}{h_1}}(x_1) G_{\frac{q_2}{h_2}}(x_2) dx_1 dx_2 \frac{d\hat{\sigma}}{d\hat{t}} dy, \quad (6)$$

$$\frac{d\sigma}{dy} = \int (-t) G_{\frac{q_1}{h_1}}(x_1) G_{\frac{q_2}{h_2}}(x_2) dx_1 dx_2 \frac{d\hat{\sigma}}{d\hat{t}}. \quad (7)$$

$$\frac{d\sigma}{dp_T^2} = \int (\frac{-t}{2p_T^2}) G_{\frac{q_1}{h_1}}(x_1) G_{\frac{q_2}{h_2}}(x_2) dx_1 dx_2 \frac{d\hat{\sigma}}{d\hat{t}} \quad (8)$$

Similarly obtained $\frac{d\sigma}{dx_T^2}$, and $\frac{d\sigma}{dCos(\theta)}$

$$\frac{d\sigma}{dx_T^2} = \int dx_1 dx_2\, G_{q_1/h_1}(x_1)\, G_{q_2/h_2}(x_2)\, \frac{d\hat{\sigma}}{d\hat{t}}\, \frac{\hat{s}}{4|Cos(\theta)^*|} \tag{9}$$

$$\frac{d\sigma}{dCos(\theta)} = \int dx_1 dx_2\, G_{q_1/h_1}(x_1)\, G_{q_2/h_2}(x_2)\, \frac{d\hat{\sigma}}{d\hat{t}}\, \frac{\hat{s}}{2}. \tag{10}$$

To ensure consistency in the evaluation of PDFs under scale variation, we used LHAPDF version 6.5.5 to access modern PDFs sets with built-in support for scale-dependent evolution [26]. This allows us to study how the variation in scale impacts both the normalization and shape of differential distributions.

### 2.2 Radiative corrections of the 1. $q\bar{q} \to gq\gamma$ 2. $q\bar{q} \to gg\gamma$ and 3. $q\bar{q} \to g\gamma\gamma$ to annihilation of quark-antiquark pair $q\bar{q} \to g\gamma$ at collisions longitudinally polarized protons

The radiative corrections: 1. $q\bar{q} \to g\gamma\gamma$, 2. $q\bar{q} \to gg\gamma$ and 3. $q\bar{q} \to q\bar{q}\gamma$ to the annihilation of quark-antiquark pair $q\bar{q} \to g\gamma$ subprocess taking into account the longitudinal polarization of the colliding protons have analogous Feynman diagrams, matrix elements, and Mandelstam invariants to the process without taking into account the polarization of the colliding protons.

The longitudinal polarization of colliding particles is taken into account as [27]:

$$U(p_1)\bar{U}(p_1) = \frac{1}{2}(1 - \lambda_1\gamma_5)(\hat{p}_1 + m_1),$$

$$v(p_2)\bar{v}(p_2) = \frac{1}{2}\hat{p}_2(1 + \lambda_2\gamma_5)$$

where, $\lambda_1$, and $\lambda_2$ are the helicity of colliding protons, $m_1$, $m_2$ are the masses of the quark and antiquark, $p_1$ , $p_2$ are the momenta of the quark and antiquark, correspondingly.

In the calculations of the differential cross section of processes with polarized protons, the polarized PDFs were used. The polarized PDFs have the following form [28]:

$$\Delta u_v(x, Q_0^2) = x^{-0.441}(1-x)^{3.96}(0.928 + 0.149x^{0.5} - 1.141x + 11.612x^{1.5}),$$

$$\Delta d_v(x, Q_0^2) = x^{-0.665}(1-x)^{4.46}(-0.038 - 0.43x^{0.5} - 5.260x + 8.443x^{1.5}),$$

$$\Delta G(x, Q_0^2) = x^{-1.17}(1-x)^{5.33}(0.03 - 1.71x^{0.5} + 3.01x + 43.5x^{1.5}).$$

### 2.3 Modeling of radiative corrections of the 1. $q\bar{q} \to gq\gamma$ 2. $q\bar{q} \to gg\gamma$ and 3. $q\bar{q} \to g\gamma\gamma$ to annihilation of quark-antiquark pair $q\bar{q} \to g\gamma$ in PYHIA 8.316

We used the PYTHIA 8.316 Monte Carlo (MC) event generator for numerically modulate prompt photon production in radiative corrections 1. $q\bar{q} \to q\bar{q}\gamma$, 2. $q\bar{q} \to gg\gamma$ and 3. $q\bar{q} \to g\gamma\gamma$ for determination of the dependences of cross section on the sum of energies of colliding protons $\sqrt{s}$, transverse momentum $p_T$, cosine of scattering angle $Cos(\theta)$, rapidity y of photon and $x_T$.

MC modulation details in PYTHIA 8.316 are followings:

The modulation was performed for nonpolarized proton–proton ($pp$) collisions. To obtain statistically reliable differential distributions, $N = 5 \cdot 10^5$ events were generated for a fixed energy, and to scan the energy dependence, $10^5$events were generated for each point at $\sqrt{s}$.

The main parameters and generation conditions specified in the modulation configuration included the following physical settings:

To generate signal events involving photons and QCD partons, the following classes of rigid matrix elements of the leading order (LO) were included:

All processes of prompt photon production (PromptPhoton: all = on).

The main hard QCD processes for jet generation are: $gg \to q\bar{q}$, $q\bar{q} \to gg$ and $qg \to qg$ (HardQCD: gg 2 qqbar = on, HardQCD: qqbar 2 gg = on, HardQCD : qg 2 qg = on).

The built-in set with PDF identifier pSet = 13 (corresponding to the NNPDF 2.3 LO set) was used as the proton structure functions (PDF), which correctly describes the parton densities at the given energies.

A key aspect of this simulation was to study only parton dynamics to ensure proper comparison with the results of calculations in the FeynCalc. Therefore, the following settings were used:

The effects of initial and final parton radiative emission are included (PartonLevel: ISR = on, PartonLevel: FSR = on). This allows us to take into account the effects of logarithm resummation, the appearance of transverse recoil momentum ($k_T$- kick), and multiparton final states (processes $2 \to 3$ and higher).

The hadronization stage was strictly disabled (HadronLevel: all = off). The final state analysis was performed directly on partons (quarks, gluons, and photons) before their fragmentation into colorless hadrons.

To avoid infrared and collinear divergences inherent in the matrix elements of QCD, constraints on the transverse momentum in the parton rest frame were imposed on the generation level of the hard subprocess: $\hat{p}_T$

To construct differential distributions at $\sqrt{s} = 10$ GeV, the phase volume was limited to the range of $\hat{p}_T^{\min} = 1.0$ GeV/c and $\hat{p}_T^{\max} = 5.0$ GeV/c.

To regularize divergences in the low-momentum limit, the pTHatMinDiverge parameter was set to 0.5 GeV/c. When scanning the total cross section from energy 2.0 to 10.0 GeV, the generation threshold was lowered to $\hat{p}_T^{\min} = 0.5$ GeV/c.

Since in PYTHIA the final state becomes multiparticle after the processing of parton showers, topological selection at the parton level was used to separate the total cross section into the contributions of the three subprocesses under study.

From all stable particles of the final state, a leading photon (a photon with the maximum transverse momentum $p_T$ ) was selected, from which the kinematic variables were reconstructed: $p_T$, $Cos\theta$, rapidity $y$ and $x_T = 2p_T/\sqrt{s}$.

The events were then classified according to the composition of the parton residue:

Process $q\bar{q} \to g\gamma\gamma$: events containing at least two photons ($N_\gamma \geq 2$) and at least one gluon ($N_g \geq 1$).

Process $q\bar{q} \to gg\gamma$: events containing exactly one photon ($N_\gamma = 1$) and at least two gluons ($N_g \geq 2$).

Process $q\bar{q} \to q\bar{q}\gamma$: events containing exactly one photon ($N_\gamma = 1$) and at least two quarks/antiquarks ($N_q \geq 2$, $|ID| \in [1,5]$).

Differential cross-sections were calculated by normalizing the number of selected events in each kinematic bin to the total generated cross-section of subprocesses (info. sigmaGen ()).

2.4 Comparison of cross-section calculated in FeynCalc and modulated in PYTHIA 8.316

A comparison was made of the calculated differential cross sections of the radiative correction subprocesses 1. $q\bar{q} \to q\bar{q}\gamma$, 2. $q\bar{q} \to gg\gamma$ and 3. $q\bar{q} \to g\gamma\gamma$ to the annihilation of quark-antiquark pairs of prompt photon production in collisions of nonpolarized and longitudinally polarized colliding protons, as well as the modulated differential cross sections in PYTHIA 8.316 by studying the dependencies of the following ratios:

1. $R = \frac{d\sigma(q\bar{q}\to g\gamma\gamma)+d\sigma(q\bar{q}\to gg\gamma)+d\sigma(q\bar{q}\to q\bar{q}\gamma)}{d\sigma_{LO}(q\bar{q}\to g\gamma)}$;
2. $R = \frac{d\sigma_{POL}(q\bar{q}\to g\gamma\gamma)+d\sigma_{POL}(q\bar{q}\to gg\gamma)+d\sigma_{POL}(q\bar{q}\to q\bar{q}\gamma)}{d\sigma_{LO,POL}(q\bar{q}\to g\gamma)}$;
3. a) $R = \frac{d\sigma(q\bar{q}\to g\gamma\gamma)}{d\sigma(q\bar{q}\to g\gamma\gamma)+d\sigma(q\bar{q}\to gg\gamma)+d\sigma(q\bar{q}\to q\bar{q}\gamma)}$,

   b) $R = \frac{d\sigma(q\bar{q}\to gg\gamma)}{d\sigma(q\bar{q}\to g\gamma\gamma)+d\sigma(q\bar{q}\to gg\gamma)+d\sigma(q\bar{q}\to q\bar{q}\gamma)}$

   c) $R = \frac{d\sigma(q\bar{q}\to q\bar{q}\gamma)}{d\sigma(q\bar{q}\to g\gamma\gamma)+d\sigma(q\bar{q}\to gg\gamma)+d\sigma(q\bar{q}\to q\bar{q}\gamma)}$;
4. a) $R = \frac{d\sigma_{POL}(q\bar{q}\to g\gamma\gamma)}{d\sigma_{POL}(q\bar{q}\to g\gamma\gamma)+d\sigma_{POL}(q\bar{q}\to gg\gamma)+d\sigma_{POL}(q\bar{q}\to q\bar{q}\gamma)}$,

   b) $R = \frac{d\sigma_{POL}(q\bar{q}\to gg\gamma)}{d\sigma_{POL}(q\bar{q}\to g\gamma\gamma)+d\sigma_{POL}(q\bar{q}\to gg\gamma)+d\sigma_{POL}(q\bar{q}\to q\bar{q}\gamma)}$,

   c) $R = \frac{d\sigma_{POL}(q\bar{q}\to q\bar{q}\gamma)}{d\sigma_{POL}(q\bar{q}\to g\gamma\gamma)+d\sigma_{POL}(q\bar{q}\to gg\gamma)+d\sigma_{POL}(q\bar{q}\to q\bar{q}\gamma)}$
5. $R = \frac{d\sigma_{POL}(q\bar{q}\to g\gamma\gamma)}{d\sigma(q\bar{q}\to g\gamma\gamma)}$, $R = \frac{d\sigma_{POL}(q\bar{q}\to gg\gamma)}{d\sigma(q\bar{q}\to gg\gamma)}$, and $R = \frac{d\sigma_{POL}(q\bar{q}\to q\bar{q}\gamma)}{d\sigma(q\bar{q}\to q\bar{q}\gamma)}$;
6. $R = \frac{d\sigma_{FeynCalc}}{d\sigma_{PYTHIA}}$

on the sum of the energy of colliding protons $\sqrt{s}$, transverse momentum $p_T$, the cosine of the scattering angle of the produced prompt photons $Cos(\theta)$, rapidity y and parameter $x_T$ at $\sqrt{s}$=10 GeV.

## 3. NUMERICAL RESULTS AND THEIR DISCUSSION

3.1 Differential cross section of subprocesses 1. $q\bar{q} \to q\bar{q}\gamma$, 2. $q\bar{q} \to gg\gamma$, and 3. $q\bar{q} \to q\gamma\gamma$ prompt photon production at collisions of nonpolarized protons

Figure 1(a,b,c,d,e) shows the dependence of the total cross section of the subprocesses 1. $q\bar{q} \to q\bar{q}\gamma$ , 2. $q\bar{q} \to gg\gamma$ , and 3. $q\bar{q} \to g\gamma\gamma$ prompt photon

production on the sum of the energies of the colliding particles $\sqrt{s}$, and dependences of the differential cross section on the transverse momentum of the prompt photons $p_T$, cosine of scattering angle of photon $Cos(\theta)$, rapidity y, and $x_T$ in collisions nonpolarized protons at $\sqrt{s}$=10 GeV.

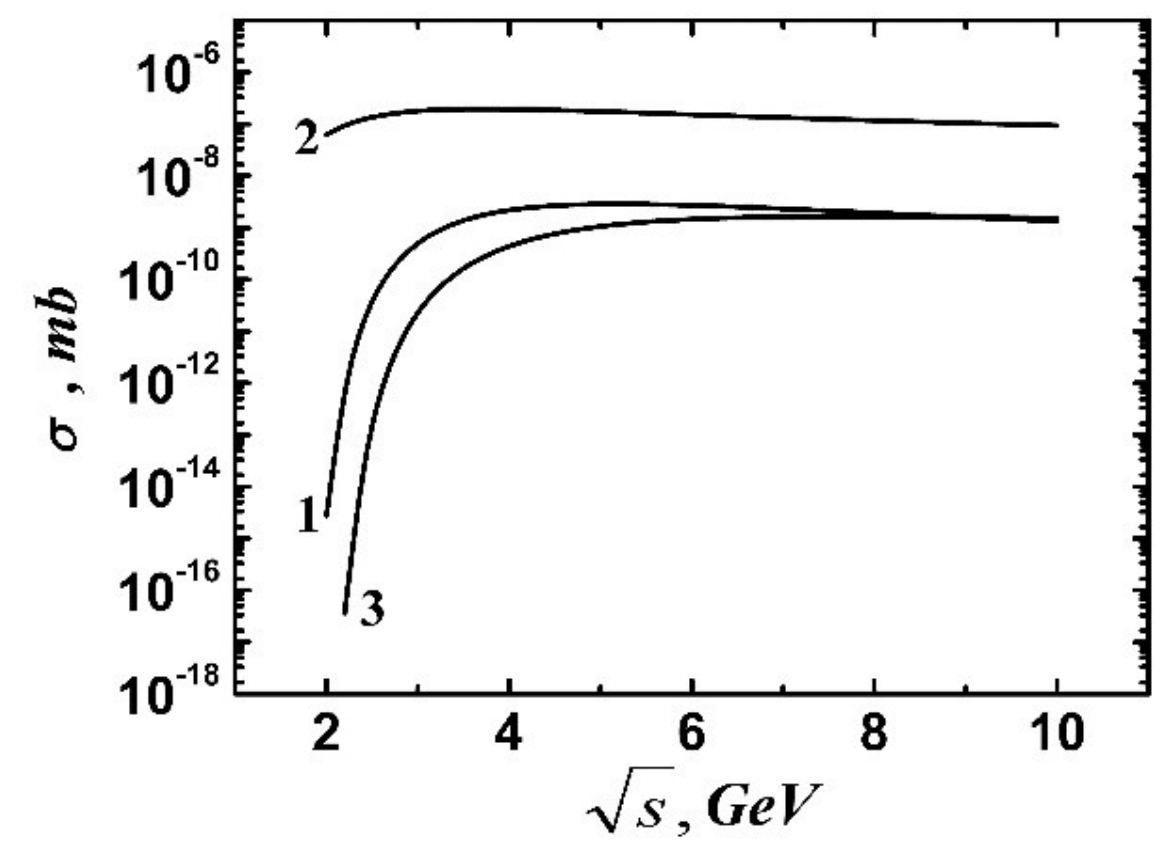

a

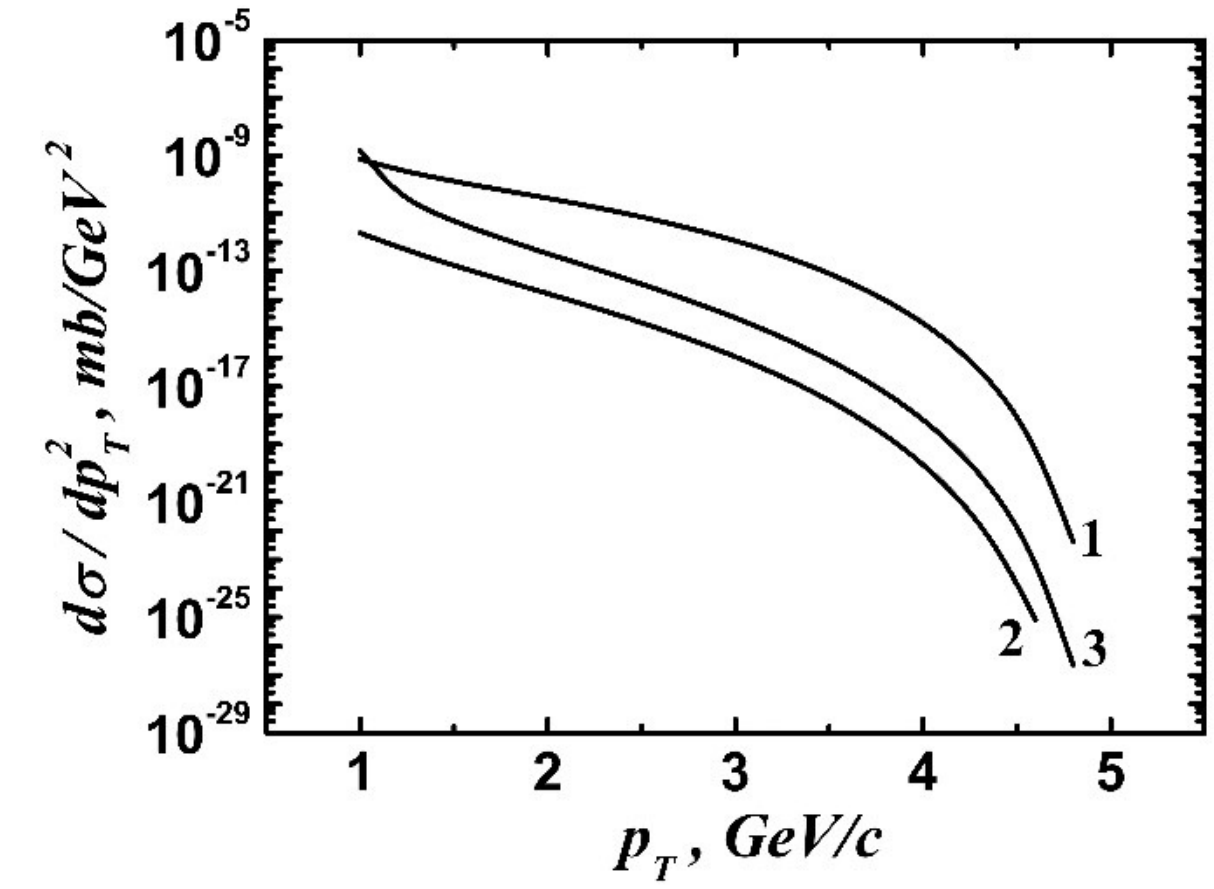

b

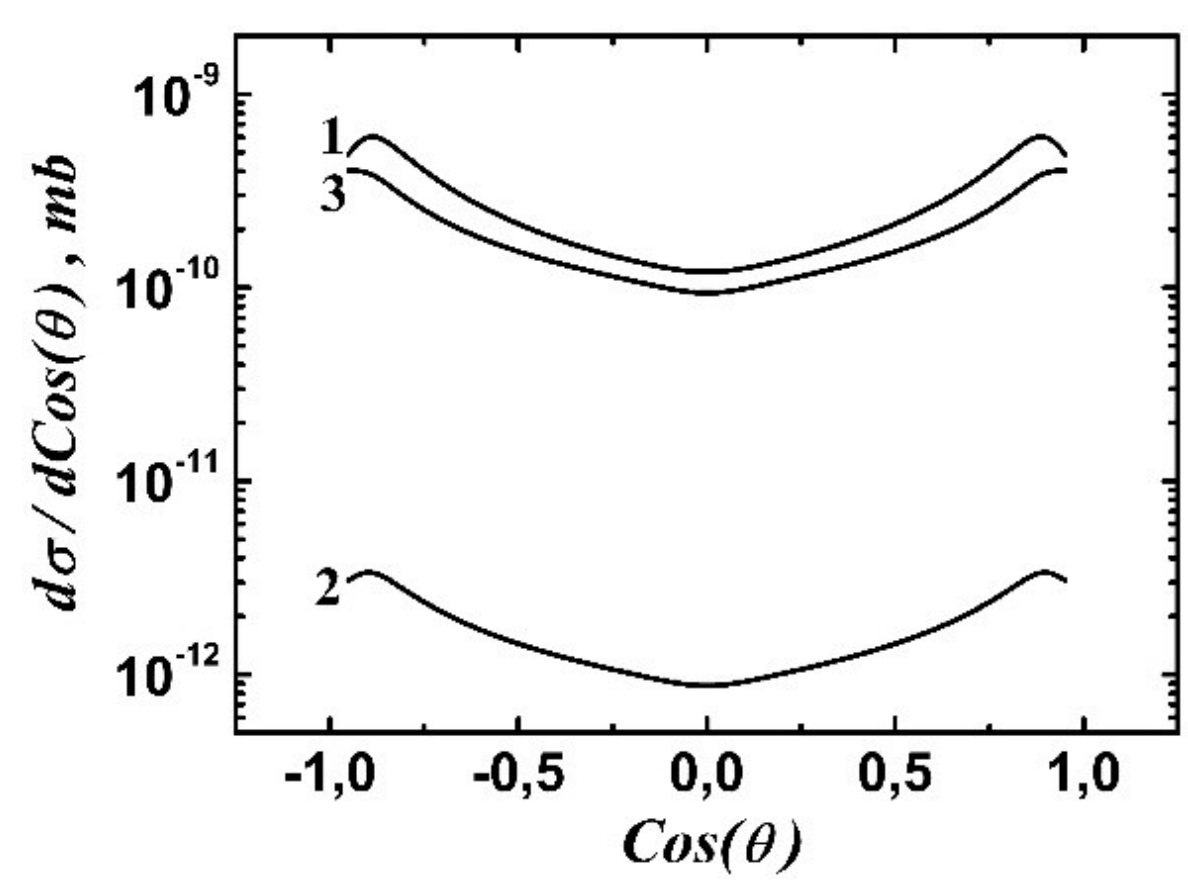

c

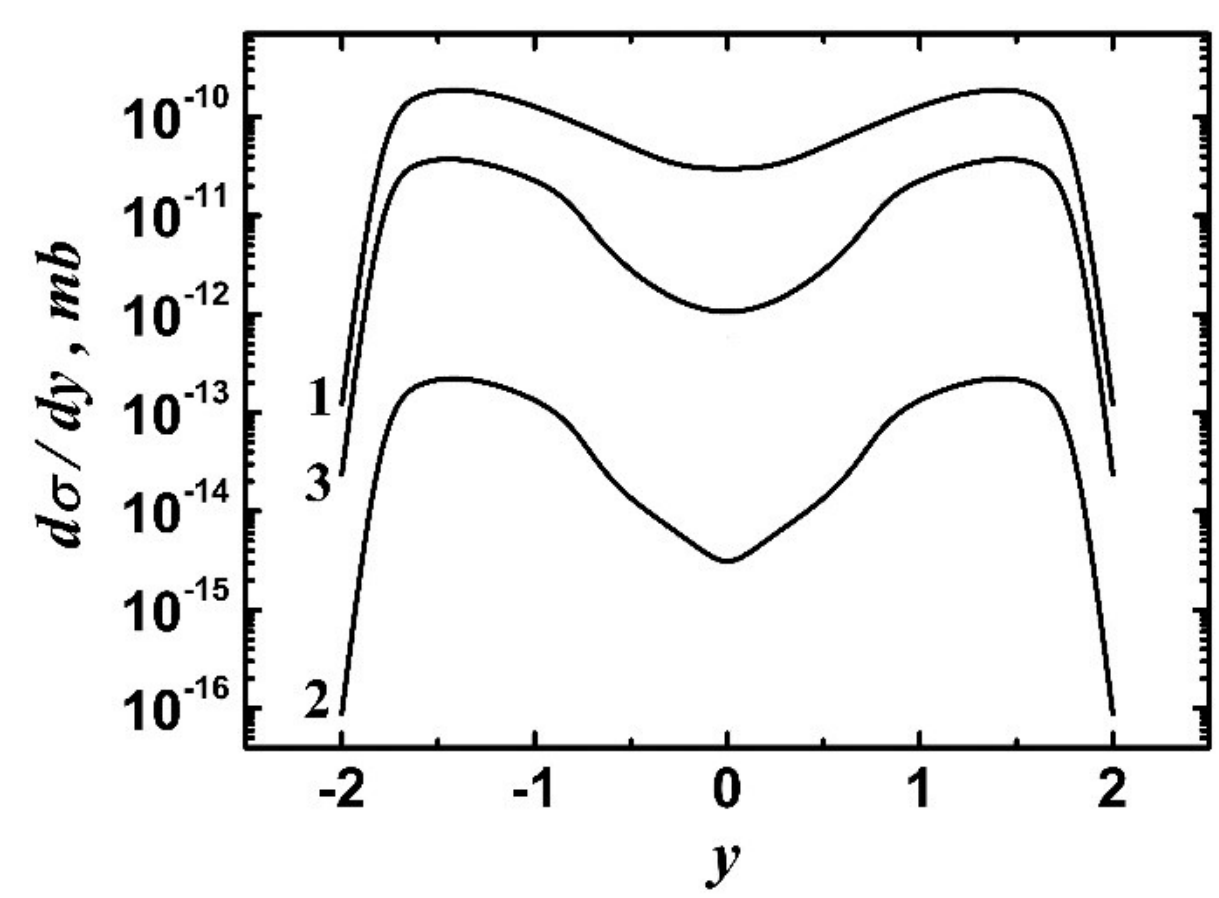

d

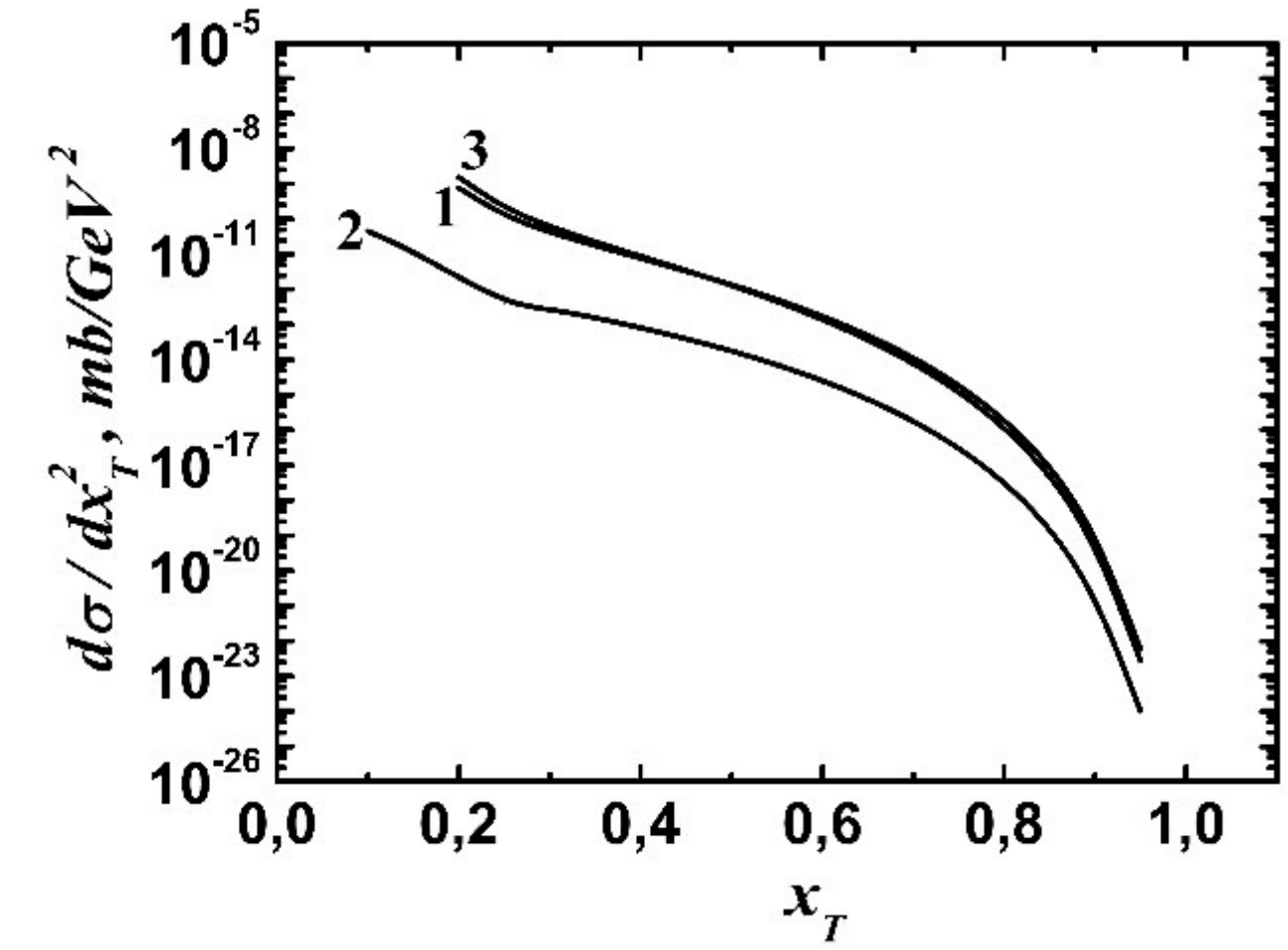

e

Figure 1(a,b,c,d,e). The dependence of the total cross section of the subprocesses 1. $q\bar{q} \to q\bar{q}\gamma$ (curve 1), 2. $q\bar{q} \to gg\gamma$ (curve 2), and 3. $q\bar{q} \to g\gamma\gamma$ (curve 3) on the sum of the energies of the colliding particles $\sqrt{s}$, and the dependences of the differential cross section on the transverse momentum of the prompt photons $p_T$, cosine of scattering angle of photon $Cos(\theta)$, rapidity y, and $x_T$ in collisions nonpolarized protons at $\sqrt{s}$=10 GeV.

The hierarchy of cross sections is determined by the orders of the coupling constants $\alpha_s$ and $\alpha_e$, as well as by the combinatorics of the color factors. The process with the production of two photons (curve 3) predictably has the smallest cross section, as it is suppressed by an additional power of the electromagnetic coupling constant compared to the QCD vertices.

As seen from Figure 1(a), all curves exhibit a sharp nonlinear increase at low values of $\sqrt{s}$ (from the threshold to approximately 4-5 GeV), after which the curves flatten out, reaching an asymptotic plateau. This behavior is classic for hard cross sections. At low energies close to the reaction threshold, the accessible phase volume is extremely small, greatly suppressing the process probability. As $\sqrt{s}$ increases, the phase volume rapidly opens up. The plateau at high energies characterizes the Bjorken scaling regime, where the parton subprocess cross section depends weakly (logarithmically) on energy, due to the property of asymptotic freedom in QCD. This indicates an insignificant contribution of this radiative correction at high energies.

As seen from Figure 1(a), the dependences total cross section on energy colliding protons $\sqrt{s}$ satisfy the following relations:

$$2.\ \sigma_{\sqrt{s}}(q\bar{q} \to gg\gamma) > 1.\ \sigma_{\sqrt{s}}(q\bar{q} \to q\bar{q}\gamma) > 3.\ \sigma_{\sqrt{s}}(q\bar{q} \to g\gamma\gamma) \qquad (11)$$

This is important for theoretical calculations that take into account prompt photons produced at all angles.

A striking disparity in the kinematic behavior of the subprocesses is evident. Curve 2. $q\bar{q} \to gg\gamma$ represents the dominant production channel, exhibiting a relatively high cross-section (approximately $10^{-7}$ mb) with a very weak energy dependence across the entire observed range. This plateau indicates that the kinematic threshold for this specific subprocess is situated well below the 2 GeV mark. Consequently, its cross-section is primarily governed by the high density of the participating partons (typically gluons in the Compton scattering $qg \to \gamma q$ process) rather than phase-space limitations.

Conversely, curves 1. $q\bar{q} \to q\bar{q}\gamma$ and 3. $q\bar{q} \to g\gamma\gamma$ exhibit pronounced threshold behavior at lower energies ($\sqrt{s} < 4$ GeV). The sharp decline in their total cross-sections as $\sqrt{s} \to 2$ GeV signifies a severe phase-space suppression. As the center-of-mass energy approaches the kinematic limit required to produce the final state particles, the available phase volume tends to zero, heavily suppressing the reaction probability. These curves likely correspond to subprocesses with lower initial parton densities or stricter kinematic constraints, such as annihilation of quark-antiquark pair $q\bar{q} \to \gamma g$.

As $\sqrt{s}$ increases beyond 4–6 GeV, the phase space for three subprocesses 1. $q\bar{q} \to q\bar{q}\gamma$ and 3. $q\bar{q} \to g\gamma\gamma$ opens up rapidly, leading to an increase in their cross-sections by several orders of magnitude. At higher energies ($\sqrt{s} > 6$ GeV), all three curves 1. $q\bar{q} \to q\bar{q}\gamma$ and 3. $q\bar{q} \to g\gamma\gamma$ demonstrate an asymptotic behavior, leveling off into a plateau. In this high-energy regime, mass effects and threshold limitations become negligible, and the cross-sections are predominantly determined by the strong coupling constant and the evolution of the parton distribution functions within the colliding protons.

The observed cross-section hierarchy remains intact, with the gluon channel 2. $q\bar{q} \to gg\gamma$ exceeding the quark channel 1. $q\bar{q} \to q\bar{q}\gamma$ and the two-photon channel $q\bar{q} \to g\gamma\gamma$ being heavily suppressed. For collision energies $\sqrt{s} > 4.5$ GeV, the energy dependence of the subprocesses 1. $q\bar{q} \to q\bar{q}\gamma$ and 3. $q\bar{q} \to g\gamma\gamma$ becomes identical.

Figure 1(b) shows a strictly monotonic decline the dependence dufferential cross section on transverse momentum $p_T$. At low $p_T$ values, the decline is smoother, but as $p_T$ approaches 5 GeV/c, the cross section drops off sharply (exponentially). Transverse momentum $p_T$ arises only as a result of hard scattering of partons at large angles. In QCD/QED perturbation theory, the matrix elements of such processes are proportional to $\sim 1/\hat{t}^2$, leading to a steep power-law decay. The sharp cutoff in the region $p_T \to \sqrt{s}/2$ (5 GeV/c) corresponds to the absolute kinematic limit of the phase volume: the produced photon carries away the maximum possible transverse momentum $p_T$, leaving zero “longitudinal” energy for the remaining particles. The contribution of corrections to the main processes is significant at small values of the transverse momentum $p_T$.

As seen from Figure 1(b), the dependences differential cross section on transverse momentum prompt photon $p_T$ satisfy the following relations:

$$1.\ \frac{d\sigma(q\bar{q}\to q\bar{q}\gamma)}{dp_T^2} > 3.\ \frac{d\sigma(q\bar{q}\to g\gamma\gamma)}{dp_T^2} > 2.\ \frac{d\sigma(q\bar{q}\to gg\gamma)}{dp_T^2}. \quad (12)$$

This relation is differs from (11). This is important for experimenters because, it primarily accounts for photons produced at an angle to the protons collision axis, science detecting prompt photons produced along that axis presents a difficult - if not impossible – experimentall challange.

As seen from Figure 1(c) the curves have a characteristic symmetrical $U$-shape. The deep minimum is located at $Cos(\theta) = 0$ ($\theta = 90°$), and at the edges ($Cos(\theta) \to \pm 1$) the cross section increases sharply. The symmetry of the graph reflects the symmetry of the initial state during $q\bar{q}$-annihilation in the center-of-mass system, and at the edges ($Cos(\theta) \to \pm 1$) the cross section increases sharply. The increase in the cross section for "forward" and "backward" scattering ($\theta \to 0$ and $\theta \to 180°$) is directly related to the presence of poles in the $t$- and $u$-channels of the Feynman diagrams. At small scattering angles, the propagators of virtual fermions $\sim 1/(p-k)^2$ tend to zero, leading to collinear divergences (infrared singularities). The minimum at the center means that "hard" scattering strictly in the transverse direction is least likely.

In Figure 1(d), the curves are symmetrical with respect to y=0. A complex "two-humped" structure is observed: local maxima at $|y| \approx 1 - 1.5$ and a small "dip" in the central region ($y = 0$). At the edges ($|y| \approx 2$), the cross section sharply drops to zero.

The curves in Figure 1(e) almost completely repeat the shape of the dependence on $p_T$ (Figure 1(b)). A smooth decline is observed at small $x_T$ and a sharp drop in the differential cross-section value to zero as $x_T \to 1$.

### 3.2 Differential cross section and double spin asymmetry $A_{LL}$ of subprocesses 1. $q\bar{q} \to q\bar{q}\gamma$, 2. $q\bar{q} \to gg\gamma$, and 3. $q\bar{q} \to q\gamma\gamma$ at collisions of longitudinally polarized protons

#### 3.2.1 Differential cross section of 1. $q\bar{q} \to q\bar{q}\gamma$, 2. $q\bar{q} \to gg\gamma$, and 3. $q\bar{q} \to q\gamma\gamma$

The dependence of the total cross section of the subprocesses 1. $q\bar{q} \to q\bar{q}\gamma$, 2. $q\bar{q} \to gg\gamma$, and 3. $q\bar{q} \to g\gamma\gamma$ production prompt photon on the sum of the energies of the colliding particles $\sqrt{s}$, and dependences of the differential cross section on the transverse momentum of the prompt photons $p_T$, cosine of scattering angle of photon $Cos(\theta)$, rapidity y, and $x_T$ in collisions longitidunally polarized protons with polarization order $P_1P_2 = \pm 0.81$ at $\sqrt{s}$=10 GeV has been investigated.

Figure 2(a,b,c,d,e) shows the dependence of the total cross section of the subprocesses 1. $q\bar{q} \to q\bar{q}\gamma$, 2. $q\bar{q} \to gg\gamma$, and 3. $q\bar{q} \to g\gamma\gamma$ production prompt photon on the sum of the energies of the colliding particles $\sqrt{s}$, and dependences of the differential cross section on the transverse momentum of the prompt photons $p_T$, cosine of scattering angle of photon $Cos(\theta)$, rapidity y, and $x_T$ in collisions longitidunally polarized protons with polarization order $P_1P_2 = \pm 0.81$ at $\sqrt{s}$=10 GeV.

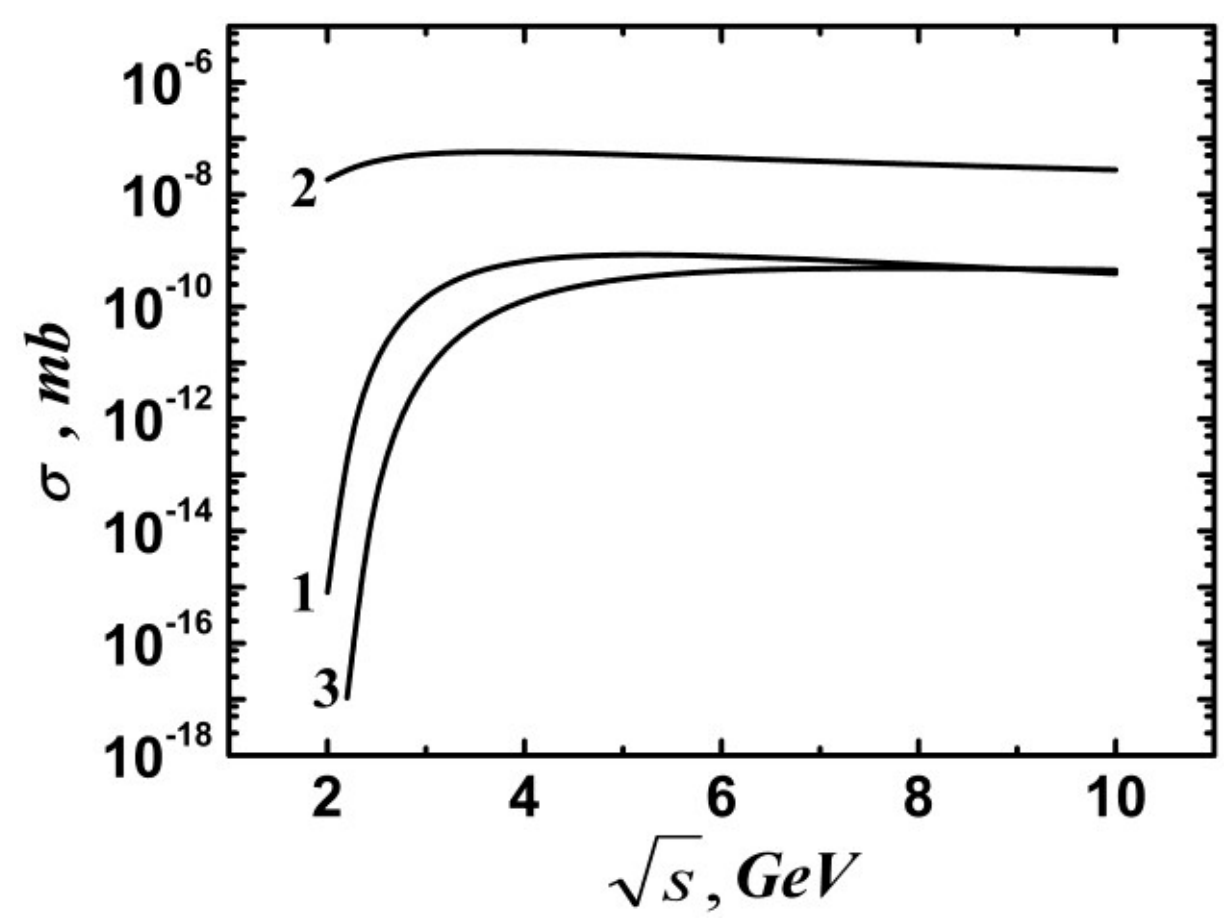

a

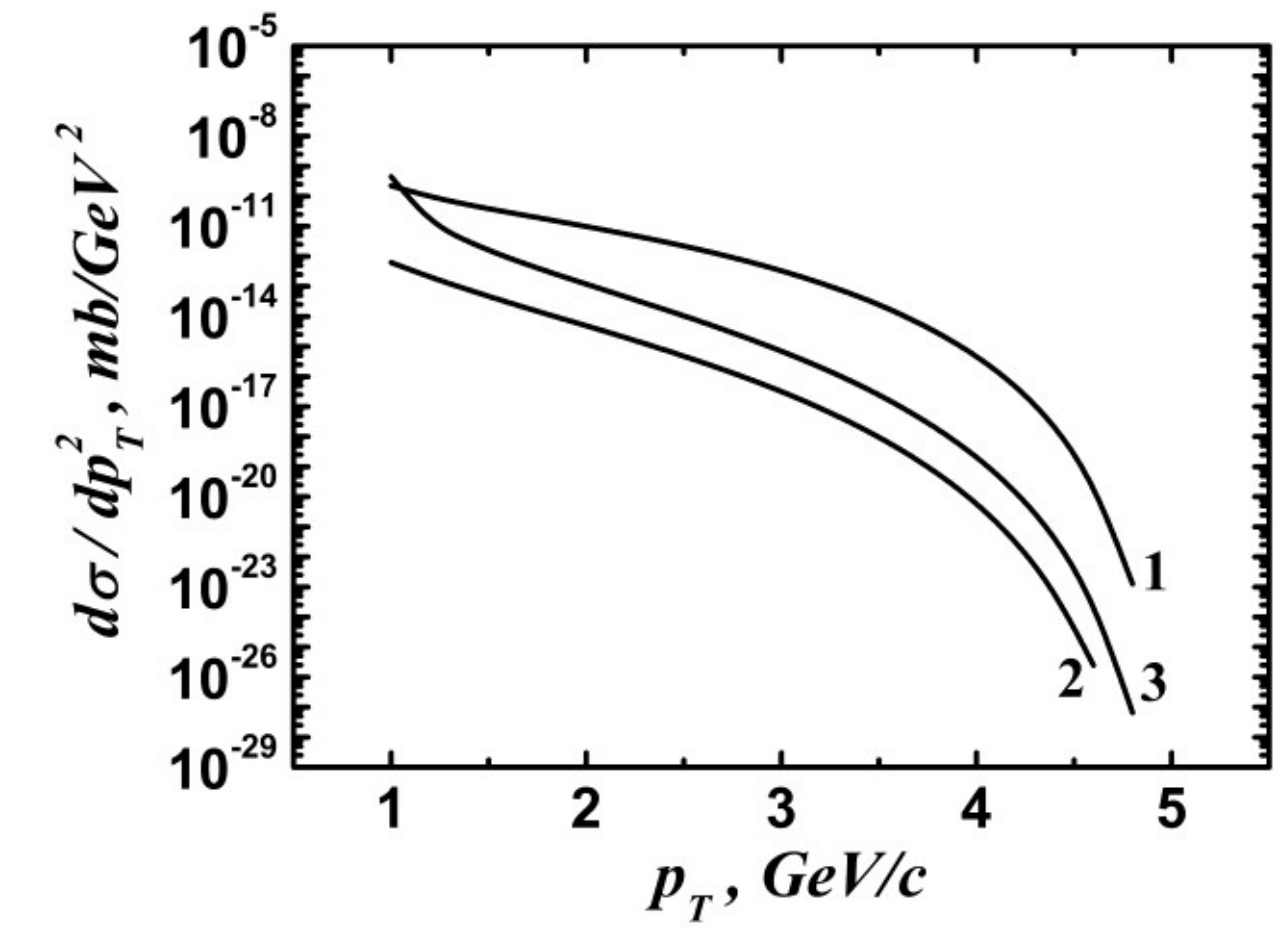

b

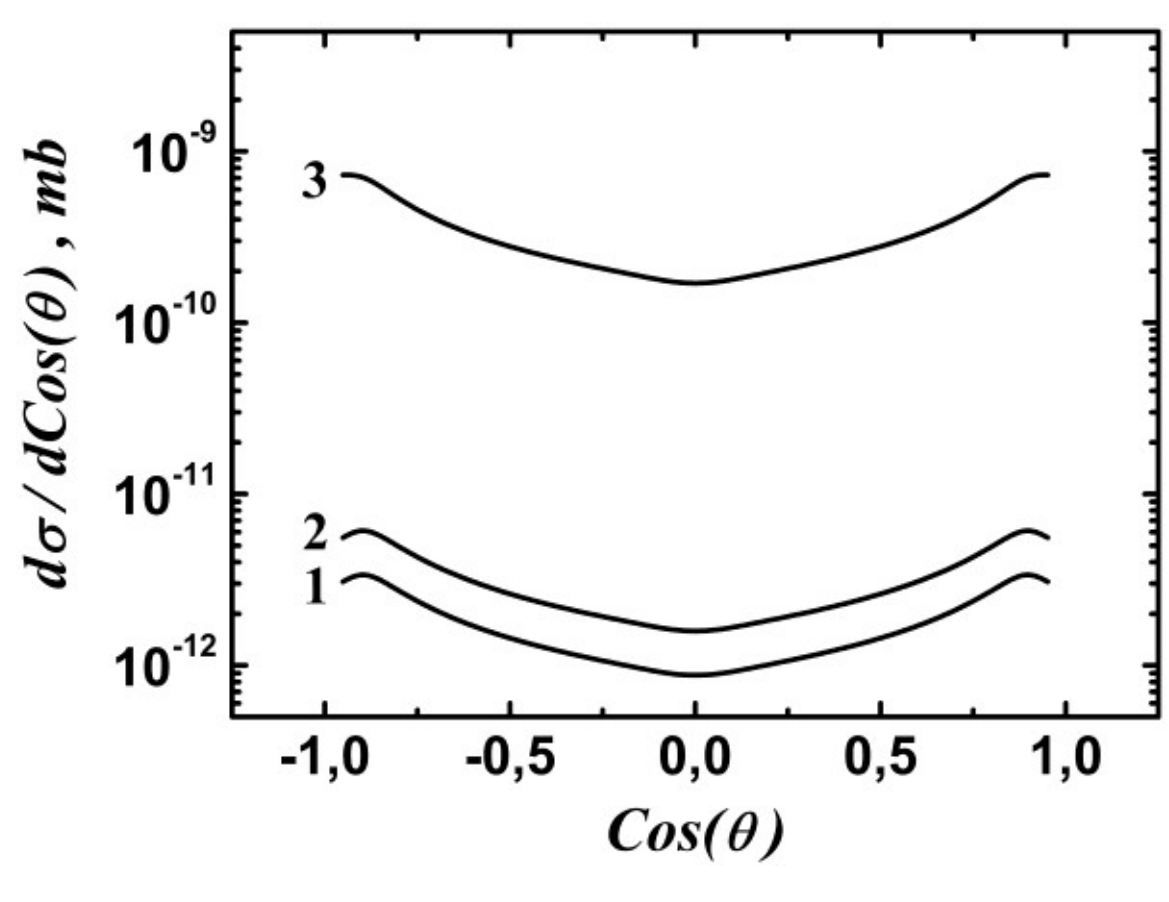

c

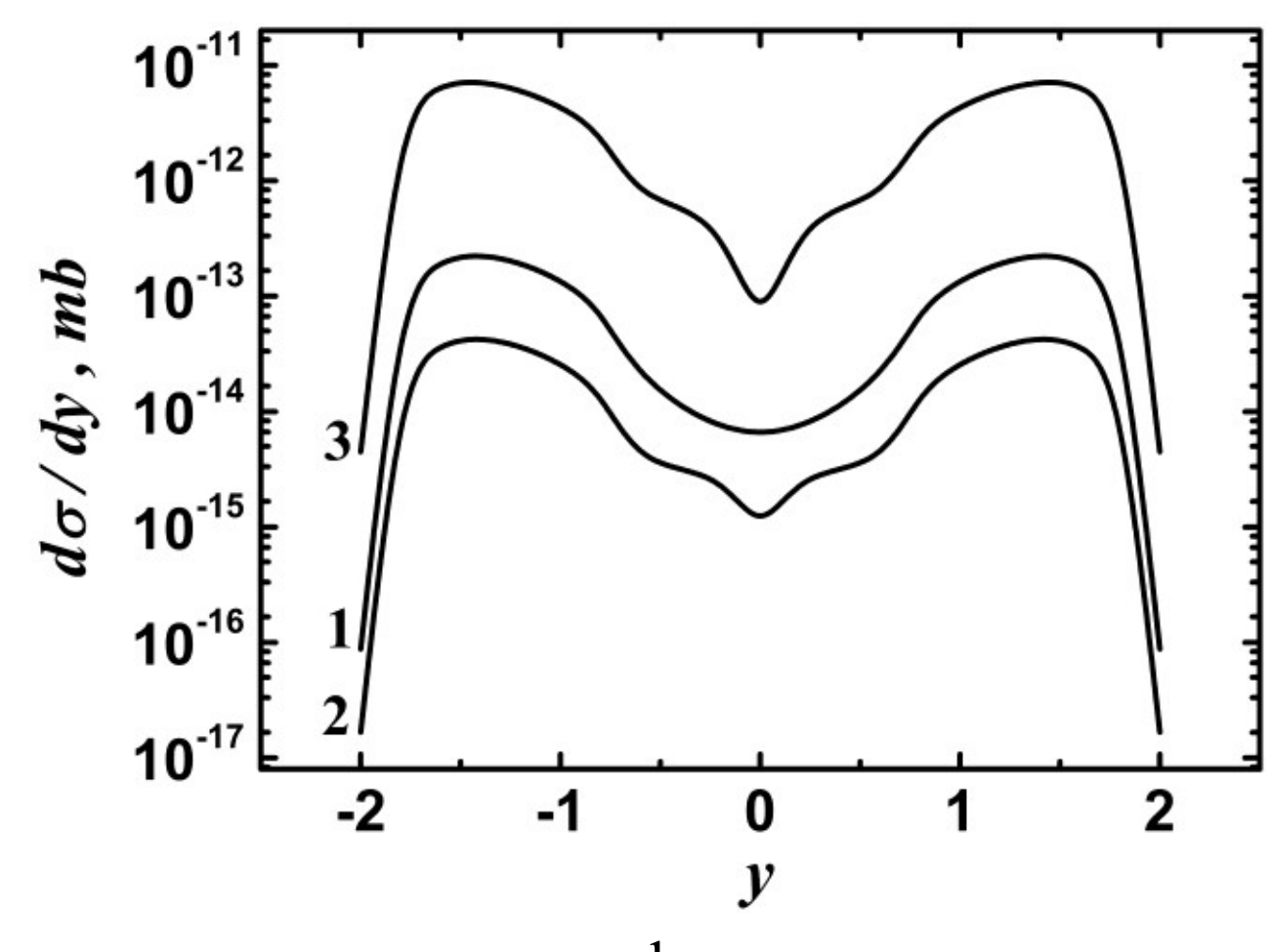

d

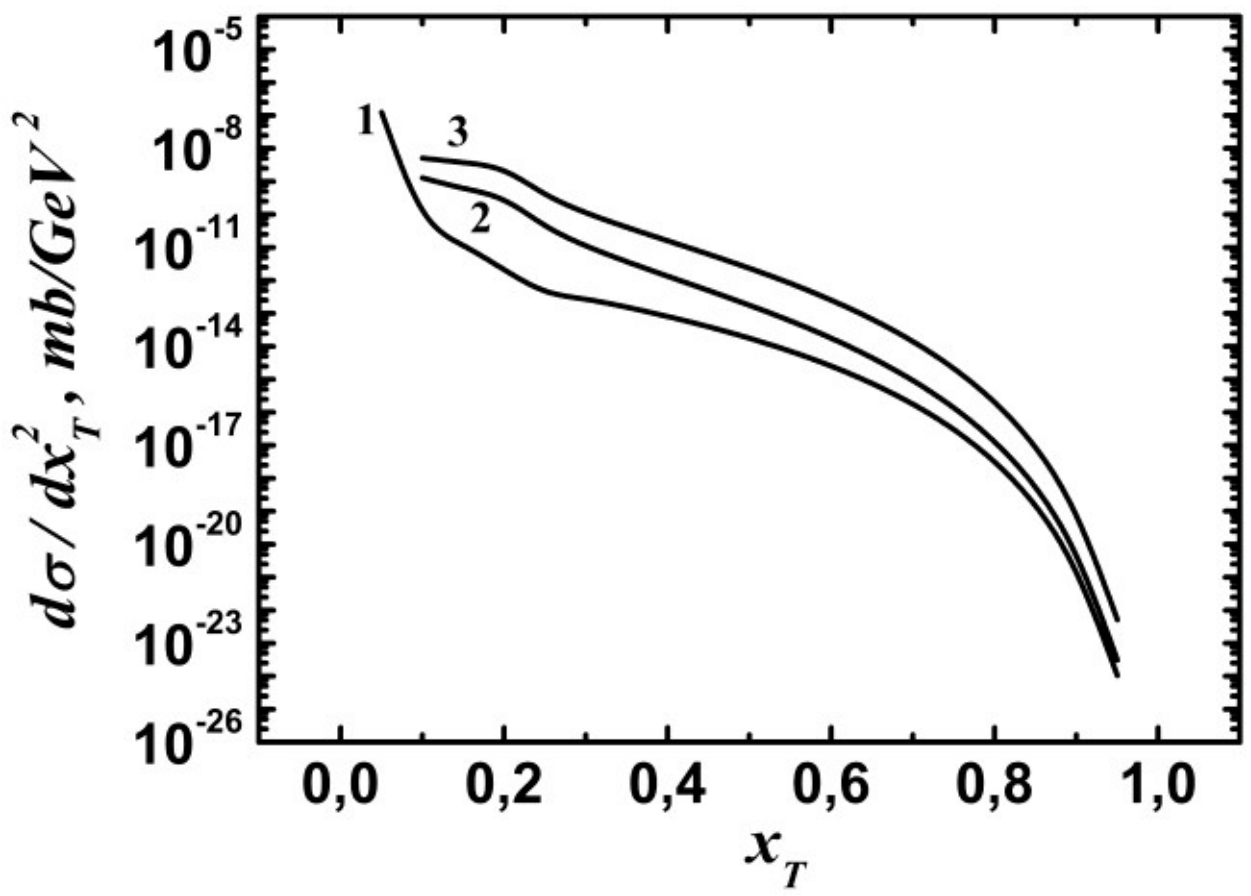

e

Figure 2(a,b,c,d,e). The dependence of the total cross section of the subprocesses 1. $q\bar{q} \to q\bar{q}\gamma$ (curve 1), 2. $q\bar{q} \to gg\gamma$ (curve 2), and 3. $q\bar{q} \to g\gamma\gamma$ (curve 3) on the sum of the energies of the colliding particles $\sqrt{s}$, and the dependences of the differential cross section on the transverse momentum of the prompt photons $p_T$, cosine of scattering angle of photon $Cos(\theta)$, rapidity y, and $x_T$ in collisions longitudinally polarized protons ($P_1P_2$=-0.81) at $\sqrt{s}$=10 GeV.

Similar to the nonpolarized case, a rapid increase from the threshold is observed, followed by a plateau in the high-energy region ($P_1P_2 = -0.81$). At opposite polarization values of the colliding particles, the total and differential cross sections increase. Similar curves were obtained for polarization $P_1P_2$=0.81. These curves lie below the curves obtained for nonpolarized protons. This is explained by the interaction of spins.

The polarization of colliding protons has almost the same effect on processes 1-3. In a collision of polarized protons the following relations (11), (12) are also satisfied.

### 3.2.2 Double spin asymmetry $A_{LL}$ of subprocesses $q\bar{q} \to q\bar{q}\gamma$, $q\bar{q} \to gg\gamma$, and $q\bar{q} \to q\gamma\gamma$

Figure 3 shows the dependence double spin asymmetry $A_{LL}$ of subprocesses 2. $q\bar{q} \to gg\gamma$ and 3. $q\bar{q} \to g\gamma\gamma$ on products of the orders of longitudinal polarization of colliding particles $P_1P_2$.

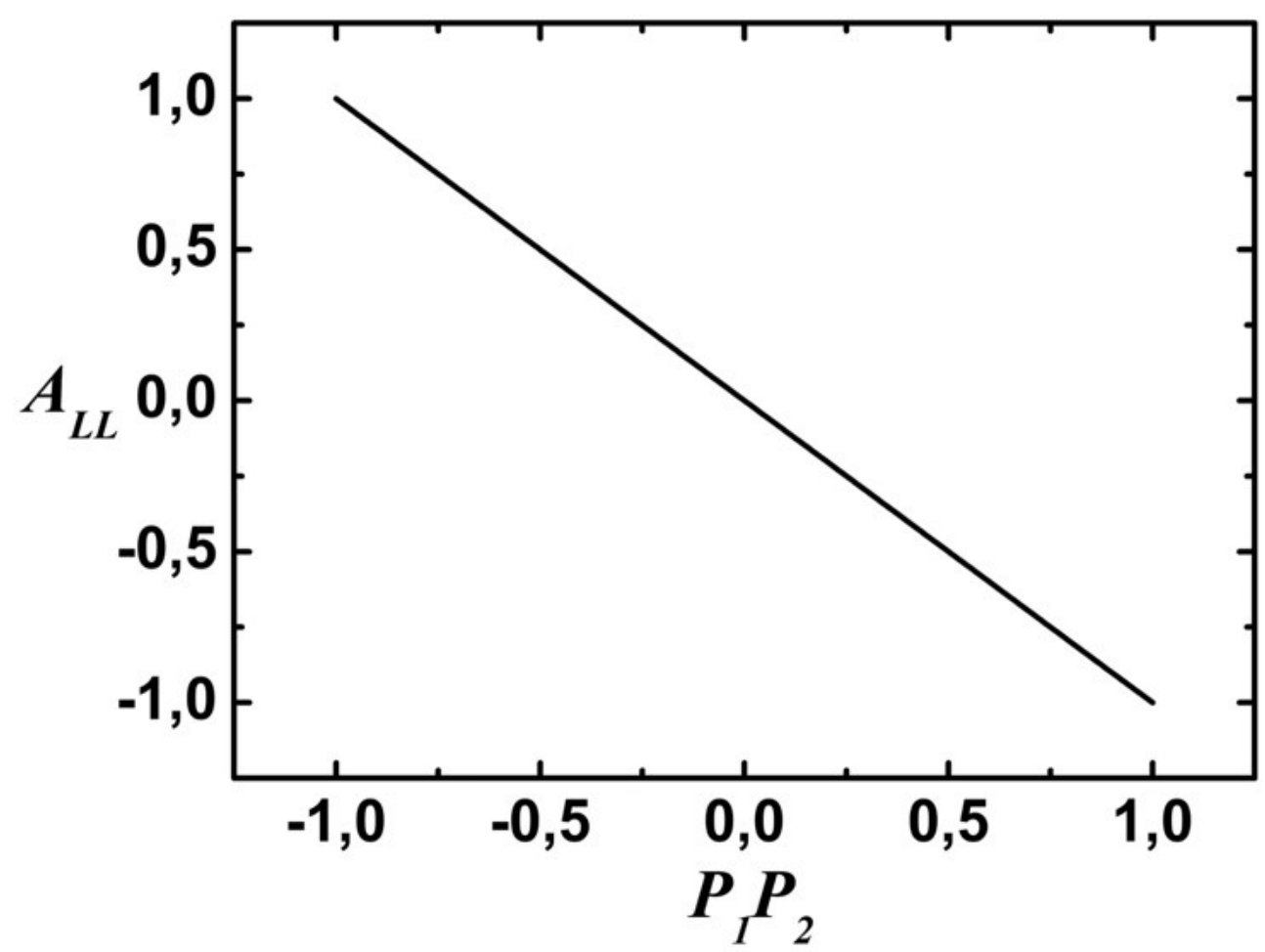

Figure 3. Dependence of the double spin asymmetry $A_{LL}$ of the subprocesses 2. $q\bar{q} \rightarrow gg\gamma$ and 3. $q\bar{q} \rightarrow g\gamma\gamma$ on the product of the orders of longitudinal polarization $P_1P_2$ of the colliding particles.

The perfect linearity of the $A_{LL} \propto P_1P_2$ graph is a direct consequence of the quantum-mechanical description of polarized particle cross sections using density matrices. The total interaction cross section of two spin-$\frac{1}{2}$ fermions is generally decomposed into spin components:

$$\sigma = \sigma_{nonpol}(1 + P_1P_2 \cdot \hat{a}_{LL}),$$

where $\hat{a}_{LL}$ is the partonic (structural) asymmetry of the matrix element itself, depending only on the kinematics, but not on the beam polarization. If one of the beams is unpolarized ($P_i = 0$), then the double spin asymmetry is strictly zero (the line passes through (0,0)).

$$\hat{a}_{LL} = \frac{\sigma_{++} - \sigma_{+-}}{\sigma_{++} + \sigma_{+-}}$$

where, $\sigma_{++}$, and $\sigma_{+-}$ cross section for the interaction of parallel and antiparallel spins, correspondingly.

The value $A_{LL} = -1$ for $P_1P_2 = 1$ means that the cross section for the interaction of parallel spins ($\sigma_{++}$) is zero, and only the interaction of partons with antiparallel spins ($\sigma_{+-}$) remains. For a quark and antiquark to annihilate into a particle with spin 1, their helicities must be opposite ($q_L\bar{q}_R$ or $q_R\bar{q}_L$). If their spins are parallel ($q_R\bar{q}_R$), the total projection spin is 0, and annihilation into a vector boson is forbidden by the law of conservation of angular momentum. This is why $\sigma_{+-}$ dominates, pushing the graph into the negative region.

The value $A_{LL} = 1$ for $P_1P_2 = -1$ physically means that the colliding quark and antiquark are 100% polarized with opposite helicities (one is fully left-handed, the other fully right-handed). Since helicity conservation strictly requires opposite helicities for the reaction to occur, at $P_1P_2 = -1$, the initial state perfectly matches this quantum mechanical requirement. The value $A_{LL} = 1$ at this point signifies that the cross-section of the process reaches its absolute maximum relative to the unpolarized case, because 100% of the particles in the colliding beams are in the "correct" helicity state to annihilate.

Figure 4(a,b,c,d) shows the dependences of the double spin asymmetry $A_{LL}$ for the subprocess $q\bar{q} \to q\bar{q}\gamma$ with the polarization order $P_1P_2 = \pm 0.81$ on the energy of colliding protons $\sqrt{s}$, transverse momentum of the photon $p_T$ rapidity y and parameter $x_T$.

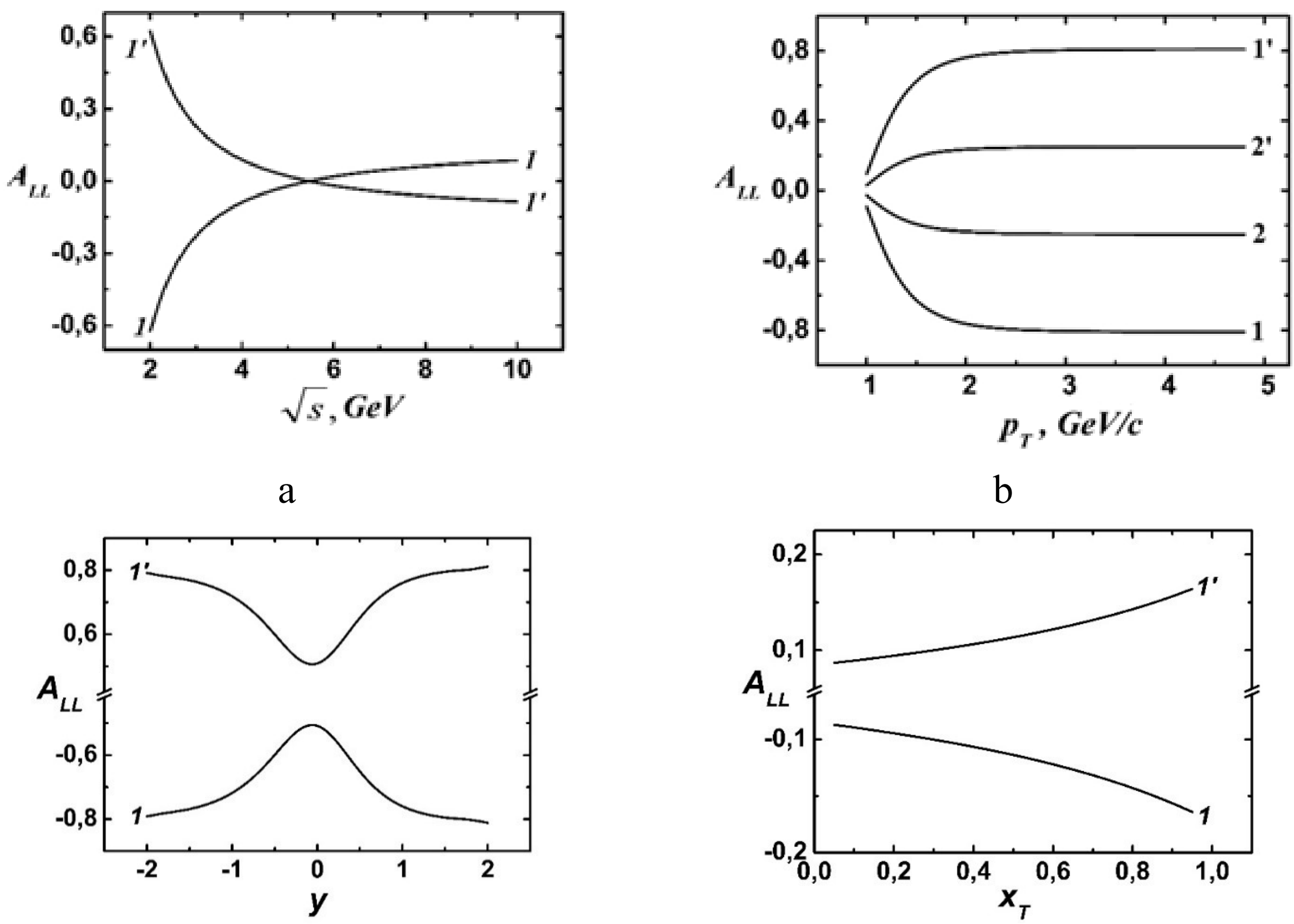

a b

c d

Figure 4(a,b,c,d). Dependence of the $A_{LL}$ subprocess $q\bar{q} \to q\bar{q}\gamma$ on the energy of colliding protons $\sqrt{s}$ for $P_1P_2$=0.81 and -0.81 curve 1 and curve 1', correspondingly (a), dependence $A_{LL}$ on the transverse momentum of the photon $p_T$ for $P_1P_2$=-0.81 and 0.81 (curve 1 and curve 1'), $P_1P_2$=-0.25 and 0.25 (curve 2 and curve 2'), correspondingly (b), dependence $A_{LL}$ on the rapidity y for $P_1P_2$=0.81 and -0.81 curve 1 and curve 1', correspondingly (c) and dependence $A_{LL}$ on the parameter $x_T$ for $P_1P_2$= 0.81 and $P_1P_2$= -0.81 curve 1 and curve 1', correspondingly (d).

As seen from Figure 4(a), at low energies, the double spin asymmetry $A_{LL}$ modulus is maximal; however, with increasing $\sqrt{s}$ (with other parameters held constant), both branches 1 and 1' ($P_1P_2$=0.81 and $P_1P_2$=-0.81 curve 1 and curve 1', correspondingly) asymptotically approach zero. As the collision energy increases, ever smaller fractions of the parton transverse momentum are required to produce a photon with a given $p_T$ ($x \sim 2p_T/\sqrt{s}$). At small x, we enter the region dominated by "sea" quarks and gluons, whose polarization ($\Delta q(x)/q(x)$) is close to zero. Thus, the nucleon's spin information is smeared out in a sea of virtual partons, and the macroscopic asymmetry disappears.

The dependence of the double spin asymmetry $A_{LL}$ on transverse momentum $p_T$ (Figure 4(b)) shows that in the region of small $p_T$, the asymmetry $A_{LL}$ is close to zero. As $p_T$ increases, the curves rapidly diverge, reaching constant asymptotes exactly equal to the given product of polarizations ($P_1P_2$=±0.81 for curves 1, 1' and $P_1P_2$=±0.25 for curves 2, 2').

The dependences of the double spin asymmetry $A_{LL}$ on rapidity $y$ (Figure 4(c)) are $X$-shaped. The maximum asymmetry $A_{LL}$ amplitude is achieved precisely at the center ($y = 0$). As one moves toward the edges of the phase space ($|y| \to 2$), the double spin asymmetry $A_{LL}$ collapses to zero.

The dependence of the double spin asymmetry $A_{LL}$ on $x_T$ (Figure 4(d)) shows a smooth increase in the asymmetry modulus from small values at small $x_T$ to maximum values (determined by the polarization of the beams) at $x_T \to 1$.

3.3 The dependences of differential cross sections of the subprocesses of radiative corrections 1. $q\bar{q} \to q\bar{q}\gamma$, 2. $q\bar{q} \to gg\gamma$, and 3. $q\bar{q} \to g\gamma\gamma$ on the sum of energies of colliding protons $\sqrt{s}$, transverse momentum $p_T$, cosine of scattering angle $Cos(\theta)$, rapidity y of photon and $x_T$ modulated in PYTHIA 8.316.

Figure 5(a,b,c,d,e) shows the dependences of the total and differential unpolarized cross sections for subprocesses: $q\bar{q} \to q\bar{q}\gamma$, $q\bar{q} \to gg\gamma$, and $q\bar{q} \to g\gamma\gamma$ on the sum of energies of colliding protons $\sqrt{s}$, transverse momentum $p_T$, cosine of scattering angle $Cos(\theta)$, rapidity y of photon and $x_T$ in the PYTHIA 8.316 modulation at $\sqrt{s}$ =10 GeV.

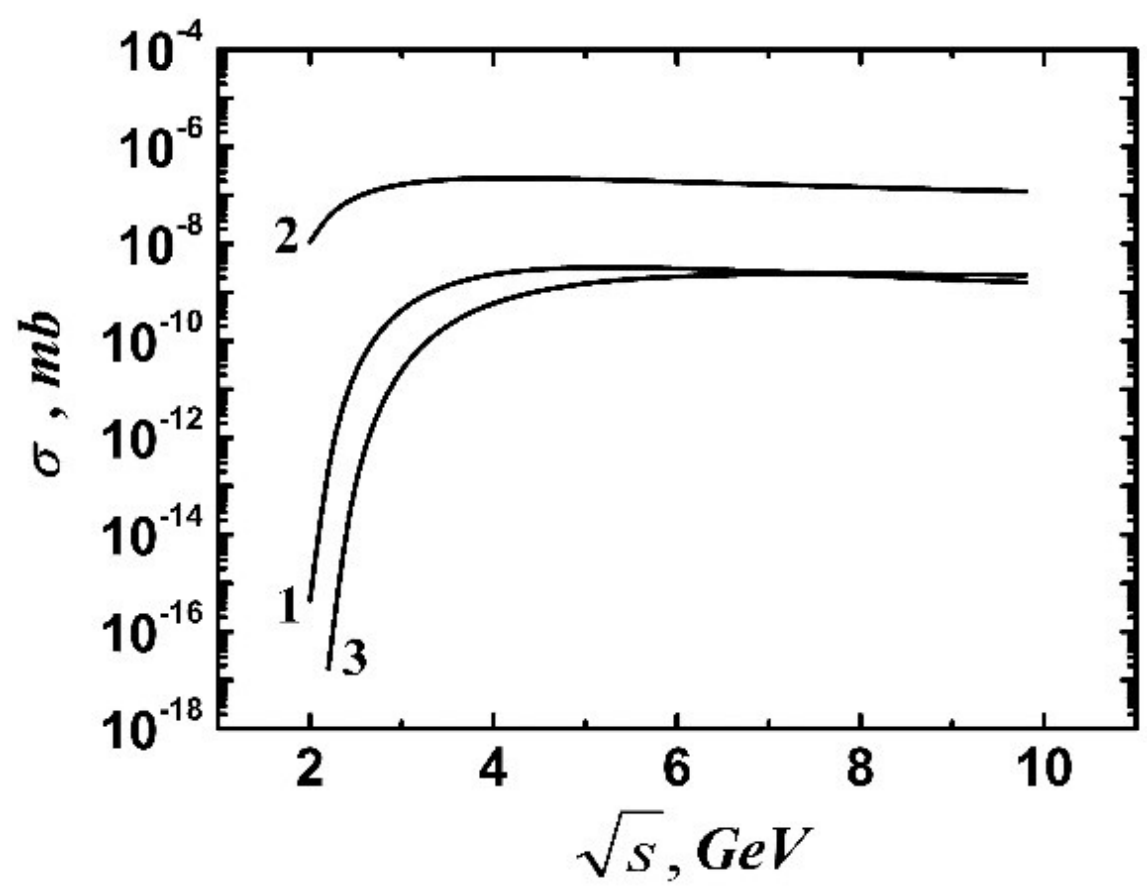

a

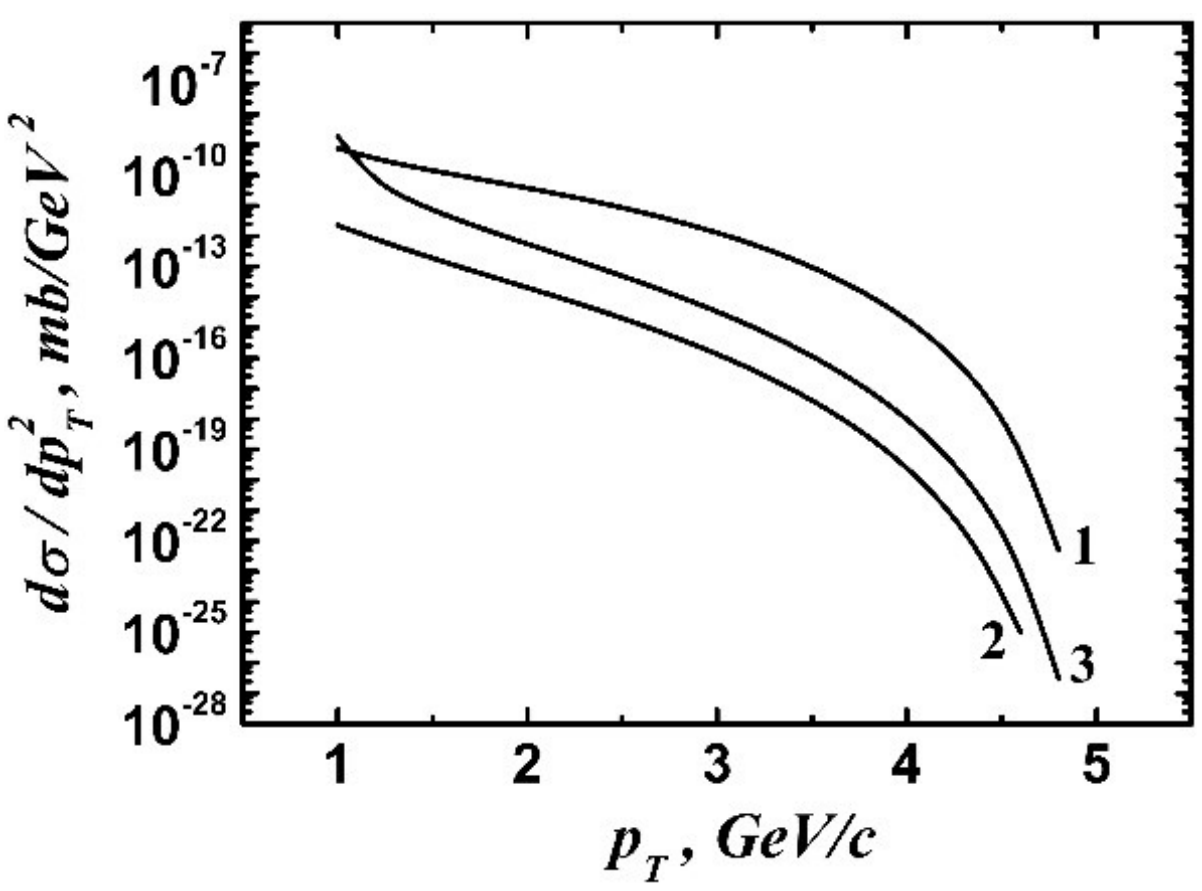

b

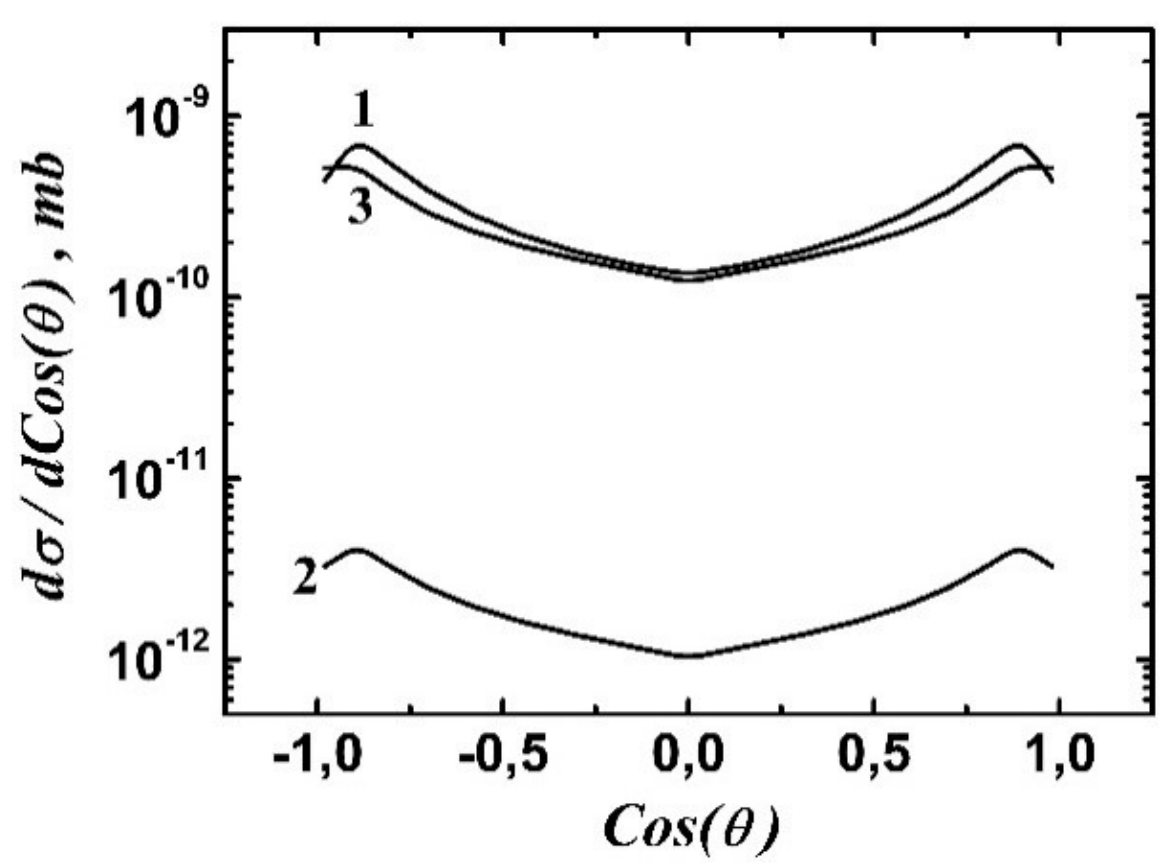

c

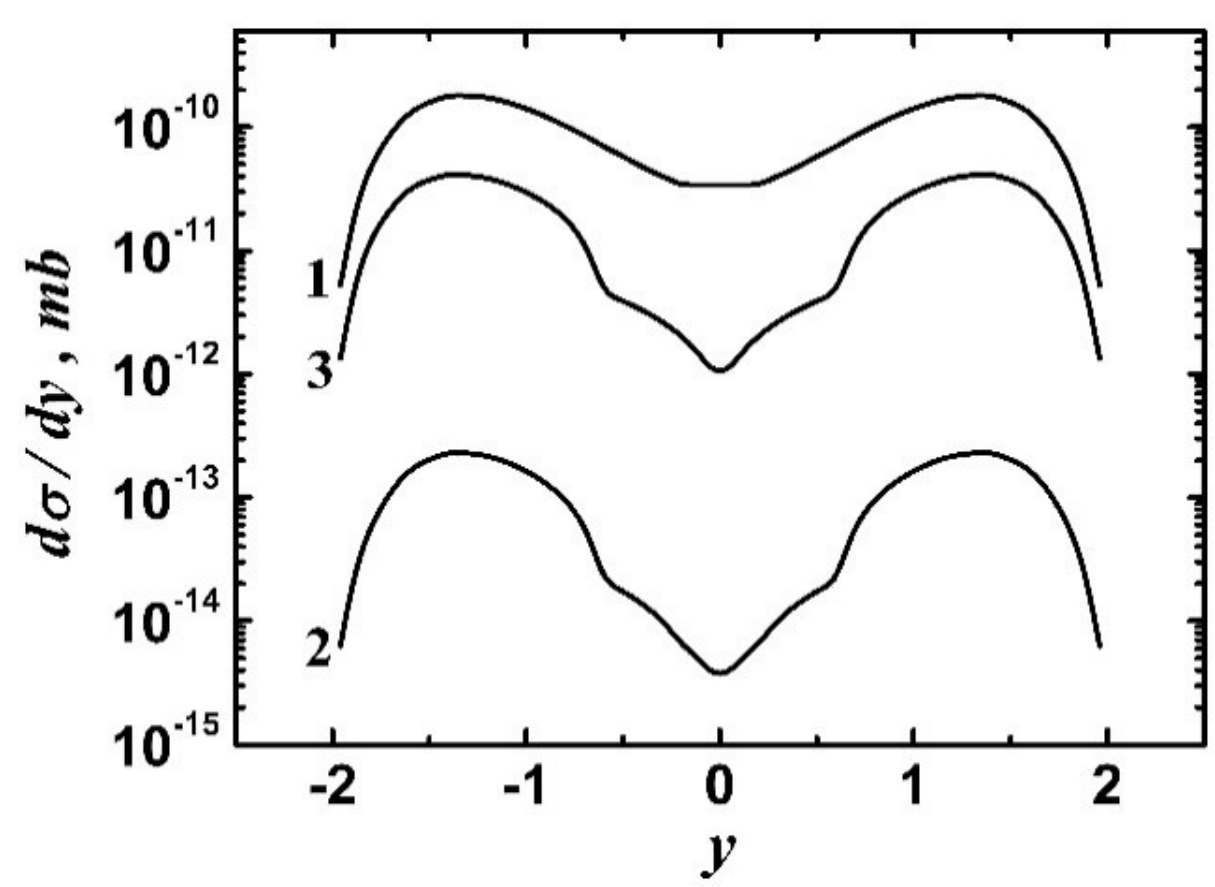

d

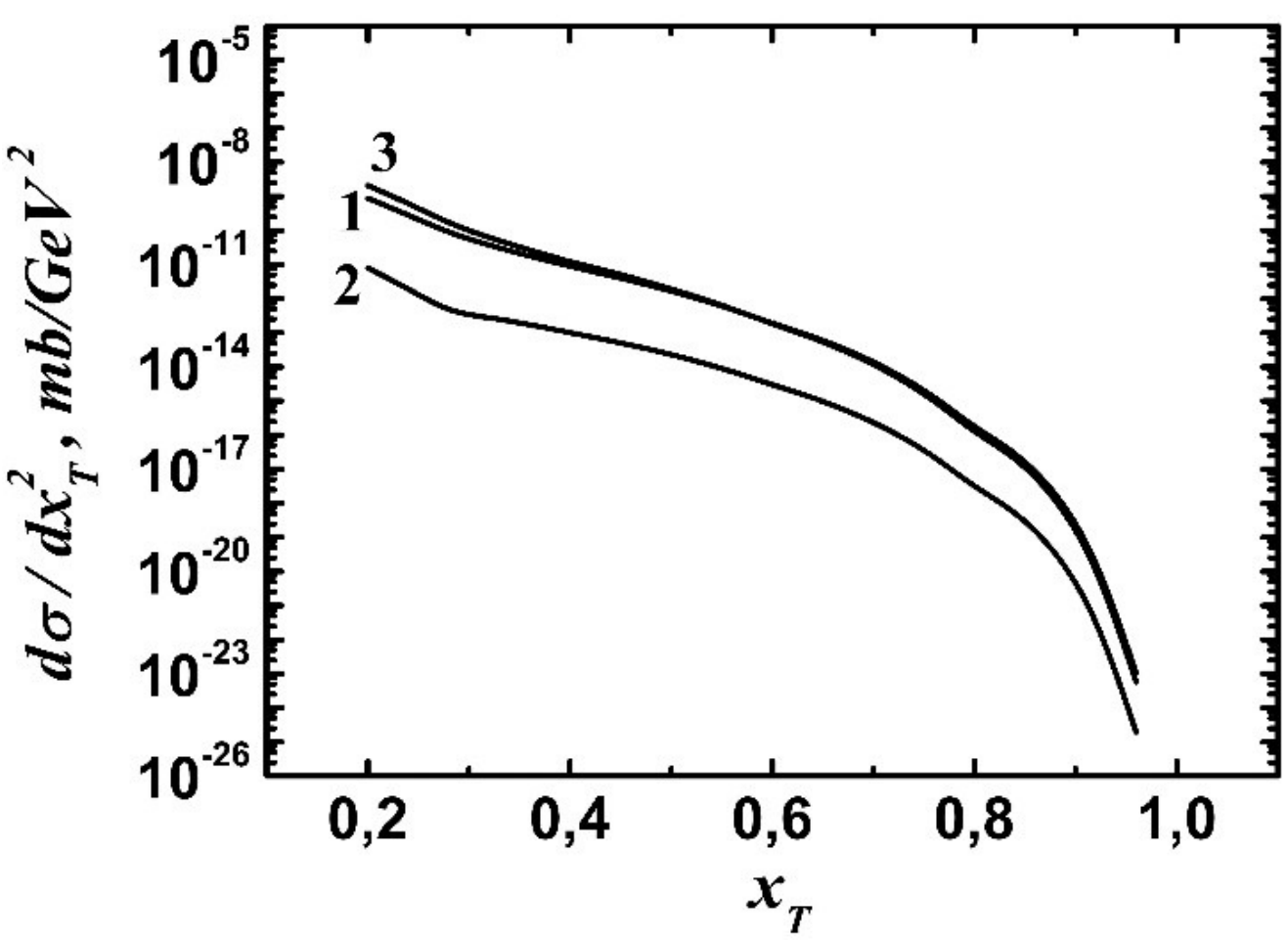

e

Figure 5(a,b,c,d,e). Dependence of the total cross section of the subprocesses 1. $q\bar{q} \to q\bar{q}\gamma$ (curve 1), 2. $q\bar{q} \to qg\gamma$ (curve 2), 3. $q\bar{q} \to g\gamma\gamma$ (curve 3) on the sum of the energy of colliding particles $\sqrt{s}$ (a) and the differential cross section of the subprocesses on the transverse momentum of prompt photons $p_T$ (b), the cosine of the scattering angle of prompt photons $Cos(\theta)$ (c), the rapidity y (d) and the parameter $x_T$ (d) in the PYTHIA 8.316 modulation at $\sqrt{s}$=10 GeV.

Figure 5(a) shows the total cross section as a function of the the sum of the energy of colliding particles $\sqrt{s}$. Similar to the analytical calculations, a threshold increase is observed, reaching a logarithmic plateau at high energies. The hierarchy of cross sections is preserved: the gluon channel 2. $q\bar{q} \to qg\gamma$ is higher than the quark channel 1. $q\bar{q} \to q\bar{q}\gamma$, and the two-photon channel 3. $q\bar{q} \to g\gamma\gamma$ is strongly suppressed. At high energies ($\sqrt{s}$ >4.5 GeV) the dependences subprocesses 1. $q\bar{q} \to q\bar{q}\gamma$ and 3. $q\bar{q} \to g\gamma\gamma$ on energy of colliding protons is match.

As seen from Figure 5(b), the dependence of the differential cross section of subprocesses on the transverse momentum of prompt photons $p_T$ shows an exponential-power-law decrease in the cross section, which stops when approaching the kinematic limit $p_T \to \sqrt{s}/2$ =5 GeV/c.

The spectrum drop is due to the hardness of the matrix element. However, at the edges of the phase space (near 5 GeV/c), unphysical singularities often arise in analytical calculations due to hard cutoffs. PYTHIA 8.316 handles this region by exponentiating the Sudakov form factors (Sudakov suppression), which describes the probability of "non-emission" of additional soft partons [25]. Furthermore, the initial radiation (ISR) imparts a small transverse recoil momentum ($k_T$-kick) to the center-of-mass system, leading to a natural kinematic "smearing" of the hard cutoff, making the spectrum smoother and more realistic for comparison with detector data.

Dependences in Figure 5(c,d,e) show similar dependencies of the differential cross section as in the case of analytical calculations.

4. Comparison of the calculated total and differential cross sections of the subprocesses of radiative corrections 1. $q\bar{q} \to q\bar{q}\gamma$, 2. $q\bar{q} \to gg\gamma$ and 3. $q\bar{q} \to g\gamma\gamma$ to annihilation of quark-antquark pair prompt photon production in collisions of polarized and longitudinally polarized colliding protons, and simulated total and differential cross sections in PYTHIA 8.316.

All ratios $R$ has been investigated in the dependences on the sum of energies of colliding protons $\sqrt{s}$, transverse momentum $p_T$, cosine of scattering angle $Cos(\theta)$, rapidity y of photon and $x_T$.

4.1 Ratio $R = \frac{d\sigma(q\bar{q}\to g\gamma\gamma)+d\sigma(q\bar{q}\to gg\gamma)+d\sigma(q\bar{q}\to q\bar{q}\gamma)}{d\sigma_{LO}(q\bar{q}\to g\gamma)}$

Figure 6(a,b,c,d,e) shows the dependences of the ratio $R$ on the sum of the energy of colliding protons $\sqrt{s}$, transverse momentum $p_T$, cosine of the scattering angle of the produced prompt photons $Cos(\theta)$, rapidity y, and parameter $x_T$ at $\sqrt{s}$=10 GeV, which, shows the relative contribution 1. $q\bar{q} \to q\bar{q}\gamma$, 2. $q\bar{q} \to gg\gamma$ and 3. $q\bar{q} \to g\gamma\gamma$ ($2 \to 3$ process) compared to the section of the LO $q\bar{q} \to g\gamma$ ($2 \to 2$ process).

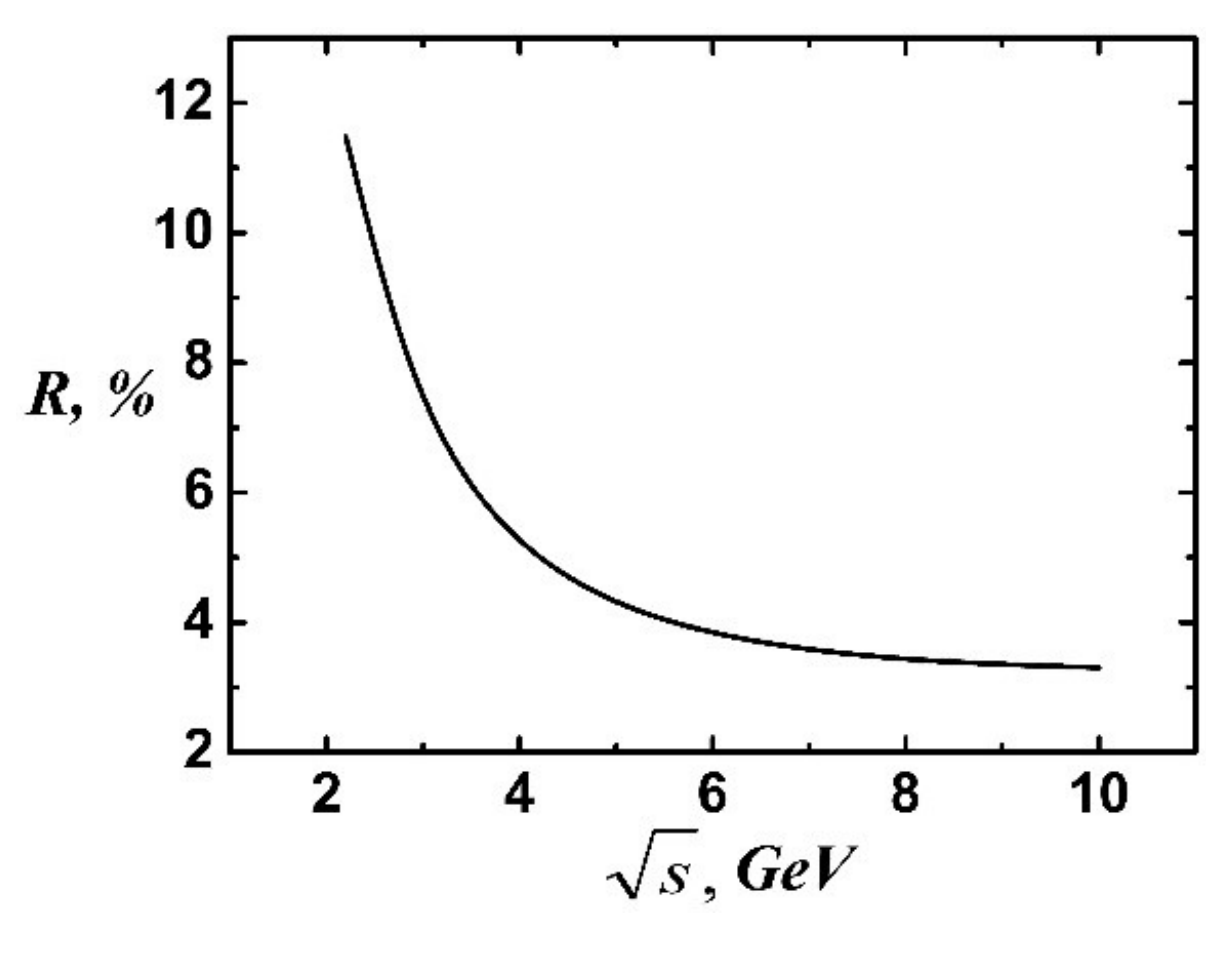

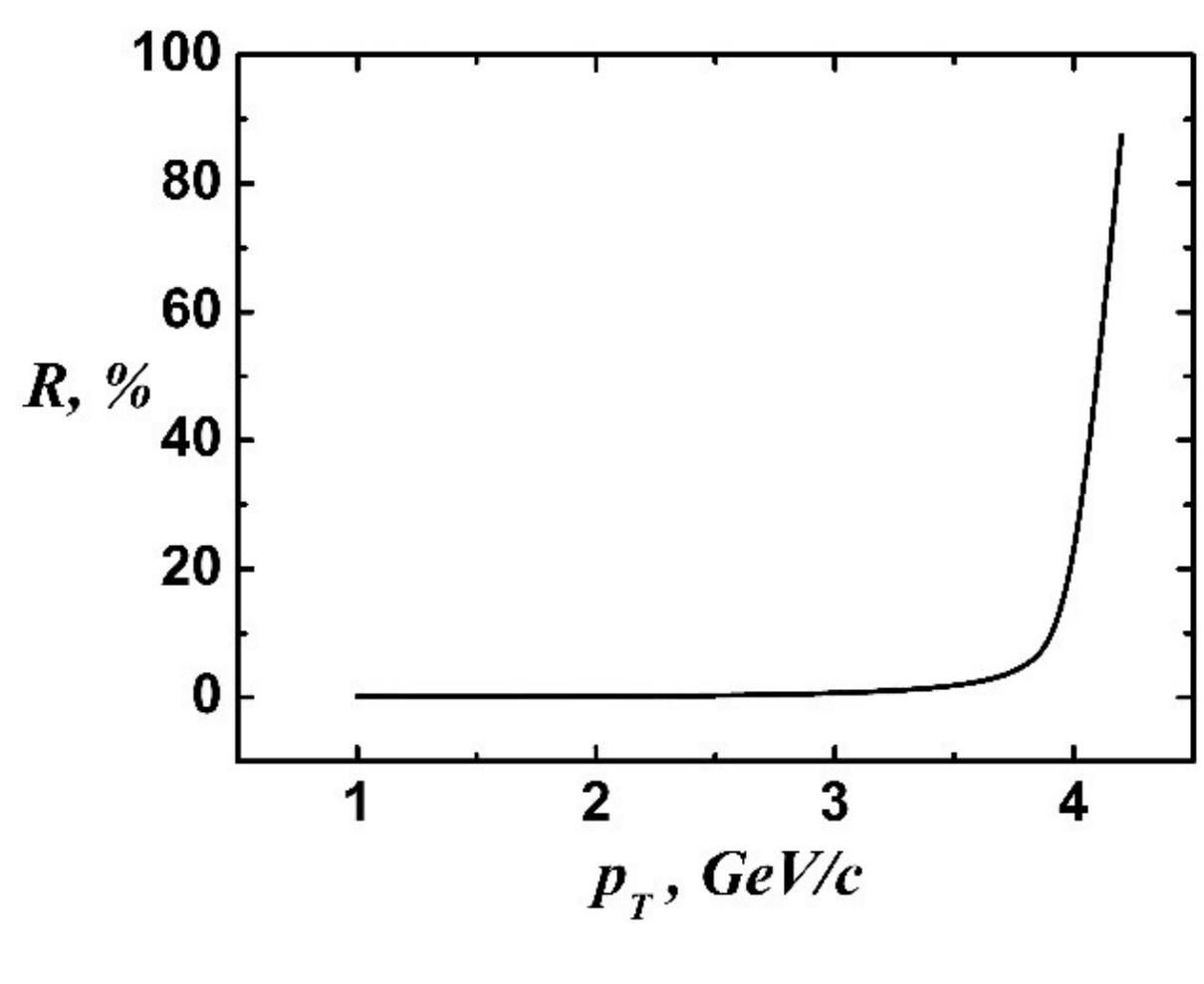

a

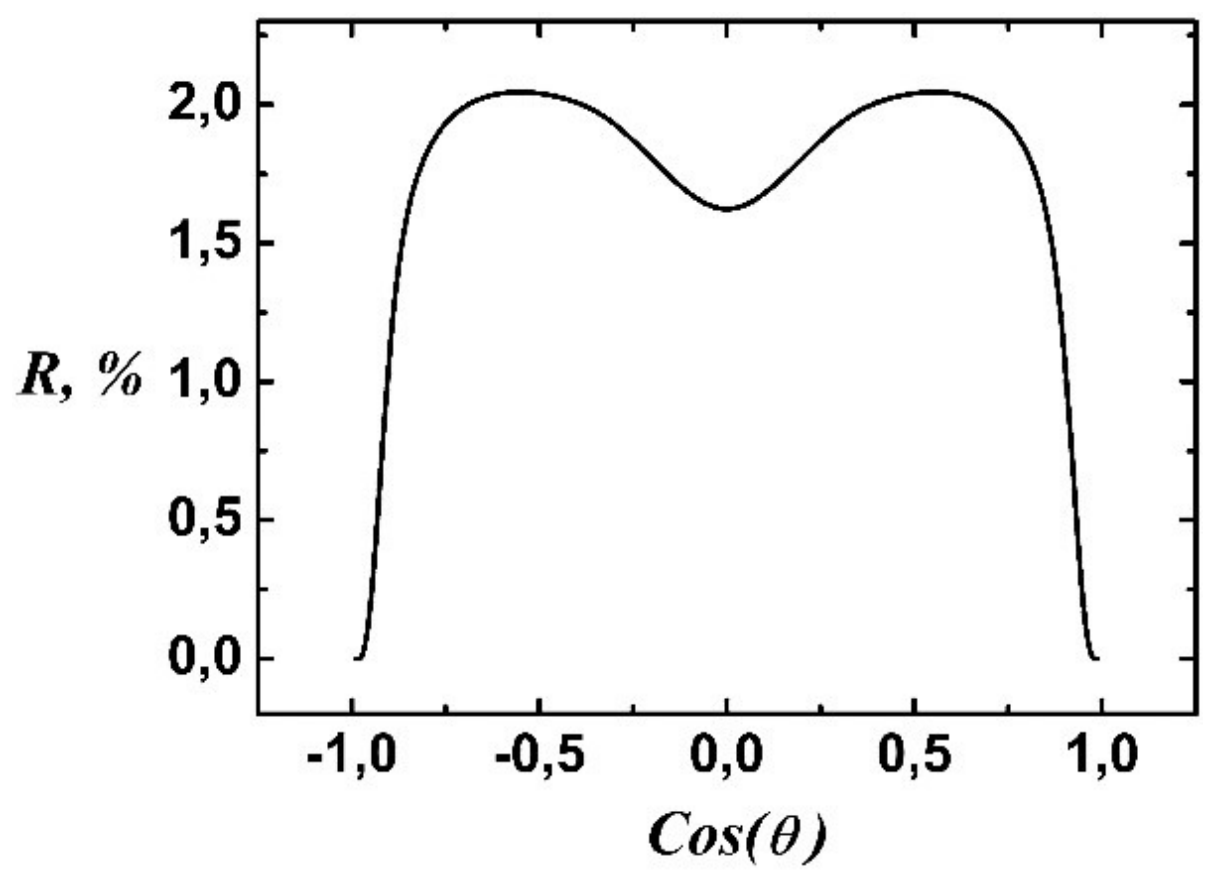

b

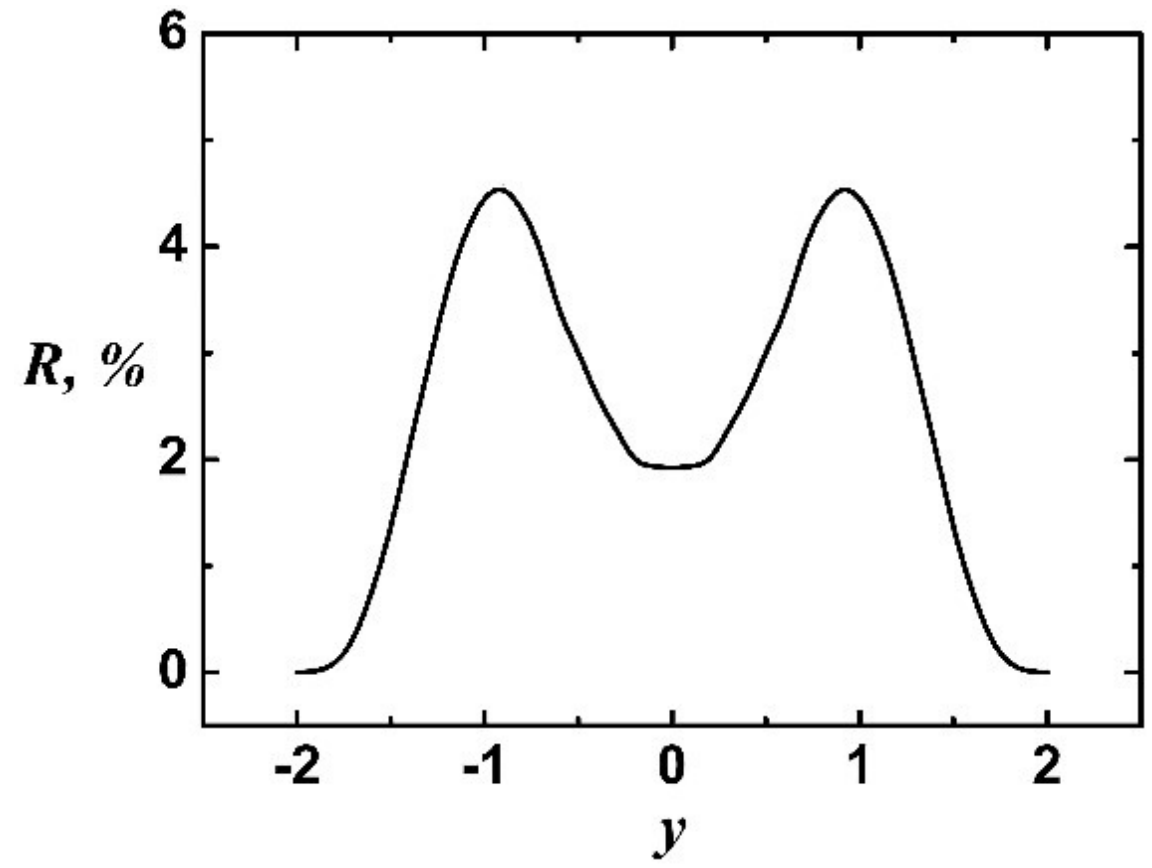

c

d

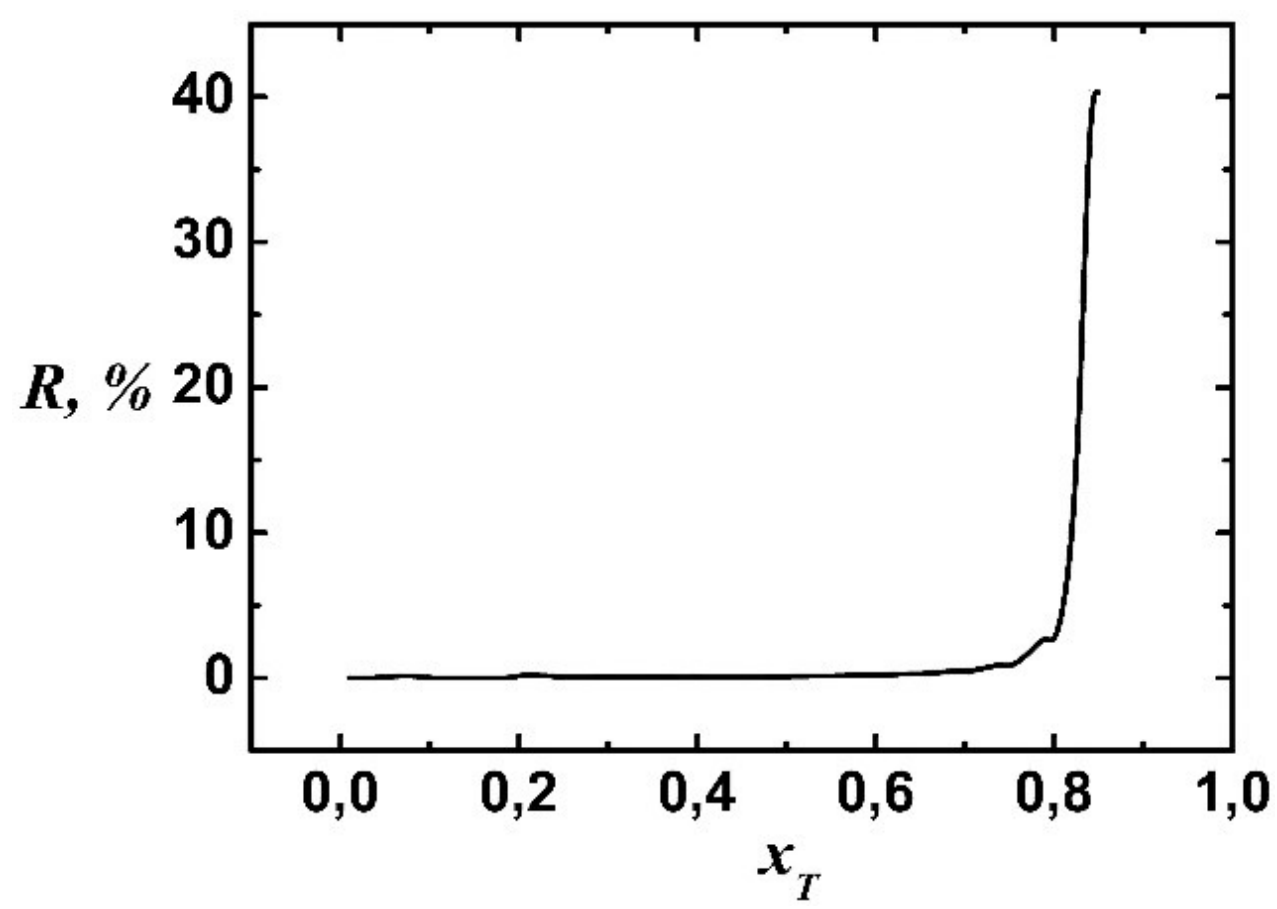

e

Figure 6(a,b,c,d,e). The dependence of ratio $R = \frac{d\sigma(q\bar{q}\to g\gamma\gamma)+d\sigma(q\bar{q}\to gg\gamma)+d\sigma(q\bar{q}\to q\bar{q}\gamma)}{d\sigma_{LO}(q\bar{q}\to g\gamma)}$ on sum energy colliding protons $\sqrt{s}$ (a), transverse momentum $p_T$ (b), cosine of scattering angle produced prompt photons $Cos(\theta)$ (c), rapidity $y$ (d) and parameter $x_T$ (e) at $\sqrt{s}$=10 GeV.

Figure 6(a,b,c,d,e) presents the kinematic dependences of the ratio $R$, which determines the total relative contribution of three radiative corrections subprocesses 1. $q\bar{q} \to q\bar{q}\gamma$, 2. $q\bar{q} \to gg\gamma$ and 3. $q\bar{q} \to g\gamma\gamma$ ($2 \to 3$ processes) compared to the Born LO cross-section $q\bar{q} \to g\gamma$ ($2 \to 2$ processes). This quantity illustrates the stability and applicability of pQCD theory in various kinematic regions.

Figure 6(a) shows that at low energies (near the reaction threshold), the contribution of radiative corrections is about $R \approx 11 - 12\%$ and mean value of contribution of radiative corrections is $R \approx$7.92%. As the energy $\sqrt{s}$ increases, the curve smoothly decreases, asymptotically approaching the level of $R \approx 3 - 4\%$. The ratio of the cross-section of a process with additional parton emission to the base cross-section is directly proportional to the strong coupling constant $\alpha_s(Q^2)$. The observed decline of the curve is a direct manifestation of asymptotic freedom in Quantum Chromodynamics: as the momentum transfer (or energy $\sqrt{s}$) increases, the "running" coupling constant $\alpha_s$ logarithmically decreases, which leads to the suppression of higher-order corrections.

Figure 6(b) shows that over a wide range of low and moderate transverse momenta $p_T$, the ratio $R$ is negligible (close to zero). However, as $p_T \to 5$ GeV/c (which corresponds to the absolute kinematic limit $p_T^{\max} = \sqrt{s}/2$ for $\sqrt{s} = 10$ GeV), the curve experiences a catastrophic growth, reaching almost $100\%$. This is explained by the fact that the cross-section of the basic $q\bar{q} \to g\gamma$ ($2 \to 2$ process) drops faster with increasing $p_T$ than the cross-sections of radiative corrections ($2 \to 3$ processes), where a more complex redistribution of momenta is possible (e.g., due to $t$-channel quark exchange). At the phase space boundary, the perturbation series diverges, indicating the need for the resummation of soft and collinear contributions.

The angular distribution (Figure 6(c)) has a characteristic $M$-shape. The absolute values of $R$ in the central region are very small ($1\% - 2\%$), indicating good convergence of perturbation theory for large-angle scattering. Peaks are observed at $Cos\theta \approx \pm 0.6$, and at the kinematic edges ($Cos\theta \to \pm 1$), the ratio drops to zero. This behavior is due to the competition of singularities: the LO differential cross-section has strict divergences $\sim 1/(1 - Cos^2\theta)$ ($t$- and $u$-channel poles), which drive the denominator to infinity faster than the numerators of the three-particle processes grow at the edges.

The rapidity profile $y$ (Figure 6(d)) completely replicates the angular dynamics. It is a symmetric $M$-shaped curve with a local minimum at $y = 0$ (transverse scattering) and maxima at $|y| \approx 1$. The drop of the curve to zero as $|y| \to 2$ corresponds to the collinear limit, where the Born $s$-channel cross-section dominates, and the relative fraction of radiative emission is suppressed by kinematics.

Figure 6(e) demonstrates the behavior of the radiative $K$-factor as a function of the dimensionless scaling variable $x_T = 2p_T/\sqrt{s}$. Similar to the $p_T$ dependence, the curve sharply shoots up (to almost $40\%$) as it approaches the exclusive limit $x_T \to 1$. At this point, the produced prompt photon carries away the maximum possible fraction of transverse energy, and the phase space for additional radiation shrinks. The huge increase in $R$ here proves that at kinematic boundaries, describing processes exclusively using LO matrix elements becomes physically incorrect.

4.2 Ratio $R = \frac{d\sigma_{POL}(q\bar{q}\to g\gamma\gamma)+d\sigma_{POL}(q\bar{q}\to gg\gamma)+d\sigma_{POL}(q\bar{q}\to q\bar{q}\gamma)}{d\sigma_{LO,\,POL}(q\bar{q}\to g\gamma)}$

Figure 7(a,b,c,d,e) shows the dependences of the ratios $R = \frac{d\sigma_{POL}(q\bar{q}\to g\gamma\gamma)+d\sigma_{POL}(q\bar{q}\to gg\gamma)+d\sigma_{POL}(q\bar{q}\to q\bar{a}\gamma)}{d\sigma_{LO,\,POL}(q\bar{q}\to g\gamma)}$ on sum energy colliding protons $\sqrt{s}$, transverse momentum $p_T$, cosine of scattering angle produced prompt photons $Cos(\theta)$, rapidity $y$ and parameter $x_T$ at $\sqrt{s}$=10 GeV, which in this case reflects the share of a specific subprocess in the total cross-section of main prompt photons production in longitudinally polarized protons.

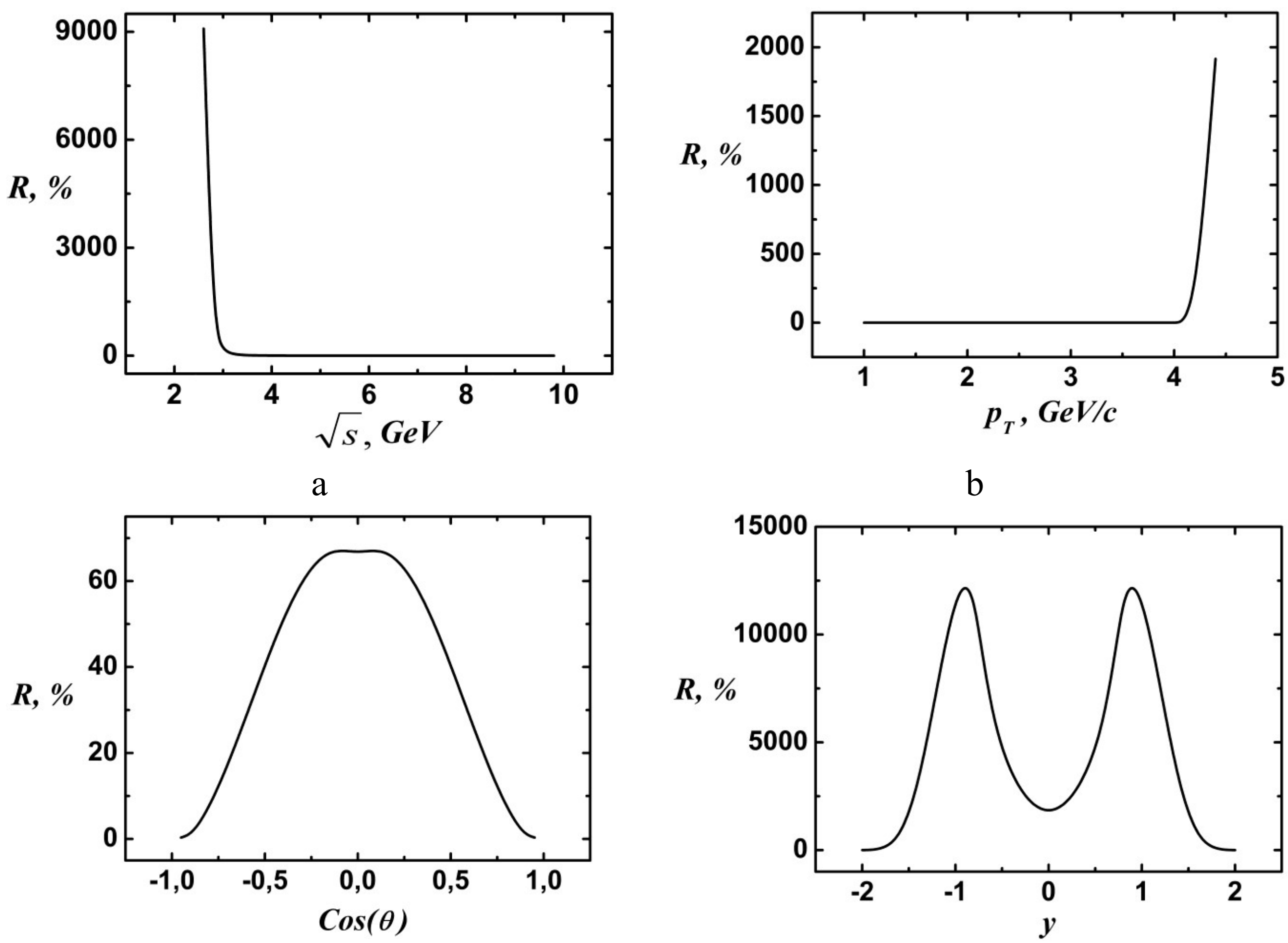

a b

c d

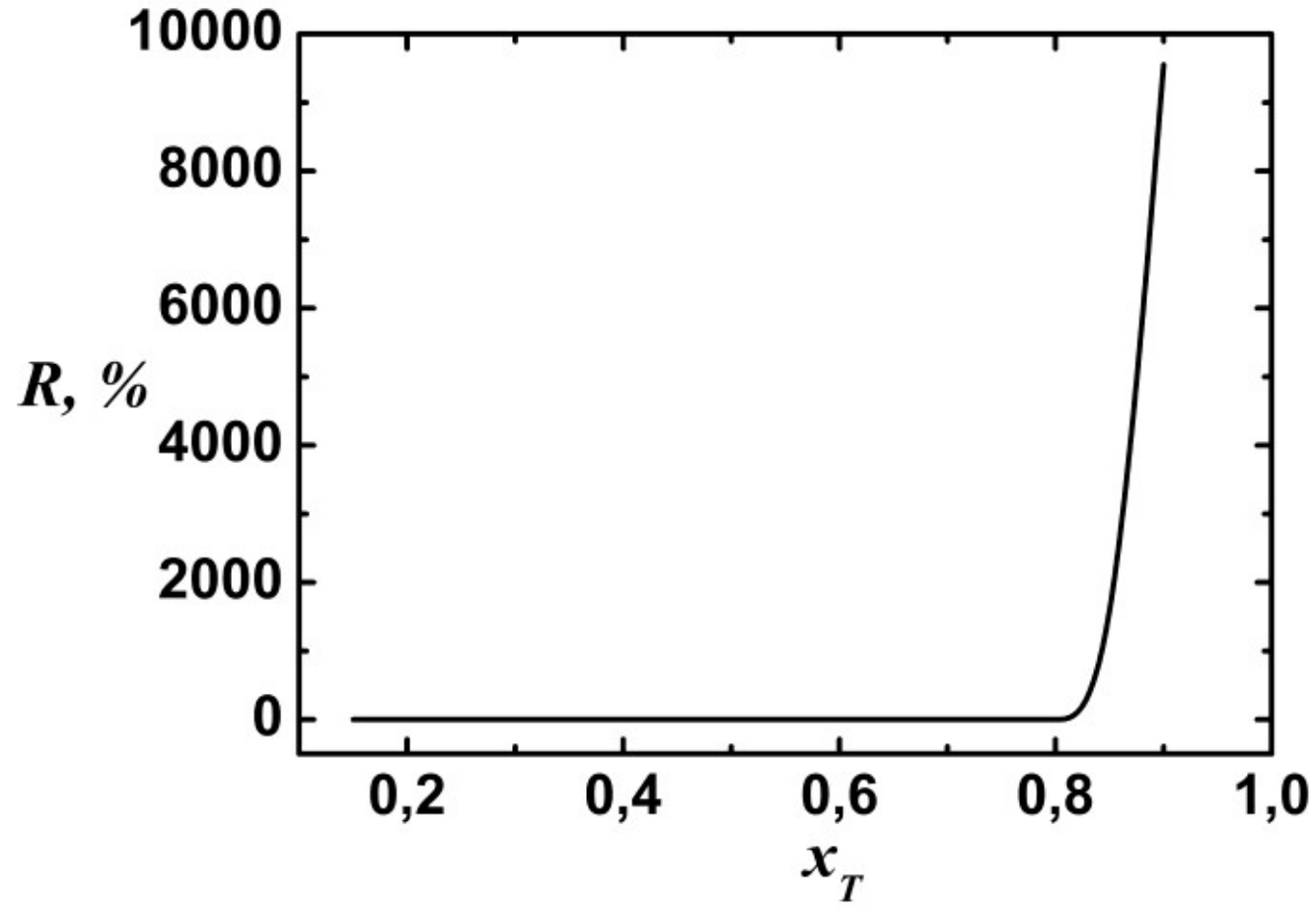

e

Figure 7(a,b,c,d,e). Dependences of the ratio $R=\frac{d\sigma_{POL}(q\bar{q}\to g\gamma\gamma)+d\sigma_{POL}(q\bar{q}\to gg\gamma)+d\sigma_{POL}(q\bar{q}\to q\bar{q}\gamma)}{d\sigma_{LO,\,POL}(q\bar{q}\to g\gamma)}$ on the sum of the energy of colliding protons $\sqrt{s}$ (a), transverse momentum $p_T$ (b), cosine of the scattering angle of the produced prompt photons $Cos(\theta)$ (c), rapidity y (d) and parameter $x_T$ (e) at $\sqrt{s}$=10 GeV.

Figure 7(a,b,c,d,e) shows the kinematic dependences of the ratio $R$, which determines the total relative contribution of three radiative corrections 1. $q\bar{q}\to q\bar{q}\gamma$, 2. $q\bar{q}\to gg\gamma$ and 3. $q\bar{q}\to g\gamma\gamma$ ($2\to 3$ process) to the polarized cross-section compared to the polarized Born LO cross-section (LO, $q\bar{q}\to g\gamma$, $2\to 2$ process).

The main feature of these graphs is the anomalously large values of the ratio. Physically, this indicates that in these specific kinematic regions, the denominator (the LO polarized cross-section) tends to zero due to strict helicity selection rules, and the dynamics of the process are entirely dominated by the radiative corrections.

As seen from Figure 7(a), the ratio $R$ have a large value of In the near-threshold region $\sqrt{s}\approx 2-3$ GeV. Physically, this means that at low energies, the annihilation of polarized quarks $q_L\bar{q}_R\to g\gamma$ is heavily suppressed by phase space kinematics. The $2\to 3$ processes, having additional final-state particles, make it easier to satisfy angular momentum conservation laws. As we move to higher energies, the asymptotic freedom of QCD takes effect, the denominator stabilizes, and the ratio drops to perturbatively reasonable values, forming a plateau.

As seen from Figure 7(b), the ratio is close to zero in the region of small and moderate $p_T$, but as it approaches the kinematic limit ($p_T\to 5$ GeV/c), there is a giant leap to $R \approx 2000\%$. At maximum $p_T$, the particles produced in a $2 \to 2$ process must scatter strictly back-to-back with no longitudinal momentum. For given initial quark helicities, such an exclusive configuration turns out to be forbidden (or highly suppressed) by the LO matrix element. Radiative emission (the numerator) lifts this strict prohibition, which leads to a mathematical divergence of the ratio.

Figure 7(c) shows the dependence ratio $R$ on scattering angle of photon $Cos\theta$. Unlike the nonpolarized case, the angular distribution has a bell shape (inverted $U$) with a maximum in the center ($Cos\theta = 0$ that is 90° scattering) at the level of $R \approx 60 - 70\%$ and a drop to zero at the edges ($Cos\theta \to \pm 1$). This shows that at transverse scattering, radiative corrections make the maximum constructive contribution to the spin asymmetry. The drop to zero at the edges means that in the strictly collinear limit, the multiparticle polarized emission vanishes faster than the Born process due to the conservation of spin projection onto the collision axis.

As seen from Figure 7(d), the dependence on rapidity has an extreme $M$-shape with giant peaks at $|y| \approx 1$. The physical nature of these peaks lies in the phenomenon known as "radiation zero" (or spin zero). At specific rapidity values, the interference of $t$- and $u$-channel amplitudes for the polarized $q\bar{q} \to g\gamma$ process leads to exact mutual cancellation of the terms (the denominator turns to zero). Since the $2 \to 3$ processes have a different kinematic topology, their cross-sections are finite at this point. Division by zero is visualized as these colossal peaks.

The dependence on $x_T = 2p_T/\sqrt{s}$ as a dimensionless analogue of the $p_T$ dependence, the curve demonstrates a sharp asymptotic surge as $x_T \to 1$. This proves once again that at the phase space boundary (in the exclusive limit), the LO polarized cross-section is completely destroyed by helicity constraints. The description of spin observables in such regions requires the mandatory use of higher-order corrections and resummation techniques.

4.3 Ratios $R = \frac{d\sigma(q\bar{q}\to q\bar{q}\gamma)}{d\sigma(q\bar{q}\to g\gamma\gamma)+d\sigma(q\bar{q}\to gg\gamma)+d\sigma(q\bar{q}\to q\bar{q}\gamma)}$,

$R = \frac{d\sigma(q\bar{q}\to gg\gamma)}{d\sigma(q\bar{q}\to g\gamma\gamma)+d\sigma(q\bar{q}\to gg\gamma)+d\sigma(q\bar{q}\to q\bar{q}\gamma)}$ and

$R = \frac{d\sigma(q\bar{q}\to g\gamma\gamma)}{d\sigma(q\bar{q}\to g\gamma\gamma)+d\sigma(q\bar{q}\to gg\gamma)+d\sigma(q\bar{q}\to q\bar{q}\gamma)}$

Figure 8(a,b,c,d,e) shows the dependences of the ratios $R$ on sum energy colliding protons $\sqrt{s}$, transverse momentum $p_T$, cosine of scattering angle produced

prompt photons $Cos(\theta)$, rapidity $y$ and parameter $x_T$ at $\sqrt{s}$=10 GeV, which in this case reflects the share of a specific subprocess in the total cross-section of the prompt photons production.

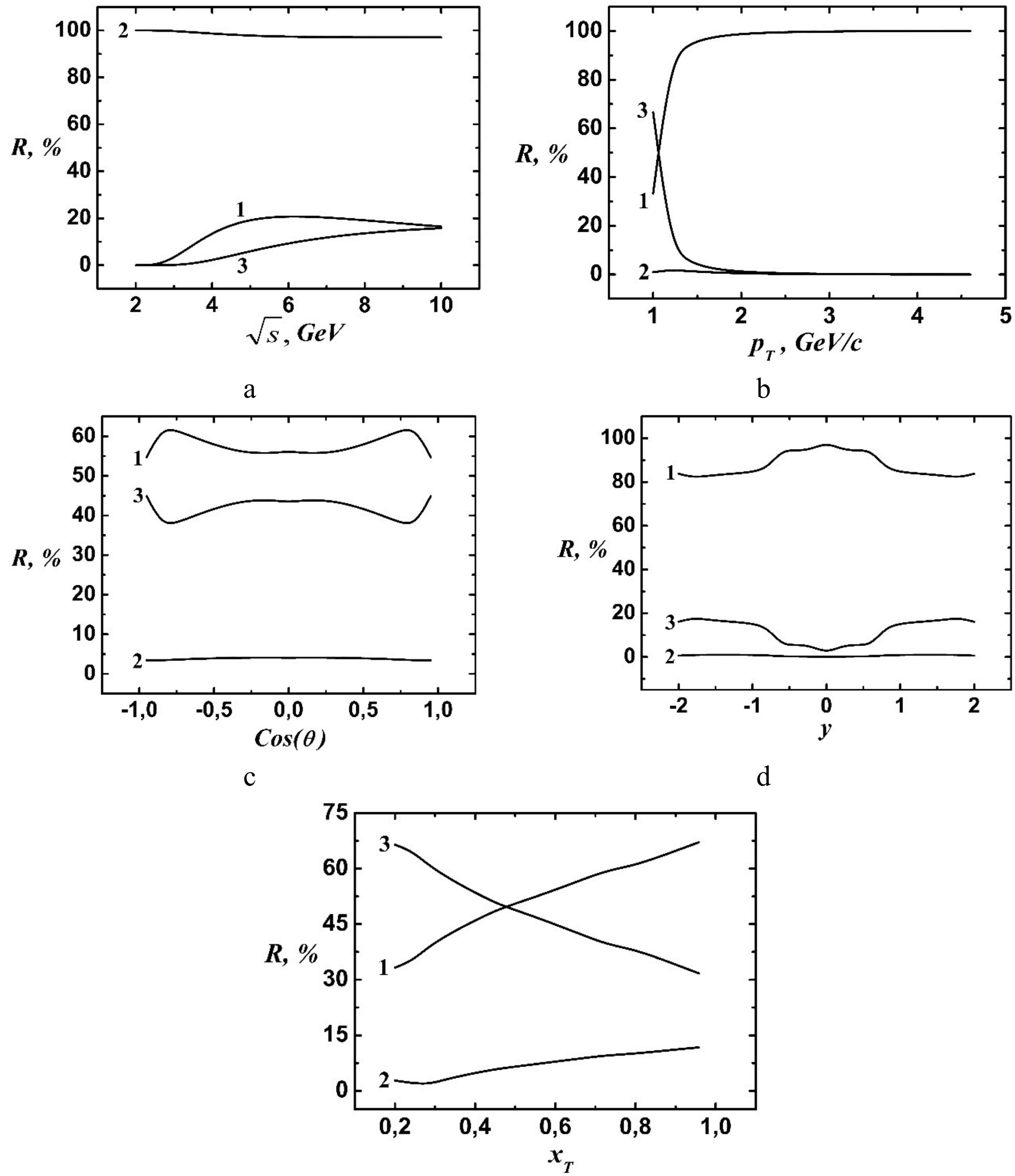

a b c d e

Figure 8(a,b,c,d,e). Dependences of the ratios
$R = \frac{d\sigma(q\bar{q}\to q\bar{q}\gamma)}{d\sigma(q\bar{q}\to g\gamma\gamma)+d\sigma(q\bar{q}\to gg\gamma)+d\sigma(q\bar{q}\to q\bar{q}\gamma)}$ (curve 1),
$R = \frac{d\sigma(q\bar{q}\to gg\gamma)}{d\sigma(q\bar{q}\to g\gamma\gamma)+d\sigma(q\bar{q}\to gg\gamma)+d\sigma(q\bar{q}\to q\bar{q}\gamma)}$ (curve 2) and
$R = \frac{d\sigma(q\bar{q}\to g\gamma\gamma)}{d\sigma(q\bar{q}\to g\gamma\gamma)+d\sigma(q\bar{q}\to gg\gamma)+d\sigma(q\bar{q}\to q\bar{q}\gamma)}$ (curve 3) on the sum of the energy of colliding protons $\sqrt{s}$ (a), transverse momentum $p_T$ (b), cosine of the scattering angle of the produced prompt photons $Cos(\theta)$ (c), rapidity y (d) and parameter $x_T$ (e) at $\sqrt{s}$=10 GeV.

Figure 8(a) shows the shares in the total cross section. Curve 2. $q\bar{q} \to gg\gamma$ consistently dominates, accounting for almost 80-100 % of the entire cross section. The share of curve 1. $q\bar{q} \to q\bar{q}\gamma$ remains at $R \approx 20\%$. Process 3. $q\bar{q} \to g\gamma\gamma$ makes a negligible contribution ($R \approx 0.5\%$). Mean value of $R$ for 1. $q\bar{q} \to q\bar{q}\gamma$ is 1- 15.1%, 3- 8.1%, correspondingly.

The process of annihilation into two gluons with photon production exhibits strong infrared and collinear enhancement (due to the strong interaction vertices and the presence of two massless gluons). Therefore, "on average" (over the entire detector), we will observe the $q\bar{q} \to gg\gamma$ channel. The fraction of $q\bar{q} \to gg\gamma$ is close to zero due to the suppression factor $\alpha_e/\alpha_s \approx 1/137$.

As $R$ depends on $p_T$ (Figure 8(b)), a radical change in the dominant process is observed. At low $p_T$ (around 1 GeV/c), processes 1. $q\bar{q} \to q\bar{q}\gamma$ and 3. $q\bar{q} \to g\gamma\gamma$ are significant. However, as $p_T$ increases, the share of the 1. $q\bar{q} \to q\bar{q}\gamma$ process rapidly increases and asymptotically approaches 100%, while the shares of the remaining processes drop to zero.

Matrix elements with gluon production 2. $q\bar{q} \to gg\gamma$ have a steeper (rapidly decaying) spectrum in $p_T$ due to radiative corrections and the kinematics of the multiparticle final state. The 1. $q\bar{q} \to q\bar{q}\gamma$ diagram, containing fermion lines in the final state, is characterized by a "harder" spectrum. At high momentum transfers $p_T$, only the hardest interaction mechanism remains.

Figure 8(c) shows the fractions in the angular distribution of $p_T$. Unlike the cross sections themselves (which had a deep $U$-shape), the $R$ fractions behave relatively flat over a wide range of angles. There are only small fluctuations at the very edges ($Cos(\theta) \to \pm 1$). The fact that the ratio of the differential cross sections is almost constant means that the structure of the kinematic singularities (the poles of the propagators in the $t$- and $u$-channels) is very similar for these subprocesses. The angular dependence cancels out upon fission. The slight change in the fractions during strictly forward/backward scattering is due to the difference in the helical amplitudes of the emission of bosons $(g, \gamma)$ and fermions $(q)$.

As with the angle, the rapidity distribution (Figure 8(d)) reflects the balance of the phase volume. The horizontality of the lines indicates that the relative probability of a particular subprocess occurring depends weakly on which region of the detector (the central barrel or the end caps) the generated photon flies to. The only exceptions are the very edges of the phase volume, where the transverse momentum conservation constraint comes into play.

4.4 Ratios $R = \frac{d\sigma_{POL}(q\bar{q}\to q\bar{q}\gamma)}{d\sigma_{POL}(q\bar{q}\to g\gamma\gamma)+d\sigma_{POL}(q\bar{q}\to gg\gamma)+d\sigma_{POL}(q\bar{q}\to q\bar{q}\gamma)}$,

$R = \frac{d\sigma_{POL}(q\bar{q}\to gg\gamma)}{d\sigma_{POL}(q\bar{q}\to g\gamma\gamma)+d\sigma_{POL}(q\bar{q}\to gg\gamma)+d\sigma_{POL}(q\bar{q}\to q\bar{q}\gamma)}$ and

$R = \frac{d\sigma_{POL}(q\bar{q}\to g\gamma\gamma)}{d\sigma_{POL}(q\bar{q}\to g\gamma\gamma)+d\sigma_{POL}(q\bar{q}\to gg\gamma)+d\sigma_{POL}(q\bar{q}\to q\bar{q}\gamma)}$

Figure 9(a,b,c,d,e) shows the dependences of the ratios $R = \frac{d\sigma_{POL}(q\bar{q}\to q\bar{q}\gamma)}{d\sigma_{POL}(q\bar{q}\to g\gamma\gamma)+d\sigma_{POL}(q\bar{q}\to gg\gamma)+d\sigma_{POL}(q\bar{q}\to q\bar{q}\gamma)}$, $R = \frac{d\sigma_{POL}(q\bar{q}\to gg\gamma)}{d\sigma_{POL}(q\bar{q}\to g\gamma\gamma)+d\sigma_{POL}(q\bar{q}\to gg\gamma)+d\sigma_{POL}(q\bar{q}\to q\bar{q}\gamma)}$ and $R = \frac{d\sigma_{POL}(q\bar{q}\to g\gamma\gamma)}{d\sigma_{POL}(q\bar{q}\to g\gamma\gamma)+d\sigma_{POL}(q\bar{q}\to gg\gamma)+d\sigma_{POL}(q\bar{q}\to q\bar{q}\gamma)}$ on the sum of energies of colliding protons $\sqrt{s}$, transverse momentum $p_T$, cosine of scattering angle $Cos(\theta)$, rapidity $y$ of photon and $x_T$ in collisions longitidunally polarized protons with polarization order $P_1 = P_2 = \pm 0.9$ that is $P_1 P_2 = -0.81$.

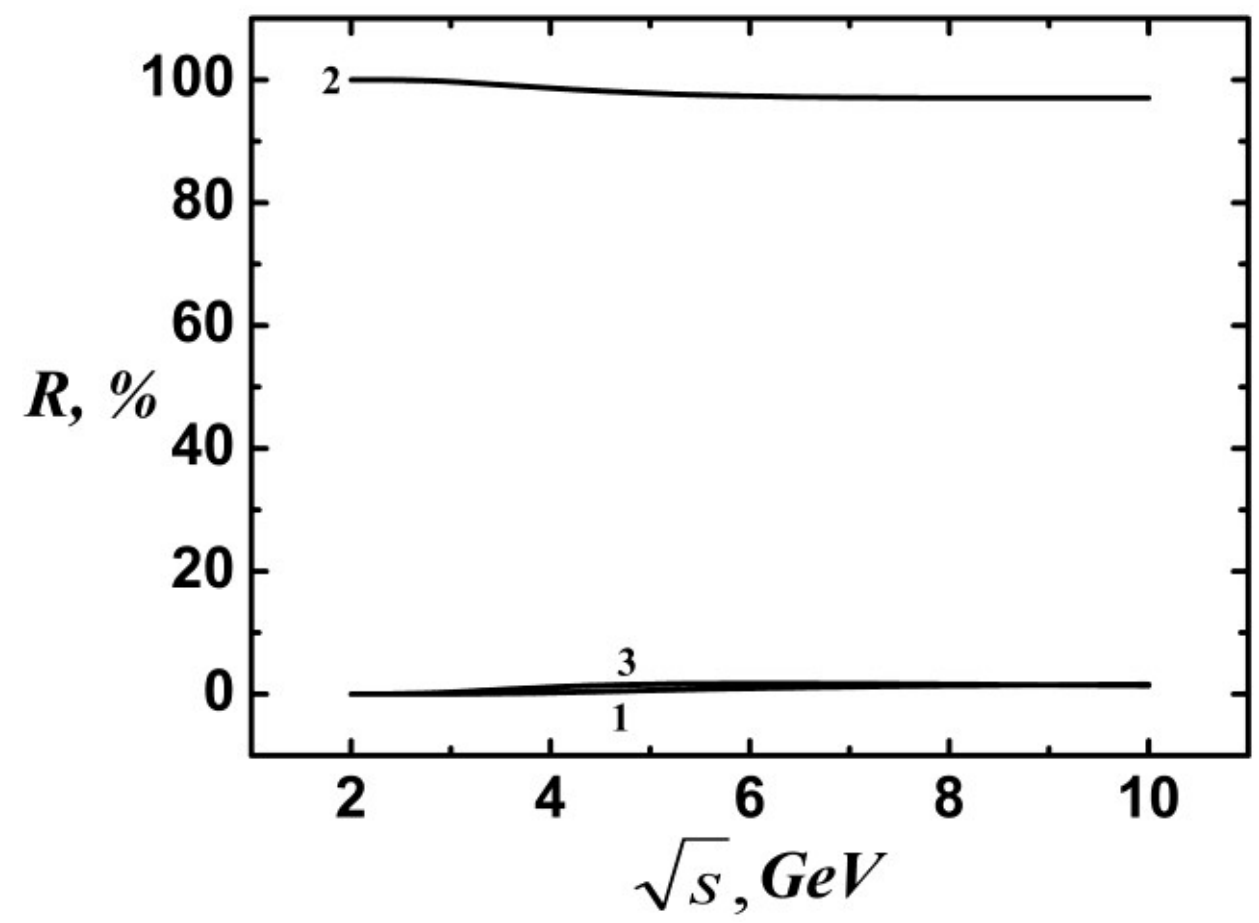
R, %
100
80
60
40
20
0
2
3
1
2
4
6
8
10
√s, GeV
a

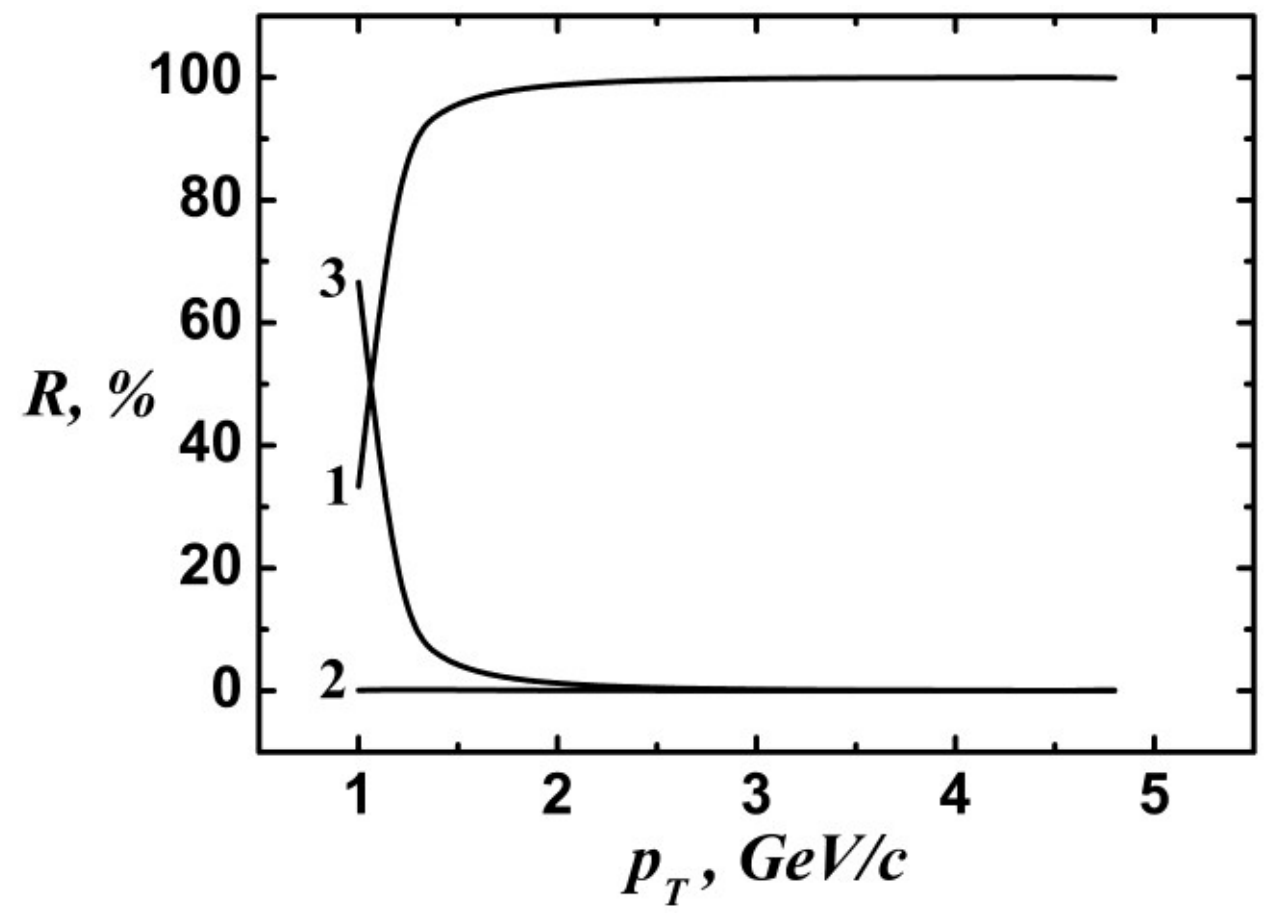
R, %
100
80
60
40
20
0
3
1
2
1
2
3
4
5
p_T, GeV/c
b

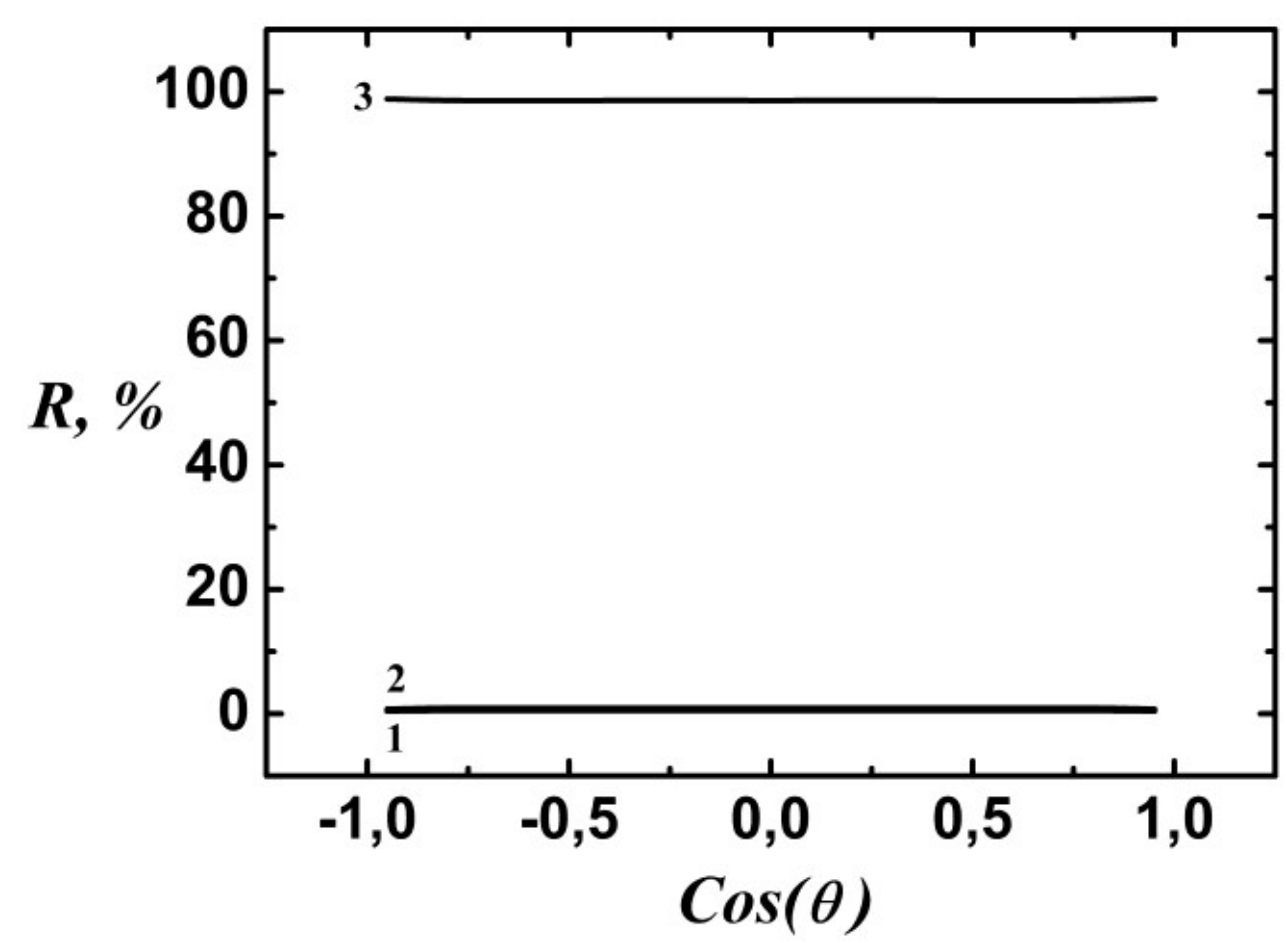
R, %
100
80
60
40
20
0
3
2
1
-1,0
-0,5
0,0
0,5
1,0
Cos(θ)
c

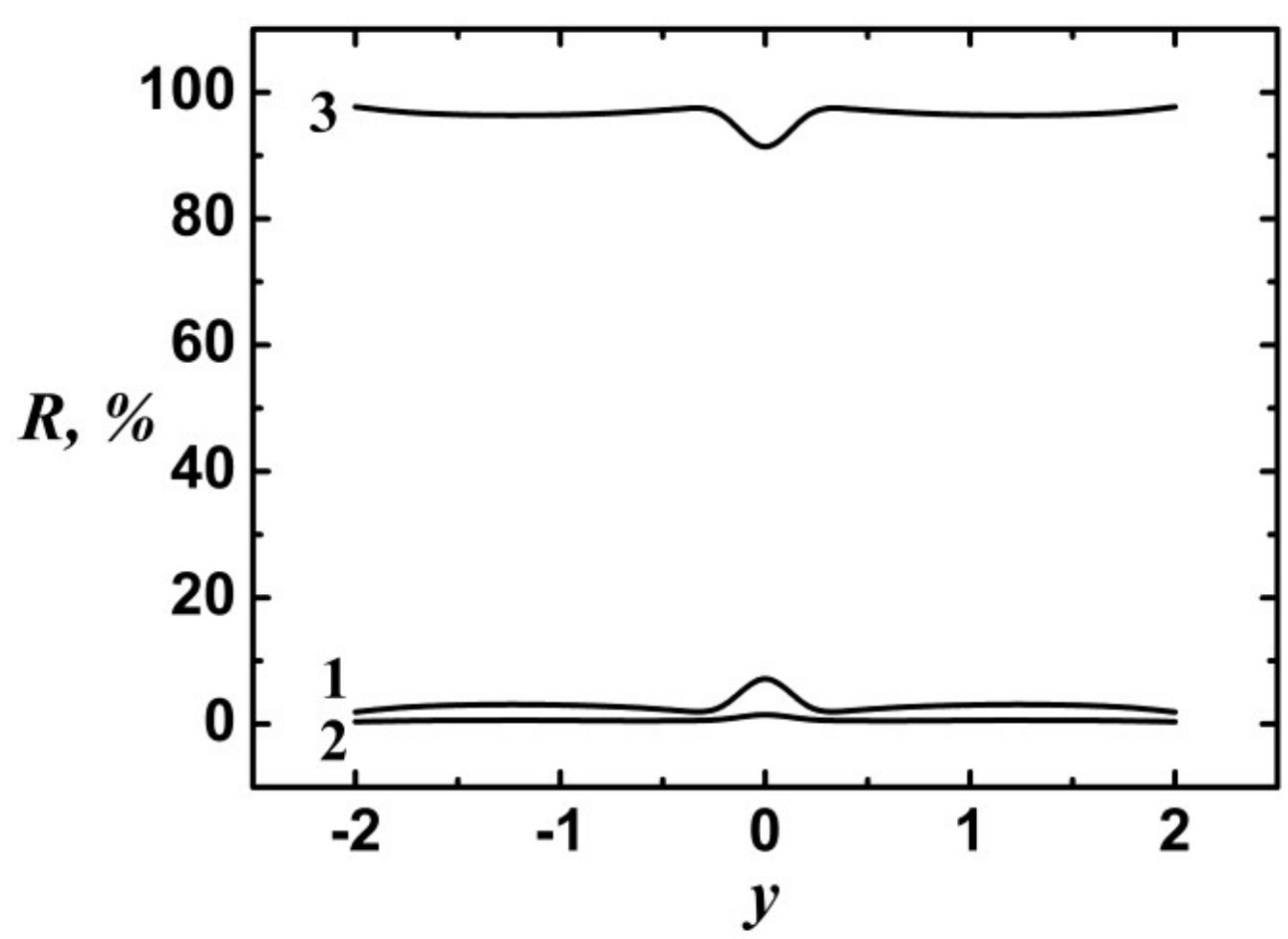
R, %
100
80
60
40
20
0
3
1
2
-2
-1
0
1
2
y
d

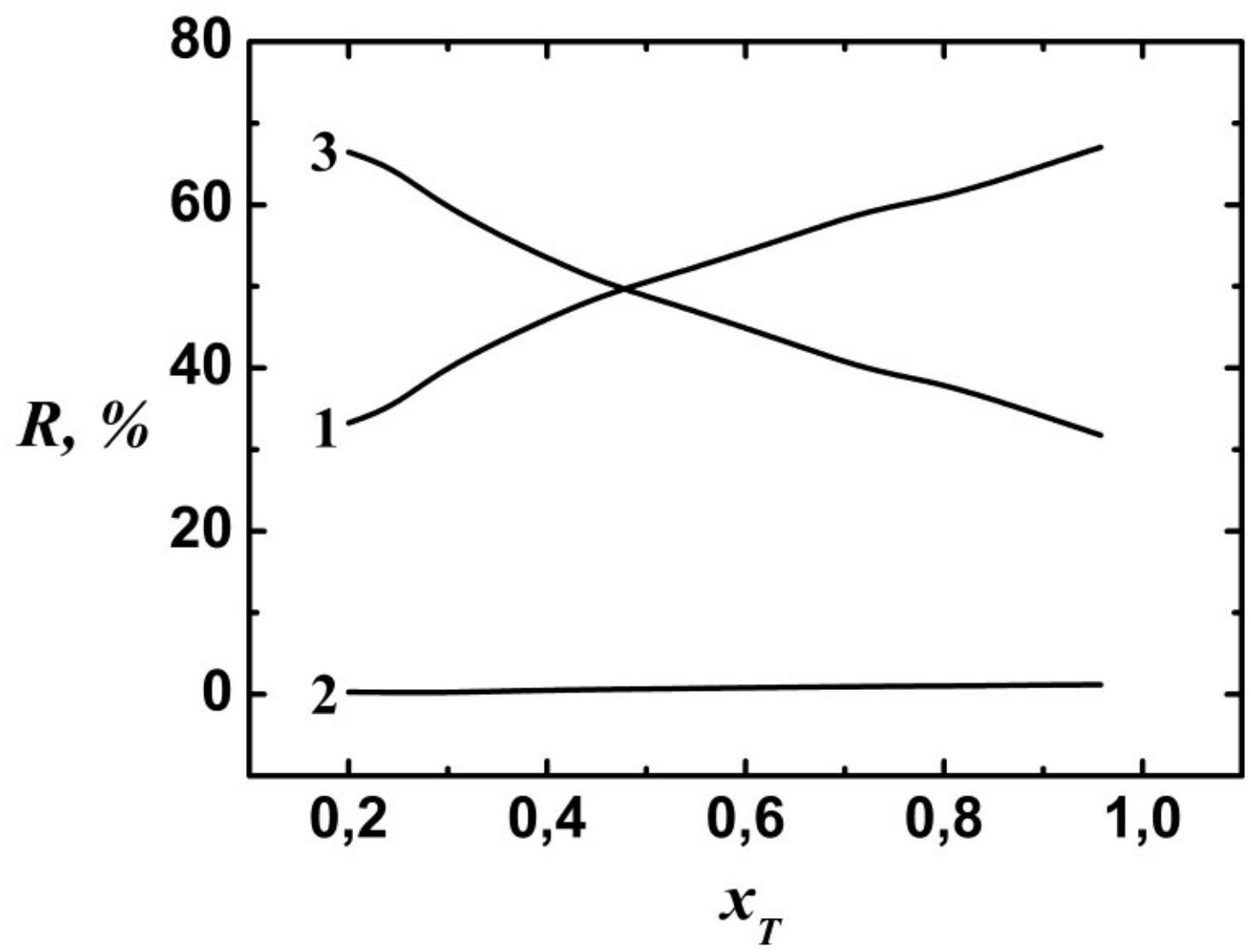
R, %
80
60
40
20
0
3
1
2
0,2
0,4
0,6
0,8
1,0
x_T
e

Figure 9(a,b,c,d,e). The dependences of the ratios $R = \frac{d\sigma_{POL}(q\bar{q}\to q\bar{q}\gamma)}{d\sigma_{POL}(q\bar{q}\to g\gamma\gamma)+d\sigma_{POL}(q\bar{q}\to gg\gamma)+d\sigma_{POL}(q\bar{q}\to q\bar{q}\gamma)}$, $R = \frac{d\sigma_{POL}(q\bar{q}\to gg\gamma)}{d\sigma_{POL}(q\bar{q}\to g\gamma\gamma)+d\sigma_{POL}(q\bar{q}\to gg\gamma)+d\sigma_{POL}(q\bar{q}\to q\bar{q}\gamma)}$ and $R = \frac{d\sigma_{POL}(q\bar{q}\to g\gamma\gamma)}{d\sigma_{POL}(q\bar{q}\to g\gamma\gamma)+d\sigma_{POL}(q\bar{q}\to gg\gamma)+d\sigma_{POL}(q\bar{q}\to q\bar{q}\gamma)}$ on the sum of the energy of colliding protons $\sqrt{s}$, transverse momentum $p_T$, cosine of the scattering angle of the produced prompt photons $Cos(\theta)$, rapidity y and parameter $x_T$ at $\sqrt{s}$=10 GeV has been investigated, which determines the share of a specific subprocess in the total polarized ($P_1 = P_2 = \pm 0.9$, that is $P_1 P_2 = -0.81$) cross section

As seen Figure 9(a), the dependence on sum energy of colliding proton as in the nonpolarized case, curve 2. $q\bar{q} \to gg\gamma$ unconditionally dominates, accounting for approximately 100% of the spin-dependent cross section, while the contributions of processes 1. $q\bar{q} \to q\bar{q}\gamma$ and 3. $q\bar{q} \to g\gamma\gamma$ tend to zero. In the region of small momentum transfers, the annihilation into two massless gluons has a huge cross section due to infrared divergences. Since vector bosons in QCD preserve helicity, this process generates the largest absolute cross section difference $\Delta\sigma$ at small $\sqrt{s}$.

The dependence on $p_T$ (Figure 9(b)) shows a radically different picture of the differential spin dynamics. At small $p_T$, the 3. $q\bar{q} \to g\gamma\gamma$ subprocess dominates, but as the transverse momentum increases, a sharp "reversal" occurs: curve 3. $q\bar{q} \to g\gamma\gamma$ rapidly drops to zero, while curve 1. $q\bar{q} \to q\bar{q}\gamma$ increases abruptly, taking up ≈100% of the polarized cross section. Process 2. $q\bar{q} \to gg\gamma$ is suppressed over the entire range. At large $p_T$, scattering occurs on valence quarks (large $x$), whose polarization in the proton is maximal. The quark scattering diagram 1. $q\bar{q} \to q\bar{q}\gamma$ conserves spin much more efficiently in hard collisions than multi-gluon channels. The decrease in the $q\bar{q} \to g\gamma\gamma$ fraction indicates that spin correlations in multiphoton/gluon vertices quickly collapse (or their amplitudes cancel each other out) at large momentum transfers $p_T$.

The distribution of the polarized cross-section fractions by angle (Figure 9(c)) showed that. curve 3. $q\bar{q} \to g\gamma\gamma$ is absolutely dominant at ≈100% for any angle θ. The fractions of processes 1. $q\bar{q} \to q\bar{q}\gamma$ and 2. $q\bar{q} \to gg\gamma$ lie near zero. Similar to the angular distribution, curve 3. $q\bar{q} \to g\gamma\gamma$ is a horizontal line at a level of ≈100%, with a small fluctuation exactly in the center ($y = 0$).

The dependence on $x_T$ (Figure 9(e)) confirms the dynamics found for $p_T$. Curve 1. $q\bar{q} \to q\bar{q}\gamma$ increases smoothly (from $R \approx 30\%$ to $R \approx 70\%$), while curve 3. $q\bar{q} \to g\gamma\gamma$ decreases symmetrically as $x_T \to 1$ approaches.

4.5 Ratios $R = \frac{d\sigma_{POL}(q\bar{q}\to q\bar{q}\gamma)}{d\sigma(q\bar{q}\to q\bar{q}\gamma)}$, $R = \frac{d\sigma_{POL}(q\bar{q}\to gg\gamma)}{d\sigma(q\bar{q}\to gg\gamma)}$ and $R = \frac{d\sigma_{POL}(q\bar{q}\to g\gamma\gamma)}{d\sigma(q\bar{q}\to g\gamma\gamma)}$

Figure 10(a,b,c,d,e) shows the dependences of the ratios $R = \frac{d\sigma_{POL}(q\bar{q}\to q\bar{q}\gamma)}{d\sigma(q\bar{q}\to q\bar{q}\gamma)}$, $R = \frac{d\sigma_{POL}(q\bar{q}\to gg\gamma)}{d\sigma(q\bar{q}\to gg\gamma)}$ and $R = \frac{d\sigma_{POL}(q\bar{q}\to g\gamma\gamma)}{d\sigma(q\bar{q}\to g\gamma\gamma)}$ on the sum of energies of colliding protons $\sqrt{s}$, transverse momentum $p_T$, cosine of scattering angle $Cos(\theta)$, rapidity $y$ of photon and $x_T$ in collisions longitidunally polarized protons with polarization order $P_1 = P_2 = \pm 0.9$ that is $P_1 P_2 = -0.81$.

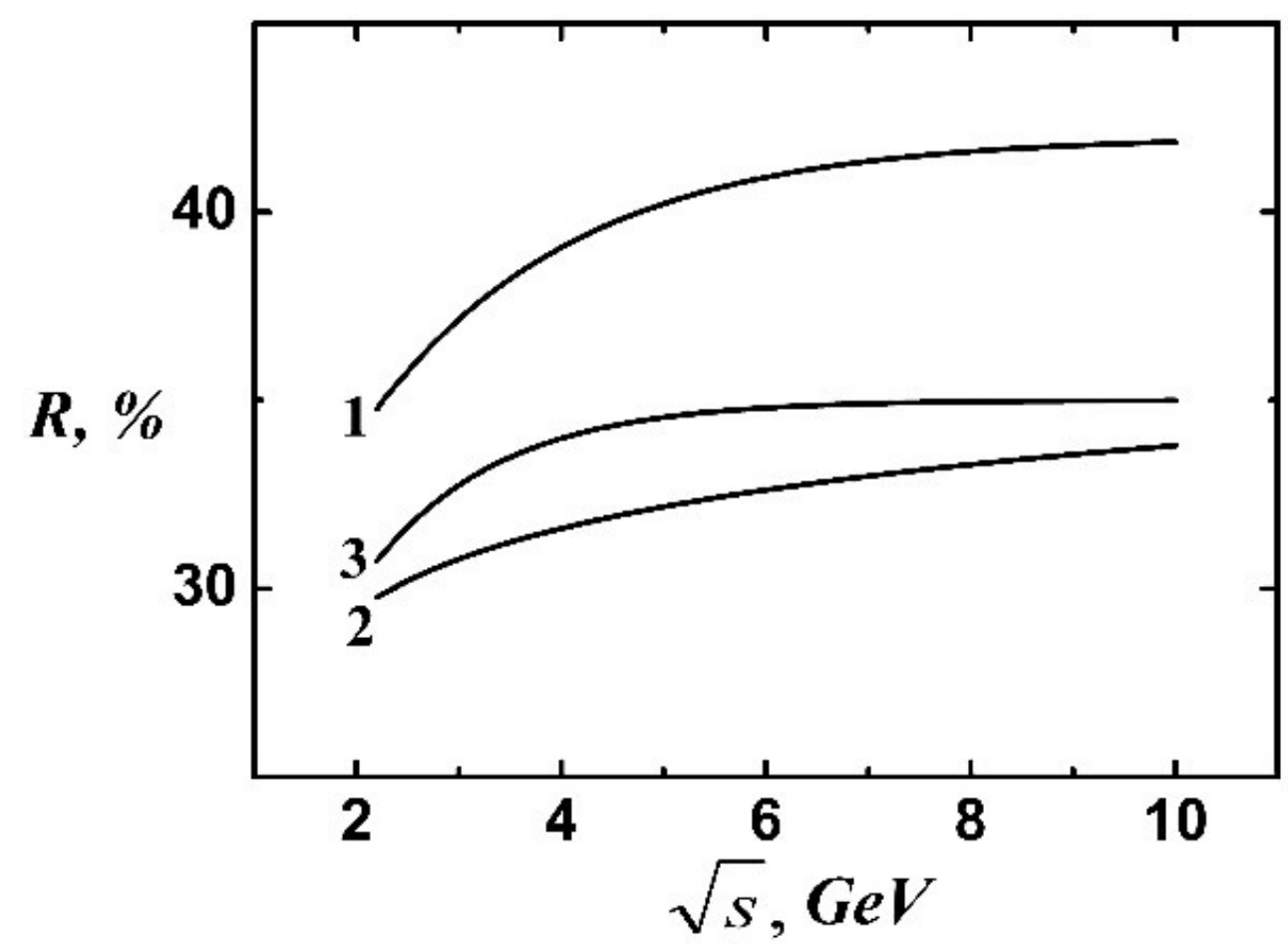

a

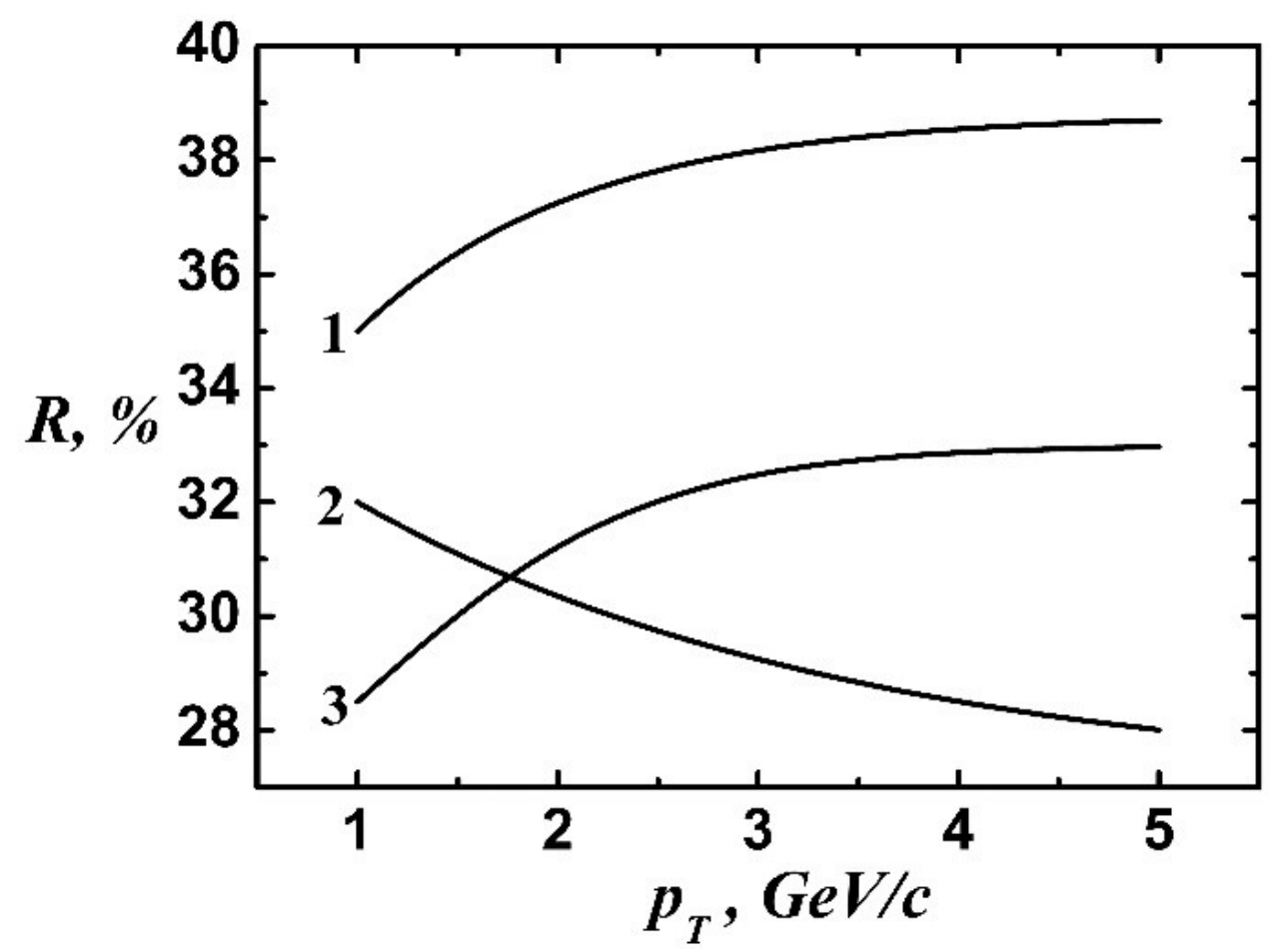

b

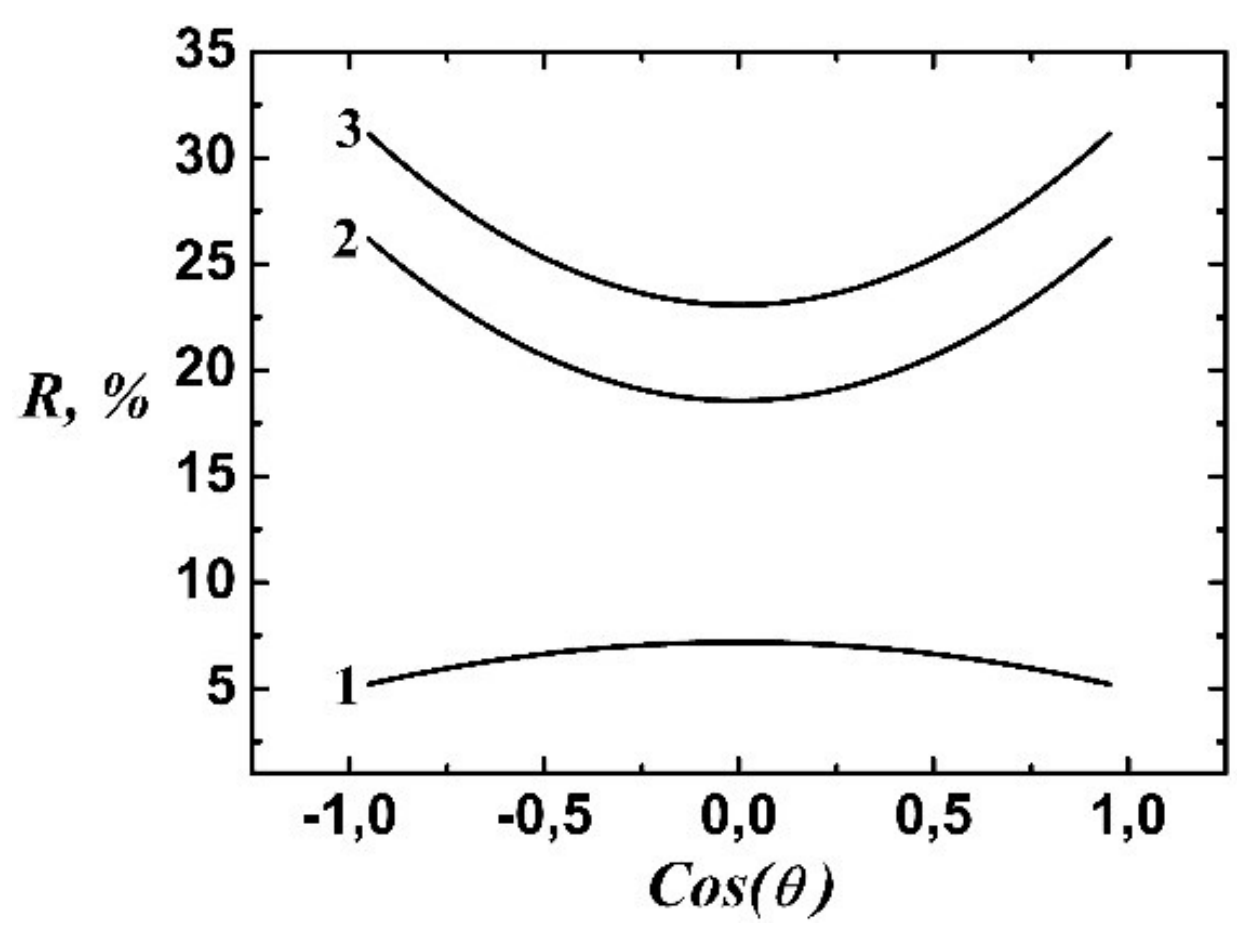

c

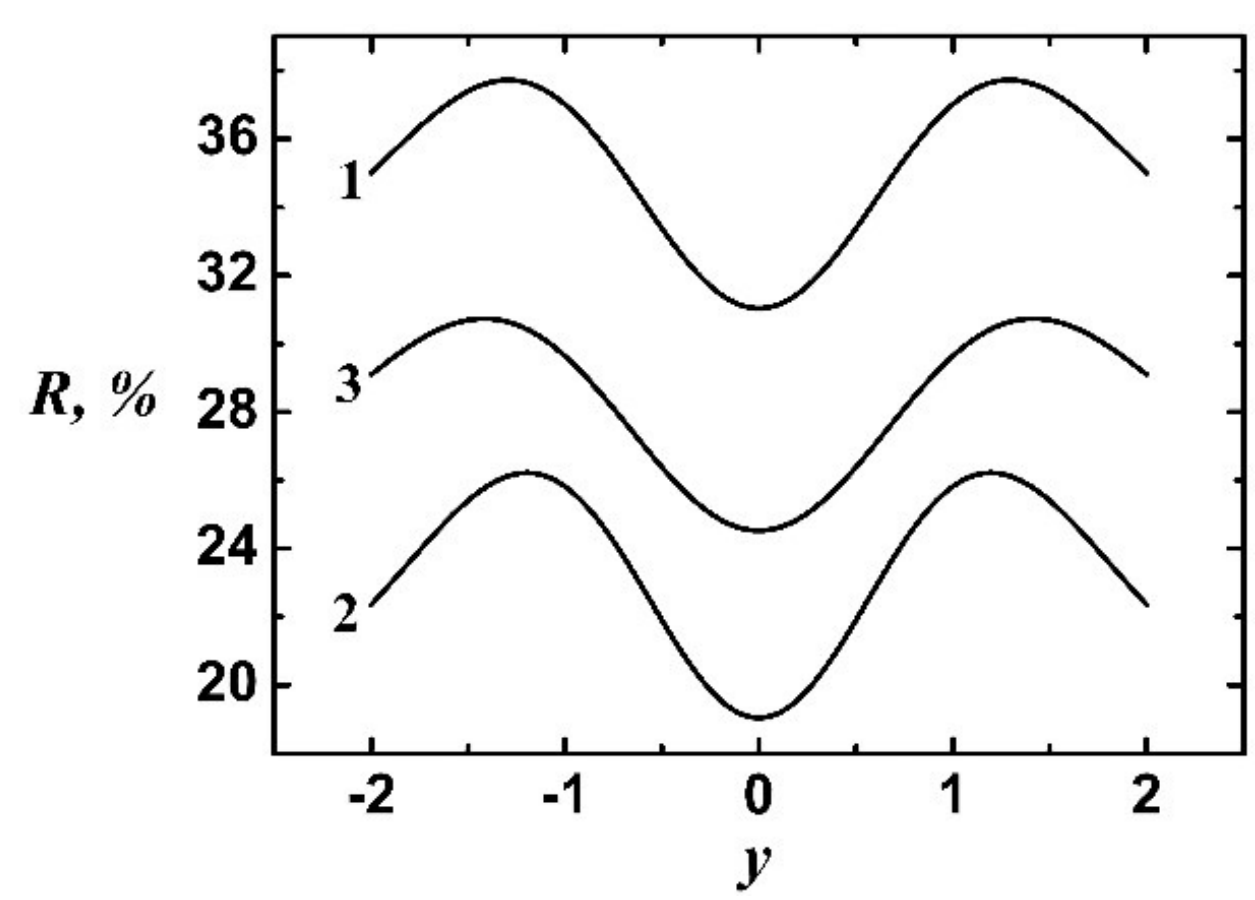

d

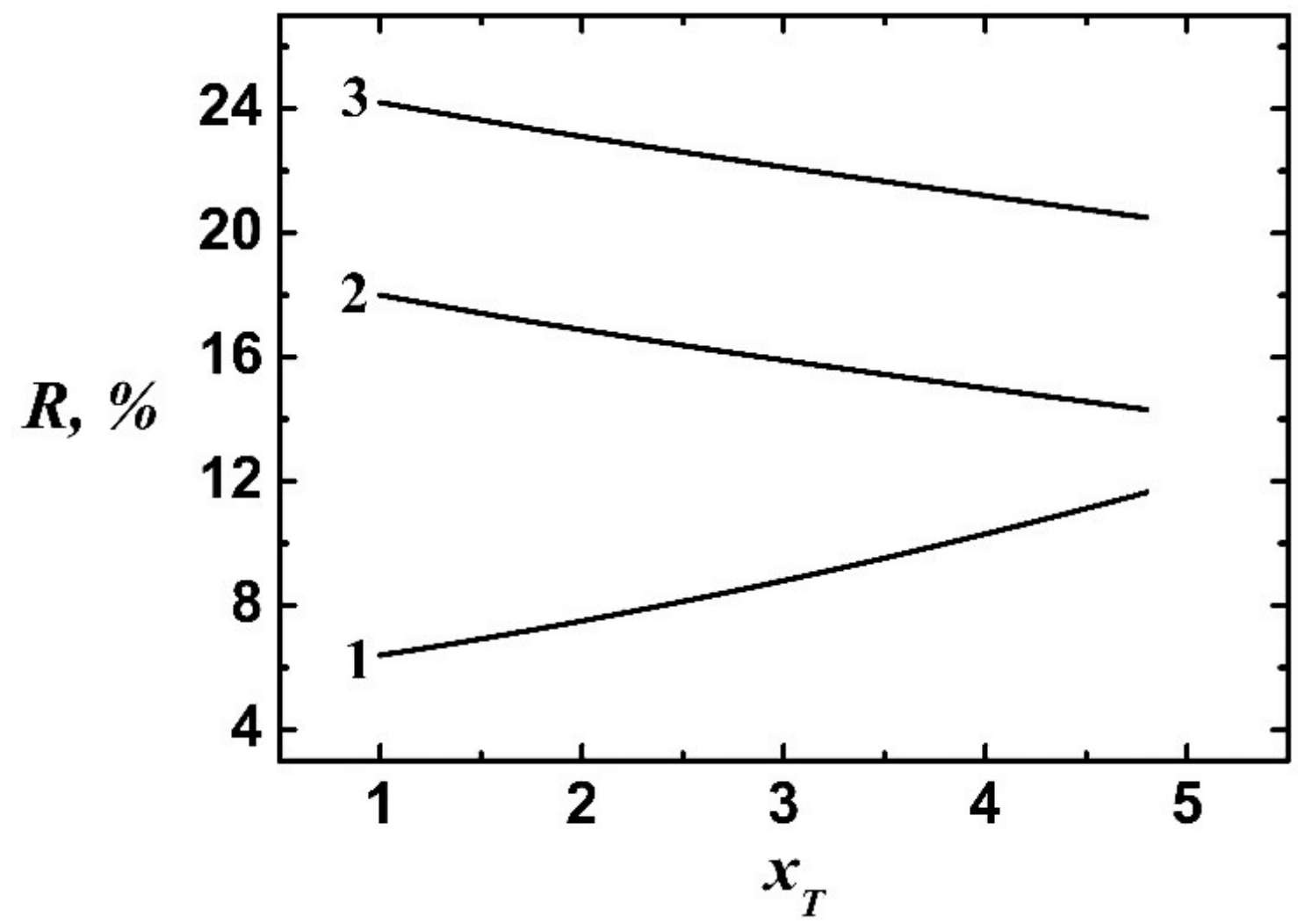

e

Figure 10(a,b,c,d,e). Dependences of the ratios $R = \frac{d\sigma_{POL}(q\bar{q}\to q\bar{q}\gamma)}{d\sigma(q\bar{q}\to q\bar{q}\gamma)}$ (curve 1), $R = \frac{d\sigma_{POL}(q\bar{q}\to gg\gamma)}{d\sigma(q\bar{q}\to gg\gamma)}$ (curve 2) and $R = \frac{d\sigma_{POL}(q\bar{q}\to g\gamma\gamma)}{d\sigma(q\bar{q}\to g\gamma\gamma)}$ (curve 3) on the sum of the energy of colliding protons $\sqrt{s}$ (a), transverse momentum $p_T$ (b), scattering angle cosine of the produced prompt photons $Cos(\theta)$ (c), rapidity y (d) and parameter $x_T$ (e) in collisions longitidunally polarized protons with polarization order $P_1 = P_2 = \pm 0.9$ (that is $P_1 P_2 = -0.81$).at $\sqrt{s}$=10 GeV.

The values of $R$ determines reflects the sensitivity of a particular subprocess to the polarization of the initial beams at $P_1 = P_2 = \pm 0.9$.

As seen from Figure 10(a), the ratio increases smoothly in the near-threshold region (up to $\sqrt{s}$≈4-6 GeV) and then reaches a horizontal asymptotic plateau. At low energies, the process is confined to a narrow phase volume, where the effects of small momentum fractions of x dominate. As the energy increases, we enter the deeply inelastic scaling regime, where the ratio of polarized to nonpolarized matrix elements stabilizes. The difference in plateau levels curve 1. $q\bar{q} \to q\bar{q}\gamma$ is higher than curves 2. $q\bar{q} \to gg\gamma$ and 3. $q\bar{q} \to g\gamma\gamma$ indicates that the subprocess with conservation of the fermion line 1. $q\bar{q} \to q\bar{q}\gamma$ has a different helical structure compared to pure annihilation into bosons 2. $q\bar{q} \to gg\gamma$.

Figure 10(b) shows the change in the ratio $R$ with increasing transverse momentum of prompt photons. Curve 1. $q\bar{q} \to q\bar{q}\gamma$ increases with increasing $p_T$, curve 2. $q\bar{q} \to gg\gamma$ increases slightly, and curve 3. $q\bar{q} \to g\gamma\gamma$ gradually decreases. The increase in curve 1. $q\bar{q} \to q\bar{q}\gamma$ directly reflects the fact that the harder the scattering, the greater the contribution of quark spin effects. The decrease in curve 3. $q\bar{q} \to g\gamma\gamma$ is due to the interference of the diagrams and the specificity of the electromagnetic peaks during the emission of two photons.

The angular dependence of the ratio $R$ is shown in Figure 10(c). The curves have a wave-like shape with a pronounced minimum strictly in the center at $Cos(\theta) = 0$ (scattering by 90°). Approaching the collinear limits, the fraction of the polarized cross section increases, but decreases slightly at the very edges. The shape of these curves is dictated exclusively by the kinematics of the matrix elements $\mathcal{M}$. In the unpolarized case, the angular distribution of the annihilation is proportional to $(1 + \text{C}os^2(\theta))$, while the interference spin terms have a different dependence (for example, linear in $Cos(\theta)$ or with other combinations of Mandelstam invariants). The ratio of two different polynomials in the angle yields the observed minimum in the orthogonal direction. This means that spin correlations are least noticeable for scattering perpendicular to the collision axis.

The dependence of $R$ on the rapidity y (Figure 10(d)) shows that the curve has an oscillating (W-shaped) profile. The minima are located in the central region ($y$ =0), and the maxima are in the intermediate rapidity regions ($|y|$ ≈1.5). Since the rapidity is kinematically related to the scattering angle $\theta$, the dip at $y = 0$ is a direct consequence of the minimum at $Cos(\theta) = 0$. However, at the rapidity edges, the effects of parton distributions are included. High rapidity implies a strong asymmetry in the momenta of colliding partons ($x_1 \gg x_2$). The behavior of the ratio $R$ here reflects the delicate balance between the polarization of the fast valence quark and the unpolarized background of the slow sea antiquark.

The dependences of $R$ on $x_T$ (Figure 10(e)) are monotonic, curve 1. $q\bar{q} \to q\bar{q}\gamma$ increases, curves 2. $q\bar{q} \to gg\gamma$ and 3. $q\bar{q} \to g\gamma\gamma$ decrease when approaching the kinematic limit $x_T \to 1$. The graph is a dimensionless analogue of the dependence on $p_T$.

4.6 Comparison of differential cross-sections calculated in FeynCalc and those obtained by simulation in PYTHIA 8.316.

Figure 11(a,b,c,d,e) shows the dependences of $R = \frac{d\sigma_{FeynCalc}(q\bar{q}\to q\bar{q}\gamma)}{d\sigma_{Pythia}(q\bar{q}\to q\bar{q}\gamma)}$, $R = \frac{d\sigma_{FeynCalc}(q\bar{q}\to gg\gamma)}{d\sigma_{Pythia}(q\bar{q}\to gg\gamma)}$, and $R = \frac{d\sigma_{FeynCalc}(q\bar{q}\to g\gamma\gamma)}{d\sigma_{Pythia}(q\bar{q}\to g\gamma\gamma)}$ on the sum of the energy of colliding protons $\sqrt{s}$, transverse momentum $p_T$, the cosine of the scattering angle of the produced prompt photons $Cos(\theta)$, rapidity y, and parameter $x_T$ at $\sqrt{s}$=10 GeV.

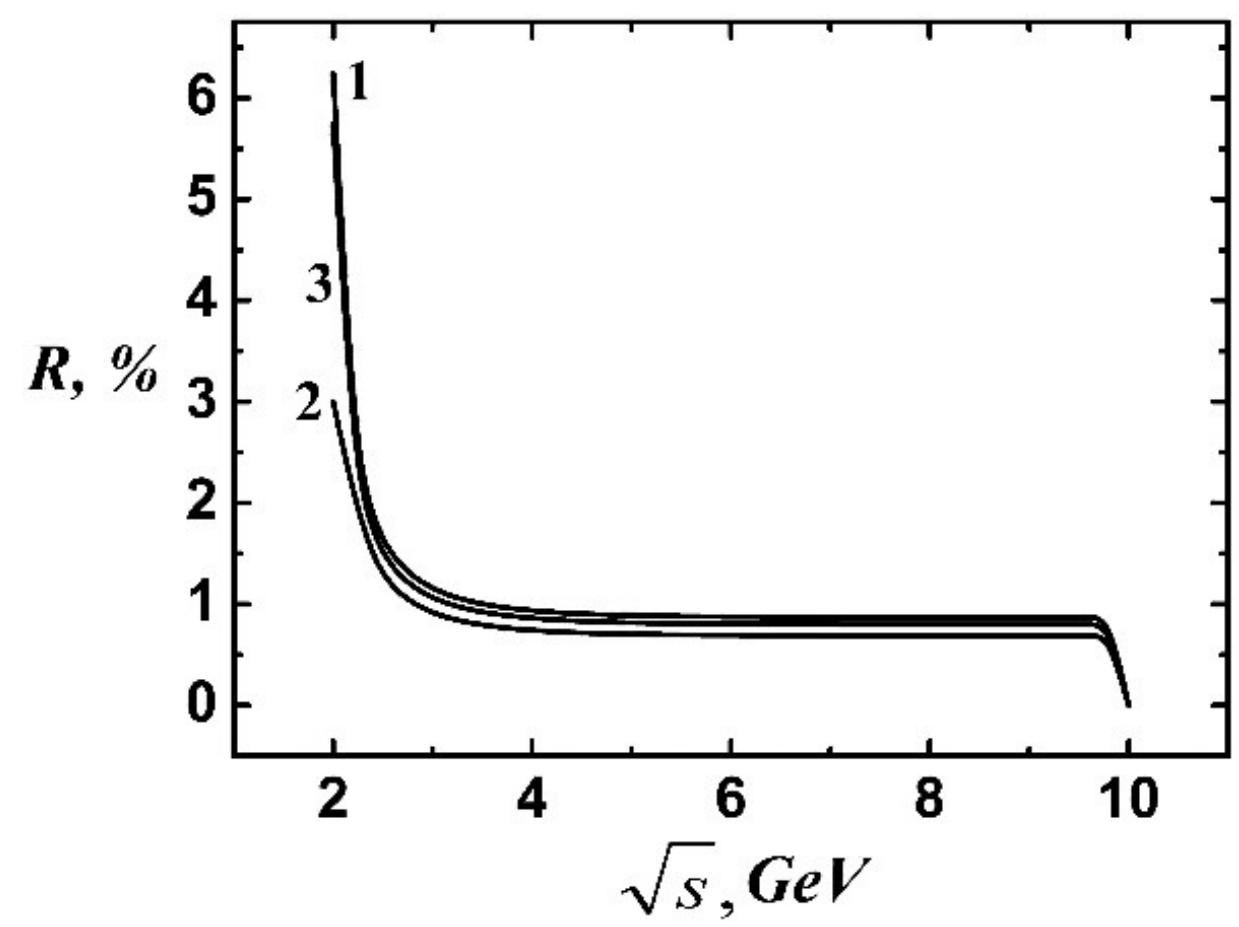

a

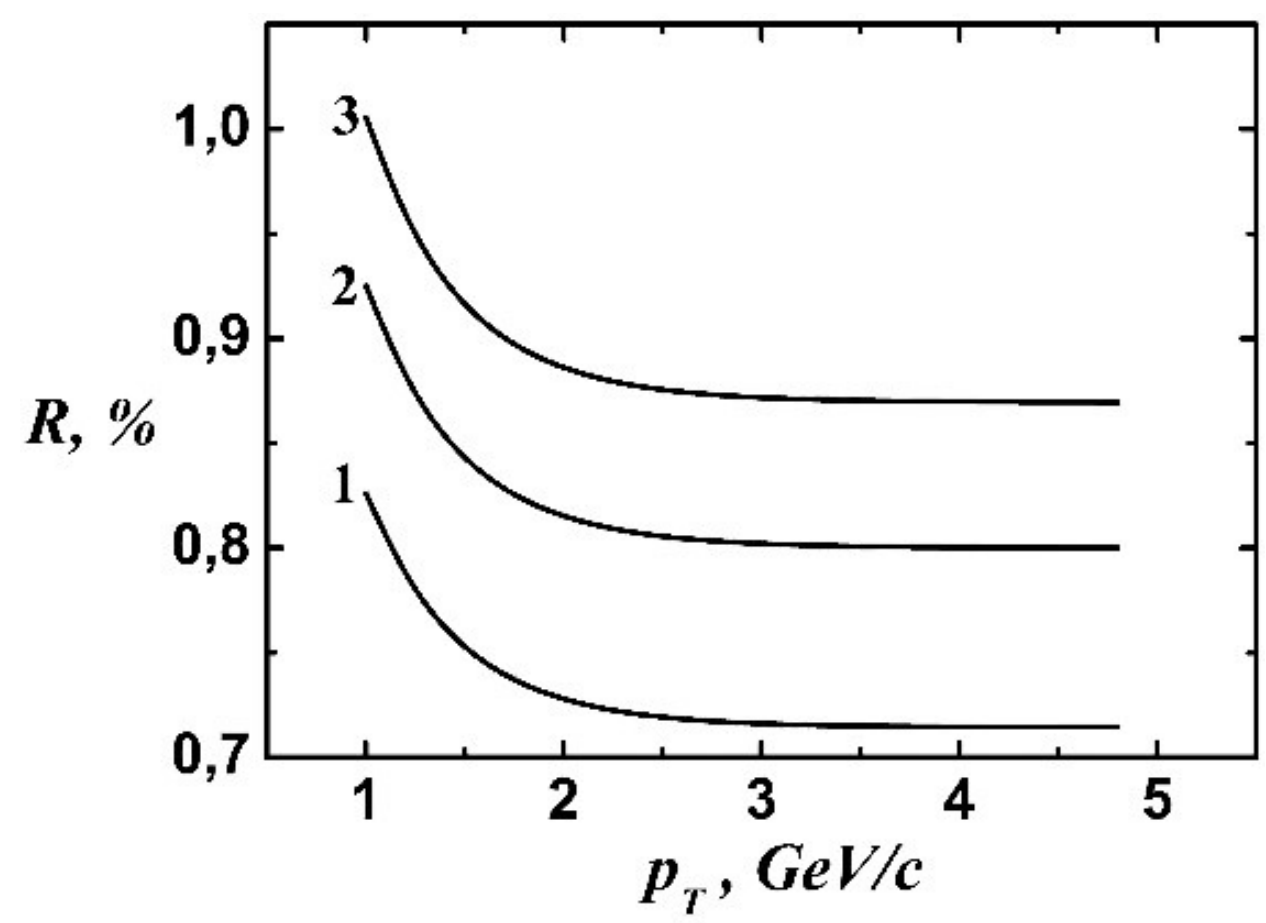

b

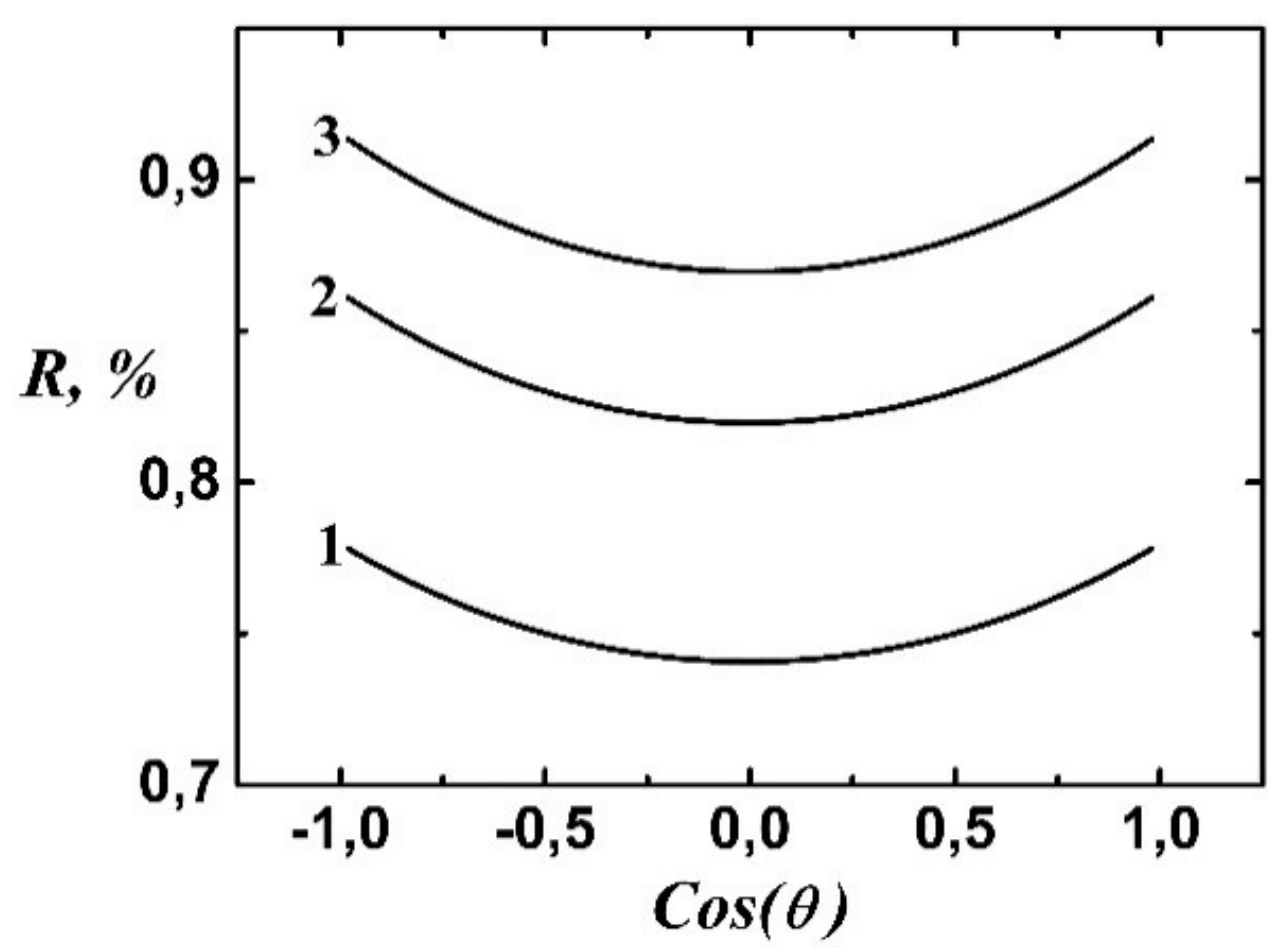

c

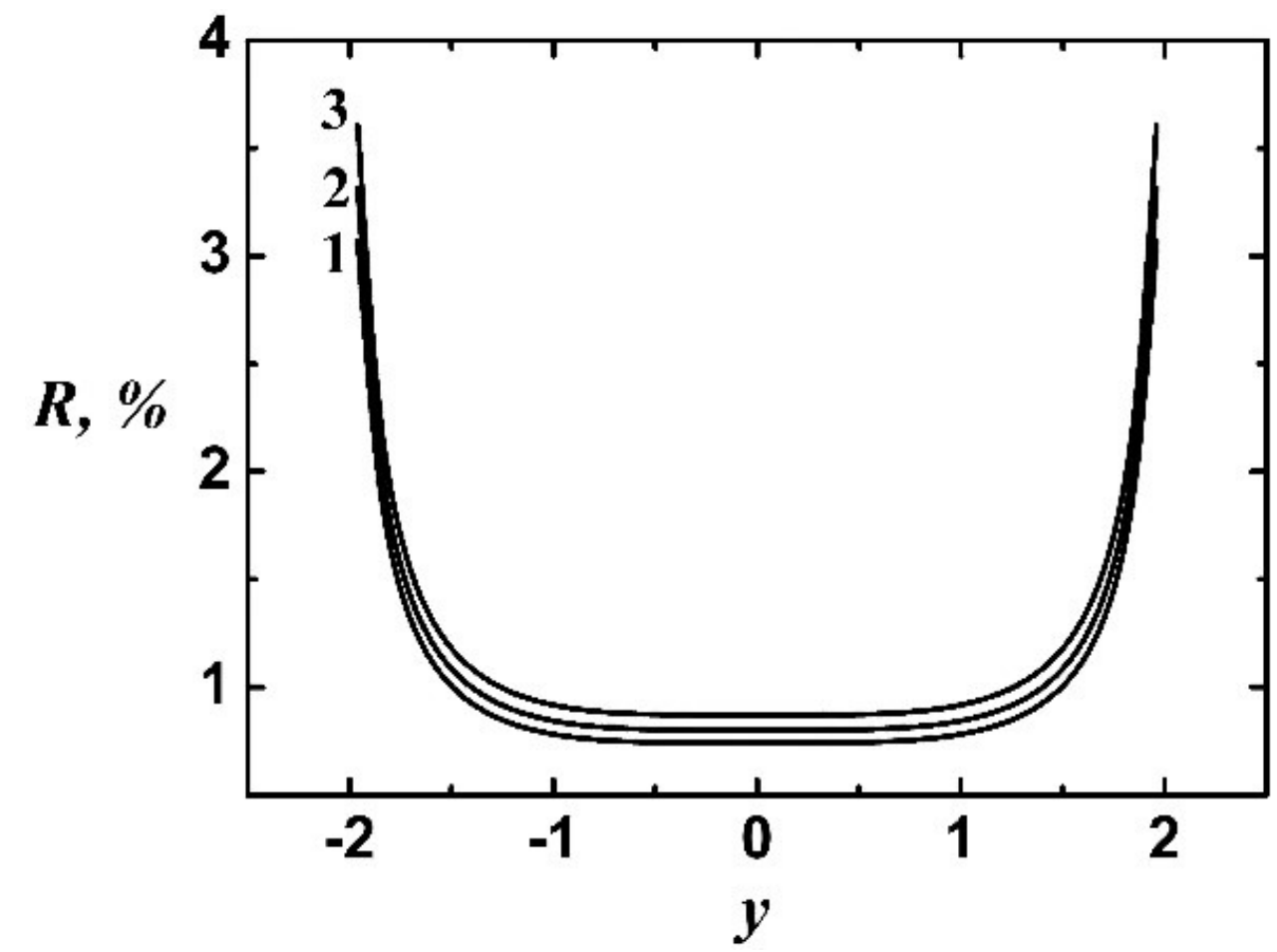

d

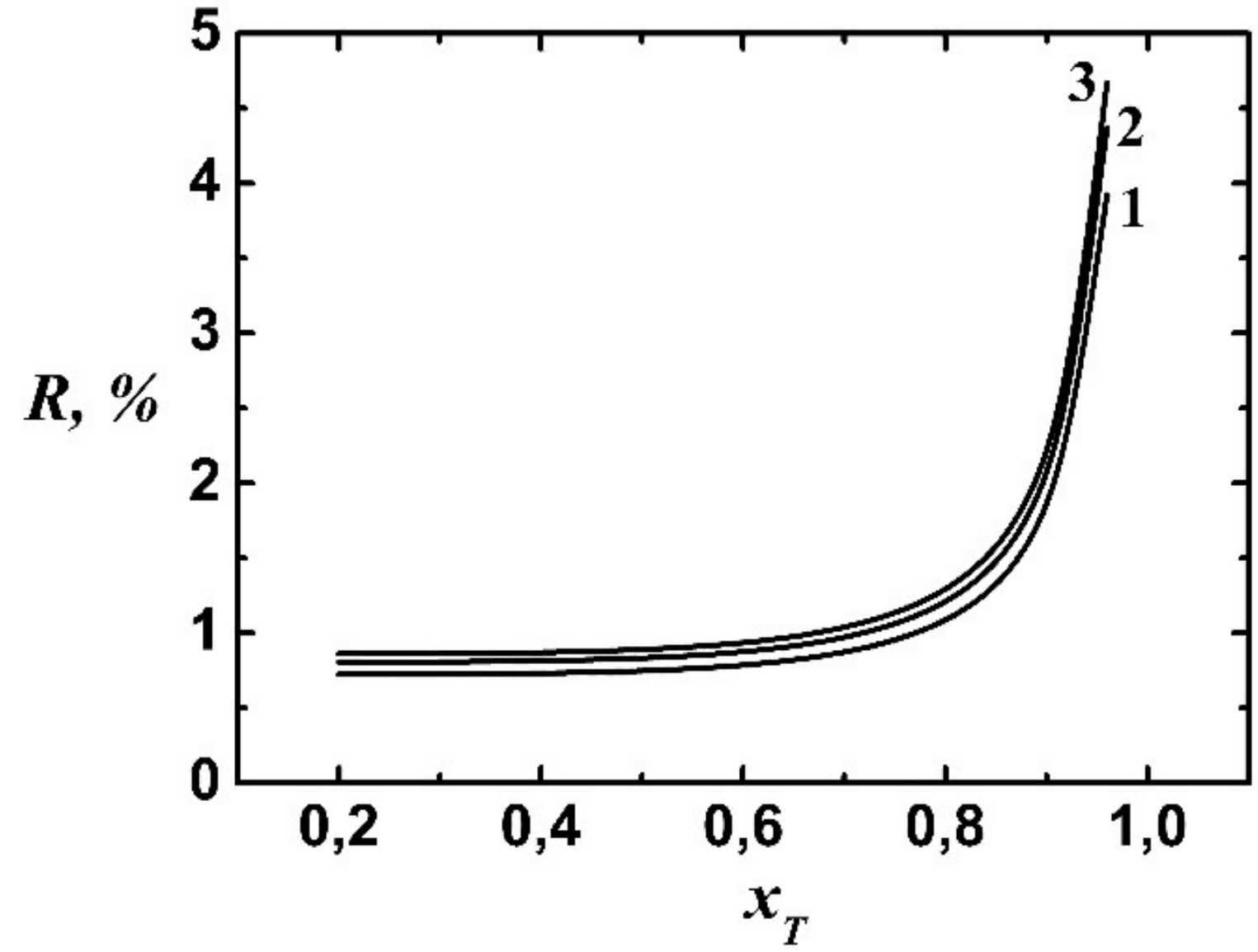

e

Figure 11(a,b,c,d,e). Dependences of ratios $R = \frac{d\sigma_{FeynCalc}(q\bar{q}\to q\bar{q}\gamma)}{d\sigma_{Pythia}(q\bar{q}\to q\bar{q}\gamma)}$ (curve 1), $R = \frac{d\sigma_{FeynCalc}(q\bar{q}\to gg\gamma)}{d\sigma_{Pythia}(q\bar{q}\to gg\gamma)}$ (curve 2) and $R = \frac{d\sigma_{FeynCalc}(q\bar{q}\to g\gamma\gamma)}{d\sigma_{Pythia}(q\bar{q}\to g\gamma\gamma)}$ (curve 3) on the sum of the energy of colliding protons $\sqrt{s}$ (a), transverse momentum $p_T$ (b), the cosine of the scattering angle of the produced prompt photons $Cos(\theta)$ (c), rapidity y (d) and parameter $x_T$ (e) at $\sqrt{s}$=10 GeV.

This ratio is an indicator of the influence of parton showers, which are present in PYTHIA 8.316 but are absent in a rigorous analytical calculation at the fixed-order level of perturbation theory (FeynCalc).

Figure 11(a,b,c,d,e) presents the kinematic dependences of the ratio $R$, which compares strict analytical leading-order calculations (LO, FeynCalc) with MC modulation that include parton showers (PYTHIA 8.316). Based on the asymptotic behavior of the curves at the phase space boundaries, this graph unequivocally demonstrates the ratio $R = \frac{d\sigma_{FeynCalc}}{d\sigma_{Pythia}}$ for the subprocesses 1. $q\bar{q} \to q\bar{q}\gamma$, 2. $q\bar{q} \to gg\gamma$, and 3. $q\bar{q} \to g\gamma\gamma$.

As seen from Figure 11(a), at near-threshold energies (small $\sqrt{s}$), the ratio starts at high values ($R \approx 5-6$ %) and rapidly drops to a plateau of $R \approx 1$. The physical reason is that near the threshold, the phase space is extremely narrow. The PYTHIA 8.316 generator requires additional energy to initiate a parton shower (ISR/FSR) and compensate for the recoil momentum. Therefore, at low energies, PYTHIA 8.316 generates significantly fewer events (the denominator is small), which causes the ratio to rise. As the phase space opens up at higher energies, the analytical and MC calculations converge.

In Figure 11(b), the ratio $R < 1\%$ (the curves lie in the range of $R \approx 0.7 - 1.0\%$). This is a consequence of Initial State Radiation (ISR) in PYTHIA 8.316. The emission of soft gluons before the hard collision imparts a transverse recoil momentum ($k_T$-kick) to the system. This kinematic smearing pushes some events from the low-$p_T$ region into the hard tail of the spectrum. As a result, the PYTHIA 8.316 cross-section at moderate and high $p_T$ is slightly larger than the "bare" FeynCalc analytics, making the fraction less than one.

As seen from Figure 11(c), the curves have a pronounced $U$-shape. In the center ($Cos(\theta) \approx 0$), $R \approx 1$. However, approaching the collinear limits ($Cos(\theta) \rightarrow \pm 1$), the ratio increases sharply. FeynCalc LO matrix elements contain $t$- and $u$-channel poles ($\sim 1/\hat{t}$) that diverge to infinity at small angles. PYTHIA 8.316 regularizes these unphysical singularities using angular ordering and minimum shower $p_T$ cutoffs. Because the numerator (FeynCalc) diverges while the denominator (PYTHIA) is regularized, the ratio tends to large values at the edges.

The rapidity dependence (Figure 11(d)) exhibits a deep $U$-shaped structure, similar to the angular distribution. In the central zone ($|y| < 1$), the analytics and simulation are in good agreement. At the kinematic edges ($|y| \rightarrow 2$), which correspond to a strong longitudinal momentum imbalance and collinear scattering, the absence of resummation effects in FeynCalc leads to an overestimated cross-section, whereas Sudakov suppression in PYTHIA keeps the cross-section finite. This causes a sharp non-linear growth of the ratio $R$ at the rapidity boundaries.

Figure 11(e) shows the dependence on $x_T = 2p_T/\sqrt{s}$. As $x_T \rightarrow 1$, the curves experience a catastrophic surge from $R \approx 1$ to $R \approx 4 - 5$. The exclusive limit $x_T \rightarrow 1$ implies that the produced photon carries away all the available transverse energy. FeynCalc analytics simply evaluate the formula at this point ($d\sigma > 0$). However, in reality (as modeled by PYTHIA 8.316), the probability of a hard strike occurring without the emission of even a single soft gluon tends to zero (suppression by the Sudakov form factor). The denominator (PYTHIA) approaches zero, causing the ratio $R \rightarrow \infty$. This graph is a brilliant illustration of the necessity of logarithmic resummation at the edges of the phase space.

## CONCLUSIONS

The sum of the contributions of radiative corrections 1. $q\bar{q} \to q\bar{q}\gamma$, 2. $q\bar{q} \to gg\gamma$, and 3. $q\bar{q} \to g\gamma\gamma$ to the subprocess annihilation of quark-antiquark pairs is approximately 7.8% of the main subprocess.. This indicates a weak interaction of the quark with the quantum of the electric field - the photon.

Radiative corrections are particularly significant at low collision energies of proton $\sqrt{s}$ and small transverse momentum of photon $p_T$.

The cross sections of the radiative corrections strictly satisfy the relation

$$2.\ \sigma_{\sqrt{s}}(q\bar{q} \to gg\gamma) > 1.\ \sigma_{\sqrt{s}}(q\bar{q} \to q\bar{q}\gamma) > 3.\ \sigma_{\sqrt{s}}(q\bar{q} \to g\gamma\gamma)$$

$$2.\quad \frac{d\sigma(q\bar{q}\to q\bar{q}\gamma)}{dp_T^2} > 3.\ \frac{d\sigma(q\bar{q}\to g\gamma\gamma)}{dp_T^2} > 2.\ \frac{d\sigma(q\bar{q}\to gg\gamma)}{dp_T^2}.$$

The gluon channel $gg\gamma$ consistently dominates, accounting for $R$ ≈80–100% of the total integrated cross-section due to strong infrared and collinear enhancements associated with massless gluons.

However, as $p_T$ approaches the absolute kinematic limit ($x_T \to 1$), the relative share of the 1. $q\bar{q} \to q\bar{q}\gamma$ process rapidly increases, reaching $R \approx 100\%$, as it represents the hardest interaction mechanism.

Longitudinal polarization of the colliding protons strongly affects the subprocesses. The double spin asymmetry $A_{LL}$ demonstrates that spin correlations in multi-gluon/photon vertices ($gg\gamma$, $g\gamma\gamma$) are quickly smeared out or cancelled out. Conversely, the 1. $q\bar{q} \to q\bar{q}\gamma$ channel efficiently conserves helicity, leading to macroscopic asymmetry values (approaching 100%) at large momentum transfers and transverse scattering angles.

A direct comparison between strict LO analytical calculations and PYTHIA 8.316 simulations revealed excellent agreement in the central kinematic regions. However, at the boundaries of the phase space (collinear limits, extreme rapidity, and $x_T \to 1$), analytical LO matrix elements suffer from unphysical infrared and collinear divergences. PYTHIA 8.316 effectively regularizes these singularities via Sudakov form factors, initial/final state radiation (ISR/FSR), and transverse recoil ($k_T$-kick).

The cross-section drops precipitously as the photon's transverse momentum $p_T$ approaches the kinematic limit $p_T \to \frac{\sqrt{s}}{2}$.

Radiative corrections at NICA modify the cross-section moderately, but they are highly sensitive to threshold resummation effects due to soft-gluon emission near the edge of the phase space.

The fragmentation contribution (where a photon is radiated collinearly from an outgoing quark) is suppressed compared to direct production.

Looking at radiative corrections to annihilation quark-antquark pair $q\bar{q} \to g\gamma$ at NICA energies provides a clean environment to test clean pQCD calculations near the kinematic threshold without the overwhelming gluon backgrounds found at the LHC.

## APPENDIX 1.
## THE FEYNMAN DIAGRAMS AND MATRIX ELEMENTS OF SUBPROCESS 1. $q\bar{q} \to q\bar{q}\gamma$

**A1_1.** Radiative correction 1. $q\bar{q} \to q\bar{q}\gamma$ with a photon propagator

In the 1. $q\bar{q} \to q\bar{q}\gamma$ subprocess, the initial quark-antiquark pair can annihilate into a photon or gluon, after which this virtual particle (photon or gluon) forms a quark-antiquark pair. Subsequently, one of the final particles (quark or antiquark) produce a photon.

The Feynman diagrams for the radiative correction 1. $q\bar{q} \to q\bar{q}\gamma$ with a photon propagator are shown in Figure A1_1(a,b,c,d,e,f,g,h).

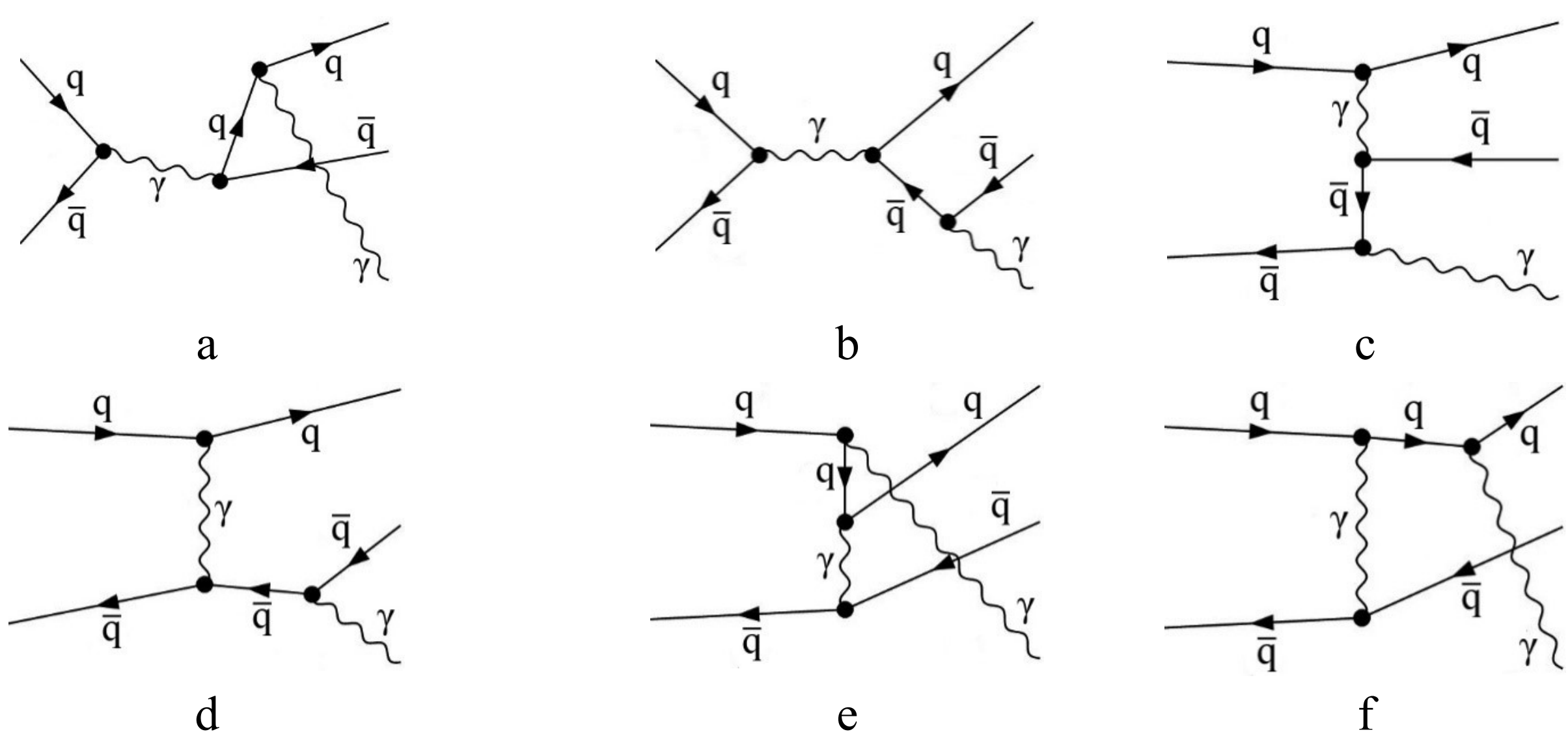

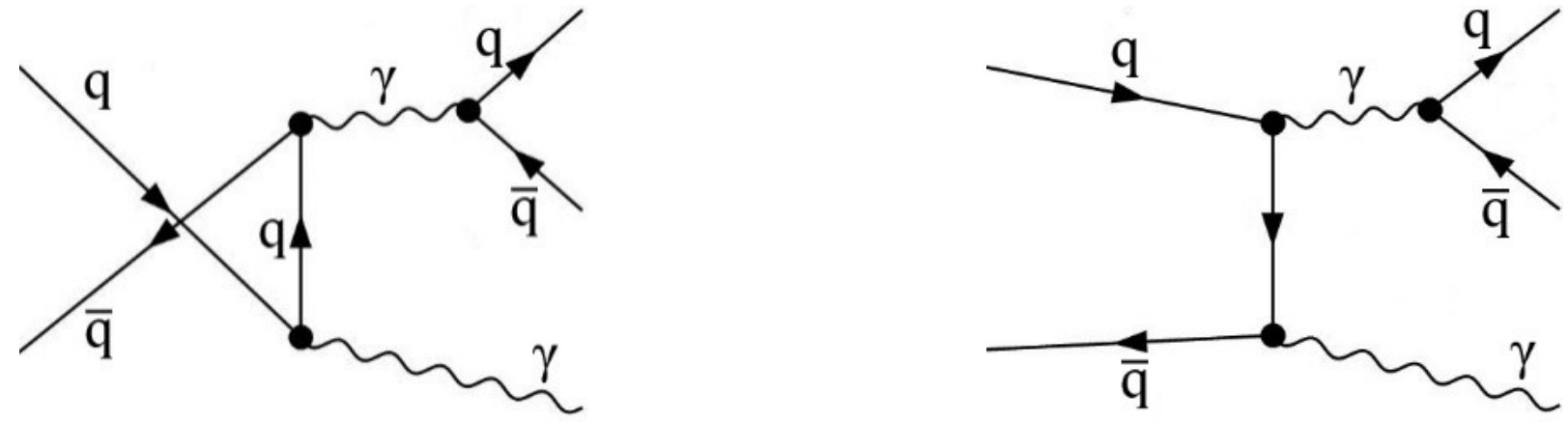

g h

Figure A1_1(a,b,c,d,e,f,g,h). Feynman diagrams of the 1. $q\bar{q} \to q\bar{q}\gamma$ subprocess with a photon propagator

The Mandelstam invariants for the subprocess $q\bar{q} \to q\bar{q}\gamma$ are given below:

$$\hat{s} = (p_1 + p_2)^2 = (k_1 + p_3 + p_4)^2, \hat{t} = (p_1 - k_1)^2 = (p_4 + p_3 - p_2)^2,$$
$$\hat{u} = (p_2 - k_1)^2 = (p_4 + p_3 - p_1)^2, \hat{q}_1 = (p_1 - p_3)^2 = (k_1 + p_4 - p_2)^2,$$
$$\hat{q}_2 = (p_2 - p_4)^2 = (k_1 + p_3 - p_1)^2, \hat{s}_1 = (p_4 + p_3)^2 = (p_1 + p_2 - k_1)^2,$$
$$\hat{t}_1 = (p_1 - p_4)^2 = (k_1 + p_1 - p_2)^2, \hat{u}_1 = (p_2 - p_2)^2 = (k_1 + p_4 - p_1)^2,$$
$$\hat{q}_3 = (p_3 + k_1)^2 = (p_1 + p_2 - p_1)^2, \hat{q}_4 = (p_4 + k_1)^2 = (p_1 + p_2 - p_3)^2.$$

The matrix elements of the subprocess $q\bar{q} \to q\bar{q}\gamma$ with photon propagator have the following form:

$$M_1 = \frac{-4e^3 e_q \delta_{Col1Col2} \delta_{Col3Col4} \bar{v}(p_2)\gamma U(p_1) \bar{U}(p_3) \hat{\varepsilon} * (k_1)\left(\hat{k}_1 + \hat{p}_3 + m_q\right)\gamma v(p_4)}{81(k_1 + p_3 + p_4)^2\left((-k_1 - p_3)^2 - m_q^2\right)},$$
$$M_2 = \frac{-4e^3 e_q \delta_{Col1Col2} \delta_{Col3Col4} \bar{v}(p_2)\gamma U(p_1) \bar{U}(p_3) \gamma\left(-\hat{k}_1 - \hat{p}_4 + m_q\right)\hat{\varepsilon} * (k_1) v(p_4)}{81(k_1 + p_3 + p_4)^2\left((k_1 + p_4)^2 - m_q^2\right)},$$
$$M_3 = \frac{4e^3 e_q \delta_{Col1Col3} \delta_{Col2Col4} \bar{U}(p_3)\gamma U(p_1) \bar{v}(p_2) \hat{\varepsilon} * (k_1)\left(\hat{k}_1 - \hat{p}_2 + m_q\right)\gamma v(p_4)}{81(k_1 - p_2 + p_4)^2\left((p_2 - k_1)^2 - m_q^2\right)},$$
$$M_4 = \frac{4e^3 e_q \delta_{Col1Col3} \delta_{Col2Col4} \bar{U}(p_3)\gamma U(p_1) \bar{v}(p_2) \gamma\left(\hat{k}_1 - \hat{p}_4 + m_q\right)\hat{\varepsilon} * (k_1) v(p_4)}{81(k_1 - p_2 + p_4)^2\left((k_1 + p_4)^2 - m_q^2\right)},$$
$$M_5 =$$
$$= \frac{4e^3 e_q \delta_{Col1Col3} \delta_{Col2Col4} \bar{v}(p_2)\gamma v(p_4) \bar{U}(p_3) \gamma\left(-\hat{p}_2 + \hat{p}_3 + \hat{p}_4 + m_q\right)\hat{\varepsilon} * (k_1) U(p_1)}{81(p_2 - p_4)^2\left((p_2 - p_3 - p_4)^2 - m_q^2\right)},$$
$$M_6 = \frac{4e^3 e_q \delta_{Col1Col3} \delta_{Col2Col4} \bar{v}(p_2)\gamma v(p_4) \bar{U}(p_3) \hat{\varepsilon} * (k_1)\left(\hat{k}_1 + \hat{p}_3 + m_q\right)\gamma U(p_1)}{81(p_4 - p_2)^2\left((-k_1 - p_3)^2 - m_q^2\right)},$$

$$M_7 =$$

$$= \frac{-4e^3 e_q \delta_{Col1Col2} \delta_{Col3Col4} \bar{U}(p_3)\gamma v(p_4)\bar{v}(p_2)\gamma\left(-\hat{p}_2 + \hat{p}_3 + \hat{p}_4 + m_q\right)\hat{\varepsilon} * (k_1)U(p_1)}{81(p_3 + p_4)^2\left((p_2 - p_3 - p_4)^2 - m_q^2\right)},$$

$$M_8 = \frac{-4e^3 e_q \delta_{Col1Col2} \delta_{Col3Col4} \bar{U}(p_3)\gamma v(p_4)\bar{v}(p_2)\hat{\varepsilon} * (k_1)\left(\hat{k}_1 - \hat{p}_2 + m_q\right)\gamma U(p_1)}{81(p_3 + p_4)^2\left((p_2 - k_1)^2 - m_q^2\right)}.$$

**A1_2.** Radiative correction 1. $q\bar{q} \to q\bar{q}\gamma$ with a gluon propagator

In the case where a quark-antiquark pair annihilates into a virtual gluon, the Feynman diagrams of the radiative correction 1. $q\bar{q} \to q\bar{q}\gamma$ will have the form shown in Figure A1_2(a,b,c,d,e,f,g,h).

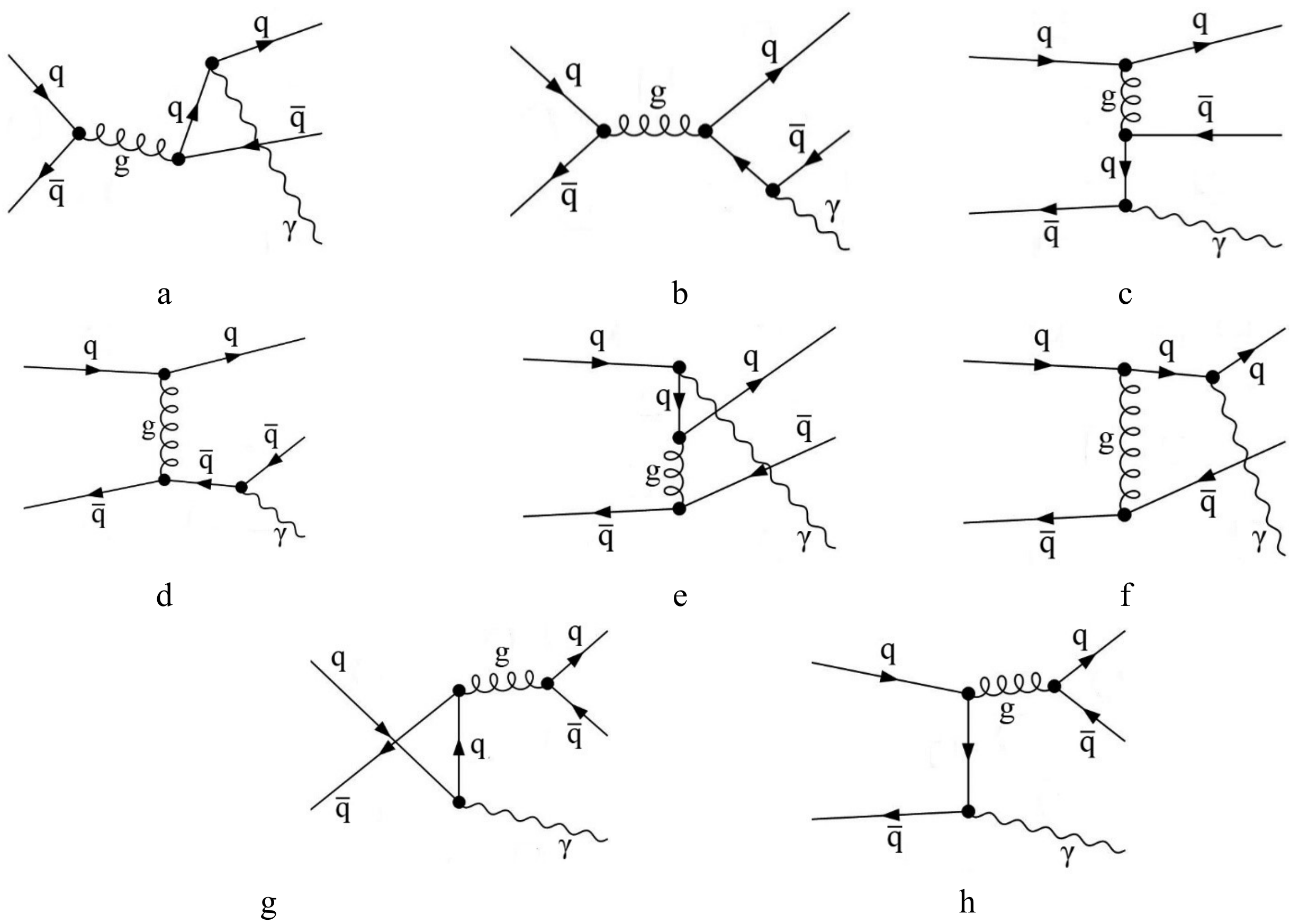

Figure A1_2(a,b,c,d,e,f,g,h). Feynman diagrams of the 1. $q\bar{q} \to q\bar{q}\gamma$ subprocess with a gluon propagator

The Mandelstam invariants in this case remain unchanged.

The matrix elements of the subprocess $q\bar{q} \to q\bar{q}\gamma$ with the gluon propagator are:

$$M_1 = \frac{-ee_q g_s^2 T_{Col2Col1}^{Glu6} T_{Col3Col4}^{Glu6} \bar{\upsilon}(p_2)\gamma U(p_1)\bar{U}(p_3)\hat{\varepsilon} * (k_1)\left(\hat{k}_1 + \hat{p}_3 + m_q\right)\gamma\upsilon(p_4)}{9(k_1 + p_3 + p_4)^2\left((-k_1 - p_3)^2 - m_q^2\right)},$$

$$M_2 = \frac{-ee_q g_s^2 T_{Col2Col1}^{Glu6} T_{Col3Col4}^{Glu6} \bar{\upsilon}(p_2)\gamma U(p_1)\bar{U}(p_3)\gamma\left(-\hat{k}_1 - \hat{p}_4 + m_q\right)\hat{\varepsilon} * (k_1)\upsilon(p_4)}{9(k_1 + p_3 + p_4)^2\left((k_1 + p_4)^2 - m_q^2\right)},$$

$$M_3 = \frac{ee_q g_s^2 T_{Col2Col4}^{Glu6} T_{Col3Col1}^{Glu6} \bar{U}(p_3)\gamma U(p_1)\bar{\upsilon}(p_2)\hat{\varepsilon} * (k_1)\left(\hat{k}_1 - \hat{p}_2 + m_q\right)\gamma\upsilon(p_4)}{9(k_1 - p_2 + p_4)^2\left((p_2 - k_1)^2 - m_q^2\right)},$$

$$M_4 = \frac{ee_q g_s^2 T_{Col2Col4}^{Glu6} T_{Col3Col1}^{Glu6} \bar{U}(p_3)\gamma U(p_1)\bar{\upsilon}(p_2)\gamma\left(\hat{k}_1 - \hat{p}_4 + m_q\right)\hat{\varepsilon} * (k_1)\upsilon(p_4)}{9(k_1 - p_2 + p_4)^2\left((k_1 + p_4)^2 - m_q^2\right)},$$

$$M_5 = \\ = \frac{ee_q g_s^2 T_{Col2Col4}^{Glu6} T_{Col3Col1}^{Glu6} \bar{\upsilon}(p_2)\gamma\upsilon(p_4)\bar{U}(p_3)\gamma\left(-\hat{p}_2 + \hat{p}_3 + \hat{p}_4 + m_q\right)\hat{\varepsilon} * (k_1)U(p_1)}{9(p_2 - p_4)^2\left((p_2 - p_3 - p_4)^2 - m_q^2\right)},$$

$$M_6 = \frac{ee_q g_s^2 T_{Col2Col4}^{Glu6} T_{Col3Col1}^{Glu6} \bar{\upsilon}(p_2)\gamma\upsilon(p_4)\bar{U}(p_3)\hat{\varepsilon} * (k_1)\left(\hat{k}_1 + \hat{p}_3 + m_q\right)\gamma U(p_1)}{9(p_4 - p_2)^2\left((-k_1 - p_3)^2 - m_q^2\right)},$$

$$M_7 = \\ = \frac{ee_q g_s^2 T_{Col2Col1}^{Glu6} T_{Col3Col4}^{Glu6} \bar{U}(p_3)\gamma\upsilon(p_4)\bar{\upsilon}(p_2)\gamma\left(-\hat{p}_2 + \hat{p}_3 + \hat{p}_4 + m_q\right)\hat{\varepsilon} * (k_1)\gamma U(p_1)}{9(p_3 + p_4)^2\left((p_2 - p_3 - p_4)^2 - m_q^2\right)},$$

$$M_8 = \frac{-ee_q g_s^2 T_{Col2Col1}^{Glu6} T_{Col3Col4}^{Glu6} \bar{U}(p_3)\gamma\upsilon(p_4)\bar{\upsilon}(p_2)\hat{\varepsilon} * (k_1)\left(\hat{k}_1 - \hat{p}_2 + m_q\right)\gamma U(p_1)}{9(p_3 + p_4)^2\left((p_2 - k_1)^2 - m_q^2\right)}.$$

## APPENDIX 2.
## THE FEYNMAN DIAGRAMS AND MATRIX ELEMENTS OF SUBPROCESS 2. $q\bar{q} \to gg\gamma$

The Feynman diagrams of the radiative correction 2. $q\bar{q} \to gg\gamma$ are shown in Figure A2(a,b,c,d,e,f,g).

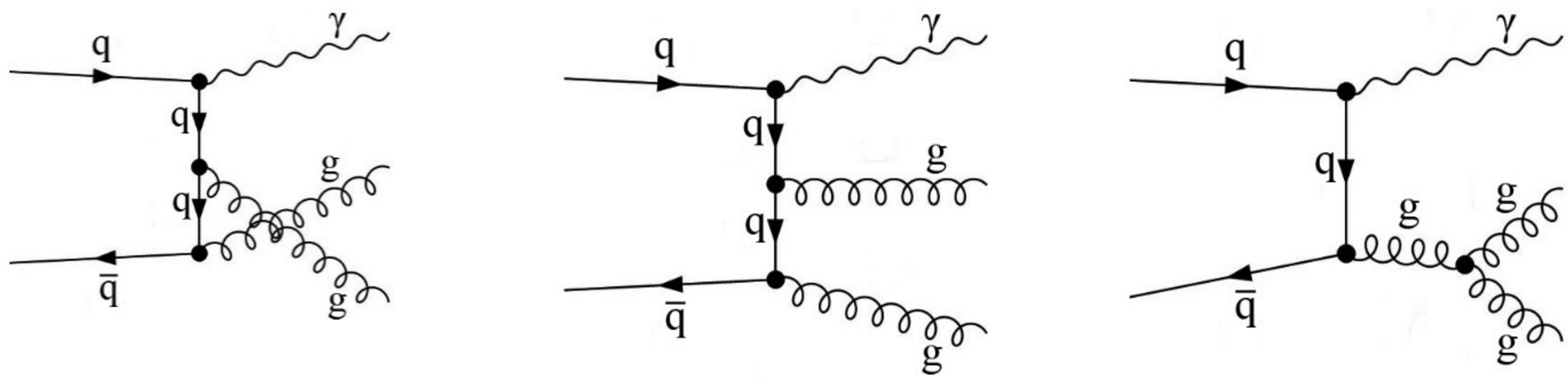

a b c

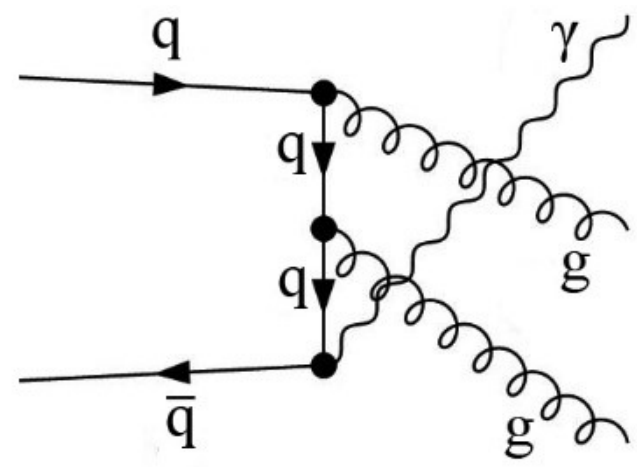

d

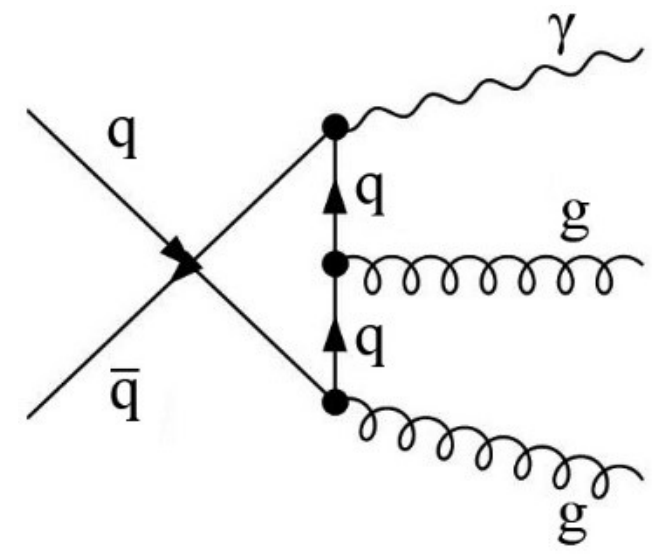

e

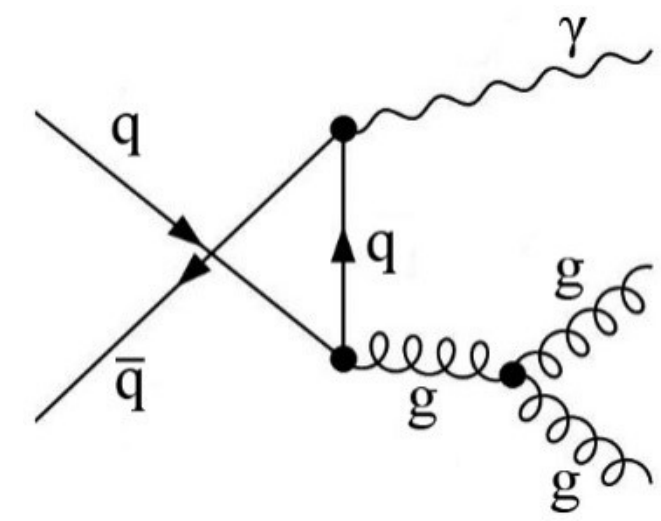

f

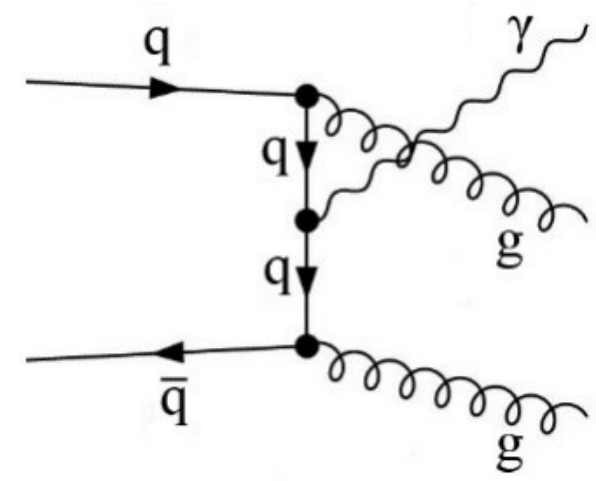

g

Figure A2(a,b,c,d,e,f,g). The Feynman diagrams of subprocess 2. $q\bar{q} \rightarrow gg\gamma$

The Mandelstam invariants for the subprocess $q\bar{q} \rightarrow gg\gamma$ are given below:

$$\hat{s} = (p_1 + p_2)^2 = (k_1 + k_2 + k_3)^2, \hat{t} = (p_1 - k_3)^2 = (k_2 + k_1 - p_2)^2,$$
$$\hat{u} = (p_2 - k_3)^2 = (k_2 + k_1 - p_1)^2, \hat{q}_1 = (p_1 - k_1)^2 = (k_3 + k_2 - p_2)^2,$$
$$\hat{q}_2 = (p_2 - k_2)^2 = (k_3 + k_1 - p_1)^2, \hat{s}_1 = (k_2 + k_1)^2 = (p_1 + p_2 - k_3)^2,$$
$$\hat{t}_1 = (p_1 - k_2)^2 = (k_3 + k_1 - p_2)^2, \hat{u}_1 = (p_2 - k_1)^2 = (k_3 + k_{21} - p_1)^2,$$
$$\hat{q}_3 = (k_1 + k_3)^2 = (p_1 + p_2 - k_2)^2, \hat{q}_4 = (k_2 + k_3)^2 = (p_1 + p_2 - k_1)^2.$$

The matrix elements for the subprocess $q\bar{q} \rightarrow gg\gamma$ have the following form:

$$M_1 = -\frac{ee_q g_s^2 T^{Glu4}_{Col2Col6} T^{Glu5}_{Col6Col1} \bar{U}(p_2)\hat{\varepsilon}*(k_2)\left(\hat{k}_2-\hat{p}_2+m_q\right)\hat{\varepsilon}*(k_3)\left(\hat{k}_2+\hat{k}_3-\hat{p}_2+m_q\right)\hat{\varepsilon}*(k_1)U(p_1)}{9\left((p_2-k_2)^2-m_q^2\right)\left((-k_2-k_3+p_2)^2-m_q^2\right)},$$

$$M_2 = -\frac{ee_q g_s^2 T^{Glu4}_{Col6Col1} T^{Glu5}_{Col2Col6} \bar{U}(p_2)\hat{\varepsilon}*(k_3)\left(\hat{k}_3-\hat{p}_2+m_q\right)\hat{\varepsilon}*(k_2)\left(\hat{k}_2+\hat{k}_3-\hat{p}_2+m_q\right)\hat{\varepsilon}*(k_1)U(p_1)}{9\left((p_2-k_3)^2-m_q^2\right)\left((-k_2-k_3+p_2)^2-m_q^2\right)},$$

$$M_3 = \frac{iee_q g_s^2 f^{Glu4Glu5Glu6}\left(k_2 \cdot \varepsilon*(k_2) + 2\left(k_3 \cdot \varepsilon*(k_2)\right)\right) T^{Glu6}_{Col2Col1} \bar{U}(p_2)\hat{\varepsilon}*(k_3)\left(\hat{k}_2+\hat{k}_3-\hat{p}_2+m_q\right)\hat{\varepsilon}*(k_1)U(p_1)}{9(k_2+k_3)^2\left((-k_2-k_3+p_2)^2-m_q^2\right)}$$

$$+$$

$$+\frac{iee_q g_s^2 f^{Glu4Glu5Glu6}\left(-2\left(k_2 \cdot \varepsilon*(k_3)\right) - \left(k_3 \cdot \varepsilon*(k_3)\right)\right) T^{Glu6}_{Col2Col1} \bar{U}(p_2)\hat{\varepsilon}*(k_2)\left(\hat{k}_2+\hat{k}_3-\hat{p}_2+m_q\right)\hat{\varepsilon}*(k_1)U(p_1)}{9(k_2+k_3)^2\left((-k_2-k_3+p_2)^2-m_q^2\right)}$$

$$+$$

$$+\frac{iee_q g_s^2 f^{Glu4Glu5Glu6}\left(\varepsilon*(k_2)\cdot\varepsilon*(k_3)\right) T^{Glu6}_{Col2Col1} \bar{U}(p_2)\left(\hat{k}_2-\hat{k}_3\right)\left(\hat{k}_2+\hat{k}_3-\hat{p}_2+m_q\right)\hat{\varepsilon}*(k_1)U(p_1)}{9(k_2+k_3)^2\left((-k_2-k_3+p_2)^2-m_q^2\right)},$$

$$M_4 = -\frac{ee_q g_s^2 T_{Col6Col1}^{Glu4} T_{Col2Col6}^{Glu5} \bar{U}(p_2)\hat{\varepsilon}*(k_1)(\hat{k}_1-\hat{p}_2+m_q)\hat{\varepsilon}*(k_3)(\hat{k}_1+\hat{k}_3-\hat{p}_2+m_q)\hat{\varepsilon}*(k_2)U(p_1)}{9((p_2-k_1)^2-m_q^2)((-k_1-k_3+p_2)^2-m_q^2)},$$

$$M_5 = -\frac{ee_q g_s^2 T_{Col2Col6}^{Glu4} T_{Col6Col1}^{Glu5} \bar{U}(p_2)\hat{\varepsilon}*(k_1)(\hat{k}_1-\hat{p}_2+m_q)\hat{\varepsilon}*(k_2)(\hat{k}_1+\hat{k}_2-\hat{p}_2+m_q)\hat{\varepsilon}*(k_3)U(p_1)}{9((p_2-k_1)^2-m_q^2)((-k_1-k_2+p_2)^2-m_q^2)},$$

$$M_6 = \frac{iee_q g_s^2 f^{Glu4Glu5Glu6}\left(k_2 \cdot \varepsilon*(k_2) + 2(k_3 \cdot \varepsilon*(k_2))\right) T_{Col2Col1}^{Glu6} \bar{U}(p_2)\hat{\varepsilon}*(k_1)(\hat{k}_1 - \hat{p}_2 + m_q)\hat{\varepsilon}*(k_3)U(p_1)}{9(k_2+k_3)^2\left((\hat{p}_2-\hat{k}_1)^2 - m_q^2\right)} +$$

$$+\frac{iee_q g_s^2 f^{Glu4Glu5Glu6}\left(-2(k_2 \cdot \varepsilon*(k_3)) - (k_3 \cdot \varepsilon*(k_3))\right) T_{Col2Col1}^{Glu6} \bar{U}(p_2)\hat{\varepsilon}*(k_1)(\hat{k}_1 - \hat{p}_2 + m_q)\hat{\varepsilon}*(k_2)U(p_1)}{9(k_2+k_3)^2\left((\hat{p}_2-\hat{k}_1)^2 - m_q^2\right)} +$$

$$+\frac{iee_q g_s^2 f^{Glu4Glu5Glu6}(\varepsilon*(k_2)\cdot\varepsilon*(k_3)) T_{Col2Col1}^{Glu6} \bar{U}(p_2)\hat{\varepsilon}*(k_1)(\hat{k}_1 - \hat{p}_2 + m_q)(\hat{k}_2 - \hat{k}_3)U(p_1)}{9(k_2+k_3)^2\left((\hat{p}_2-\hat{k}_1)^2 - m_q^2\right)},$$

$$M_7 = -\frac{ee_q g_s^2 T_{Col6Col1}^{Glu4} T_{Col2Col6}^{Glu5} \bar{U}(p_2)\hat{\varepsilon}*(k_3)(\hat{k}_3-\hat{p}_2+m_q)\hat{\varepsilon}*(k_1)(\hat{k}_1+\hat{k}_3-\hat{p}_2+m_q)\hat{\varepsilon}*(k_2)U(p_1)}{9((p_2-k_3)^2-m_q^2)((-k_1-k_3+p_2)^2-m_q^2)},$$

$$M_8 = -\frac{ee_q g_s^2 T_{Col2Col6}^{Glu4} T_{Col6Col1}^{Glu5} \bar{U}(p_2)\hat{\varepsilon}*(k_2)(\hat{k}_2-\hat{p}_2+m_q)\hat{\varepsilon}*(k_1)(\hat{k}_1+\hat{k}_2-\hat{p}_2+m_q)\hat{\varepsilon}*(k_3)U(p_1)}{9((p_2-k_2)^2-m_q^2)((-k_1-k_2+p_2)^2-m_q^2)}.$$

## APPENDIX 3.
## THE FEYNMAN DIAGRAMS AND MATRIX ELEMENTS OF SUBPROCESS 3. $q\bar{q} \to g\gamma\gamma$

The Feynman diagrams of the radiative correction for the annihilation of the quark-antiquark pair 3. $q\bar{q} \to g\gamma\gamma$ are shown in Figure A3 (a,b,c,d,e,f).

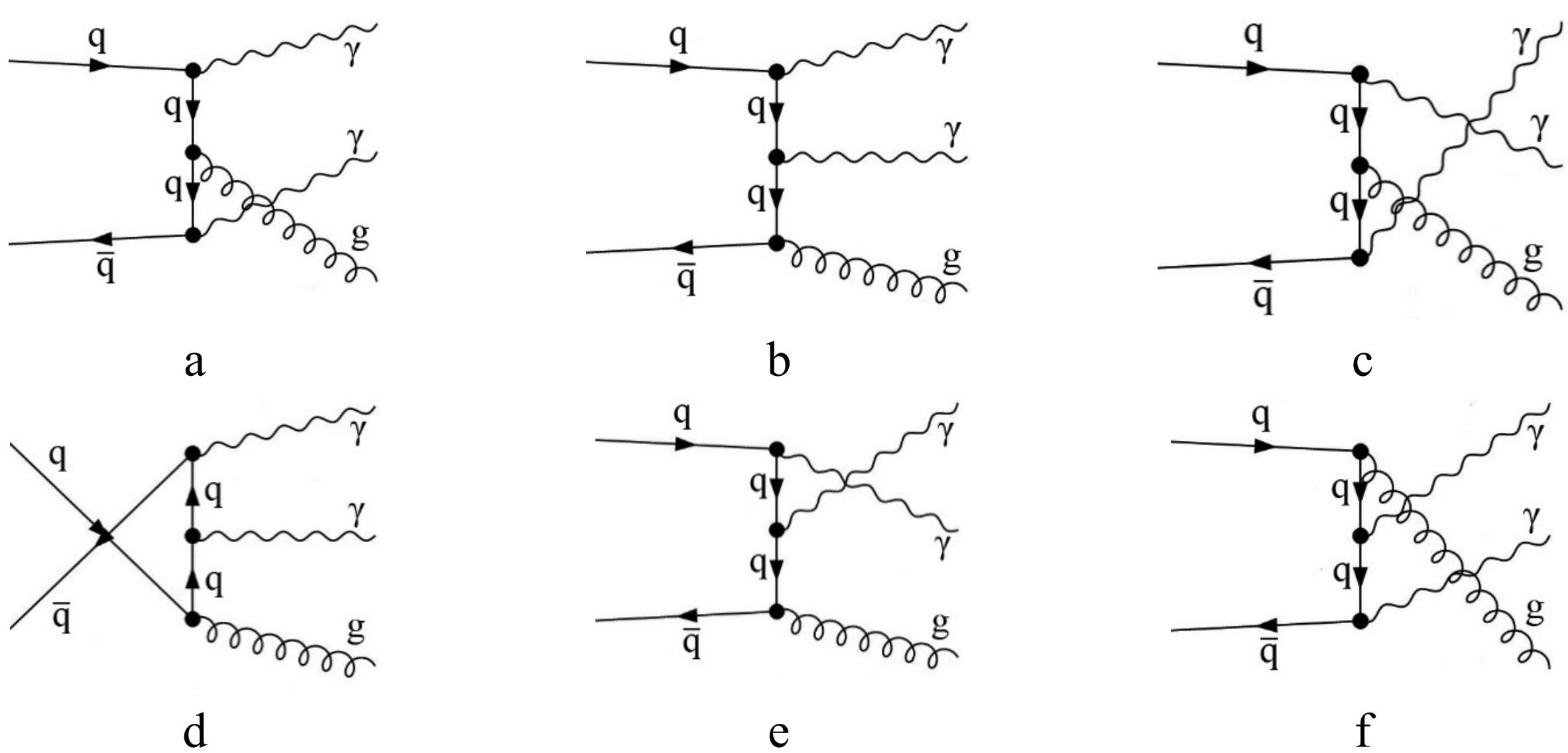

Figure A3(a,b,c,d,e,f). The Feynman diagrams of subprocess 3. $q\bar{q} \to g\gamma\gamma$

The Mandelstam invariants for the subprocess 3. $q\bar{q} \to g\gamma\gamma$ are given below:

$$\hat{s} = (p_1 + p_2)^2 = (k_1 + k_2 + k_3)^2, \hat{t} = (p_1 - k_2)^2 = (k_3 + k_1 - p_2)^2,$$
$$\hat{u} = (p_2 - k_2)^2 = (k_3 + k_1 - p_1)^2, q_1 = (p_1 - k_1)^2 = (k_3 + k_2 - p_2)^2,$$
$$q_2 = (p_2 - k_3)^2 = (k_2 + k_1 - p_1)^2, s_1 = (k_1 + k_3)^2 = (p_1 + p_2 - k_2)^2,$$
$$t_1 = (p_1 - k_3)^2 = (k_2 + k_1 - p_2)^2, u_1 = (p_2 - k_1)^2 = (k_3 + k_{21} - p_1)^2,$$
$$q_3 = (k_1 + k_2)^2 = (p_1 + p_2 - k_3)^2, q_4 = (k_2 + k_3)^2 = (p_1 + p_2 - k_1)^2.$$

The matrix elements of the subprocess 3. $q\bar{q} \to g\gamma\gamma$ have the following form:

$$M_1 = -\frac{2e^2 e_q g_s T^{Glu5}_{Col2Col1} \bar{\upsilon}(p_2)\hat{\varepsilon}^*(k_2)(\hat{k}_2 - \hat{p}_2 + m_q)\hat{\varepsilon}^*(k_3)(\hat{k}_2 + \hat{k}_3 - \hat{p}_2 + m_q)\hat{\varepsilon}^*(k_1)U(p_1)}{27((p_2 - k_2)^2 - m_q^2)((-k_2 - k_3 + p_2)^2 - m_q^2)},$$
$$M_2 = -\frac{2e^2 e_q g_s T^{Glu5}_{Col2Col1} \bar{\upsilon}(p_2)\hat{\varepsilon}^*(k_3)(\hat{k}_3 - \hat{p}_2 + m_q)\hat{\varepsilon}^*(k_2)(\hat{k}_2 + \hat{k}_3 - \hat{p}_2 + m_q)\hat{\varepsilon}^*(k_1)U(p_1)}{27((p_2 - k_3)^2 - m_q^2)((-k_2 - k_3 + p_2)^2 - m_q^2)},$$
$$M_3 = -\frac{2e^2 e_q g_s T^{Glu5}_{Col2Col1} \bar{\upsilon}(p_2)\hat{\varepsilon}^*(k_1)(\hat{k}_1 - \hat{p}_2 + m_q)\hat{\varepsilon}^*(k_3)(\hat{k}_1 + \hat{k}_3 - \hat{p}_2 + m_q)\hat{\varepsilon}^*(k_2)U(p_1)}{27((p_2 - k_1)^2 - m_q^2)((-k_1 - k_3 + p_2)^2 - m_q^2)},$$
$$M_4 = -\frac{2e^2 e_q g_s T^{Glu5}_{Col2Col1} \bar{\upsilon}(p_2)\hat{\varepsilon}^*(k_1)(\hat{k}_1 - \hat{p}_2 + m_q)\hat{\varepsilon}^*(k_2)(\hat{k}_1 + \hat{k}_2 - \hat{p}_2 + m_q)\hat{\varepsilon}^*(k_3)U(p_1)}{27((p_2 - k_1)^2 - m_q^2)((-k_1 - k_2 + p_2)^2 - m_q^2)},$$
$$M_5 = -\frac{2e^2 e_q g_s T^{Glu5}_{Col2Col1} \bar{\upsilon}(p_2)\hat{\varepsilon}^*(k_3)(\hat{k}_3 - \hat{p}_2 + m_q)\hat{\varepsilon}^*(k_1)(\hat{k}_1 + \hat{k}_3 - \hat{p}_2 + m_q)\hat{\varepsilon}^*(k_2)U(p_1)}{27((p_2 - k_3)^2 - m_q^2)((-k_1 - k_3 + p_2)^2 - m_q^2)},$$
$$M_6 = -\frac{2e^2 e_q g_s T^{Glu5}_{Col2Col1} \bar{\upsilon}(p_2)\hat{\varepsilon}^*(k_2)(\hat{k}_2 - \hat{p}_2 + m_q)\hat{\varepsilon}^*(k_1)(\hat{k}_1 + \hat{k}_2 - \hat{p}_2 + m_q)\hat{\varepsilon}^*(k_3)U(p_1)}{27((p_2 - k_2)^2 - m_q^2)((-k_1 - k_2 + p_2)^2 - m_q^2)}.$$